\documentclass[12pt]{article}
\usepackage{natbib}
 \bibpunct{(}{)}{;}{a}{,}{,}
\usepackage{graphicx}
\usepackage{amsmath,amssymb,amsthm}
\usepackage{bm}
\usepackage{rotating}
\usepackage{rotating}
\usepackage{multicol}
\usepackage{dsfont}
\usepackage{titlesec}
\usepackage{latexsym}
\usepackage{changepage}
\usepackage{multirow}
\usepackage{booktabs}
\usepackage{pbox}
\usepackage[affil-it]{authblk}
 \usepackage[latin1]{inputenc}
\usepackage[titletoc]{appendix}
\usepackage{float}
\usepackage{caption}
\usepackage[linktoc=page,colorlinks,linkcolor=blue,citecolor=blue,urlcolor=blue]{hyperref}
\usepackage{fancyhdr}
\usepackage{setspace}
\usepackage[top=1in, bottom=1in, left=1in, right=1in]{geometry}
\usepackage{enumitem}
\usepackage[normalem]{ulem}
\usepackage[table]{xcolor}
\usepackage{color}

\newcommand{\bv}{\begin{array}}
\newcommand{\ev}{\end{array}}
\newcommand{\bit}{\begin{itemize}}
\newcommand{\eit}{\end{itemize}}
\newcommand{\ben}{\begin{enumerate}}
\newcommand{\een}{\end{enumerate}}
\newcommand{\beq}{\begin{equation}}
\newcommand{\eeq}{\end{equation}}
\newcommand{\bvq}{\begin{eqnarray}}
\newcommand{\evq}{\end{eqnarray}}

\newcommand{\E}{\mathbb{E}}

\allowdisplaybreaks

\makeatletter
\newenvironment{subtheorem}[1]{%
  \def\subtheoremcounter{#1}%
  \refstepcounter{#1}%
  \protected@edef\theparentnumber{\csname the#1\endcsname}%
  \setcounter{parentnumber}{\value{#1}}%
  \setcounter{#1}{0}%
  \expandafter\def\csname the#1\endcsname{\theparentnumber.\Alph{#1}}%
  \expandafter\def\csname theH#1\endcsname{thm.\theparentnumber.\Alph{#1}}%
}{%
  \setcounter{\subtheoremcounter}{\value{parentnumber}}%
  \ignorespacesafterend
}
\makeatother
\newcounter{parentnumber}

\theoremstyle{plain}
\newtheorem{theorem}{Theorem}

\newtheorem{assumption}{Assumption}

\newtheorem{cor}{Corollary}[theorem]
\newtheorem{prop}[theorem]{Proposition}

\newcommand{\N}{\mathcal{N}}

\newcommand{\vZ}{\textbf{Z}}
\newcommand{\vY}{\textbf{Y}}
\newcommand{\vX}{\textbf{X}}
\newcommand{\vx}{\textbf{x}}

\renewcommand\refname{}
\newcommand{\ind}{\perp\!\!\!\perp}
\newcommand{\nind}{\not\!\perp\!\!\!\perp}


\begin{document}
\pagestyle{empty}

\title{Identification and estimation of treatment and interference effects in observational studies on networks}
 \author{Laura Forastiere$^*$}
 \author{Edoardo M. Airoldi$^\dagger$}
 \author{Fabrizia Mealli$^\ddagger$}
 \affil{$^*$Yale University, $^\dagger$Harvard University and $^\ddagger$University of Florence}
\date{}

{\makeatletter\renewcommand*{\@makefnmark}{}
\footnotetext{Laura Forastiere is a postdoctoral fellow at Yale University in the Yale Institute for Network Science (laura.forastiere@yale.edu). Edoardo M. Airoldi is an Associate Professor of Statistics at Harvard University (airoldi@fas.harvard.edu). Fabrizia Mealli is a Professor of Statistics at the University of Florence (mealli@disia.unifi.it).
 We thank Guido W. Imbens and Donald B. Rubin for helpful discussions.
 This work was supported, in part, by NSF awards CAREER IIS-1149662 and IIS-1409177, and by ONR awards YIP N00014-14-1-0485 and N00014-17-1-2131.}
%
% EMA is an Alfred P. Sloan Research Fellow and a Shutzer Fellow at the Radcliffe Institute for Advanced Study.}
\makeatother}

\maketitle
\thispagestyle{empty}
\begin{abstract}
Causal inference on a population of units connected through a network often presents technical challenges, including how to account for  interference. In the presence of local interference, for instance, potential outcomes of a unit depend on its treatment as well as on the treatments of other local units, such as its neighbors according to the network. In observational studies, a further complication is that the typical unconfoundedness assumption must be extended---say, to include the treatment of neighbors, and individual and neighborhood covariates---to guarantee identification and valid inference. Here, we propose new estimands that define treatment and interference effects. We then derive analytical expressions for the bias of a naive estimator that wrongly assumes away interference. The bias depends on the level of interference but also on the degree of association between individual and neighborhood treatments. We propose an extended unconfoundedness assumption that accounts for interference, and we develop new covariate-adjustment methods that lead to valid estimates of treatment and interference effects in observational studies on networks. Estimation is based on a generalized propensity score that balances individual and neighborhood covariates across units under different levels of individual treatment and of exposure to neighbors' treatment.
We carry out simulations, calibrated using friendship networks and covariates in a nationally representative longitudinal study of adolescents in grades 7-12, in the United States, to explore finite-sample performance in different realistic settings.
%\cngL{Finite sample performances in different scenarios are shown in a simulation study leveraging the friendship nomination data of Add Health, a  longitudinal study of a nationally representative sample of adolescents in grades 7-12 in the United States during the 1994-95 school year}
\bigskip
%\vfill

\end{abstract}

%% TABLE OF CONTENTS
%\singlespacing
%\newpage
%\tableofcontents
%%
%%% \setcounter{tocdepth}{4}
%%% 1 Section
%%% 2 Subsection
%%% 3 Subsubsection
%%% 4 Paragraph
%%% 5 Subparagraph
%
%\onehalfspacing

 %%% %%% %%%
 %%% %%% %%%
 %%% %%% %%%

%\newpage

\pagestyle{fancy}
\setcounter{page}{1}

\section{Introduction}
\label{sec:intro}

\subsection{Motivation}
Many experimental and observational studies are affected by the
presence of interference among units. \emph{Interference} is said
to be present when a treatment, exposure or intervention, on one
unit has an effect on the response of another unit
\citep{Cox:1958}. This phenomenon can be due to social or physical
interaction among units. For instance, widespread use of
preventive measures for infectious diseases, such as bed nets for
malaria control or vaccines for other diseases, may benefit
unprotected individuals
% because they are less likely to come into contact with infected individuals
by reducing the reservoir of infection and by affecting the vector of transmission
\citep{Binka:1998, Howard:2000, Hawley:2003}. In education,
students assigned to attend a tutoring program might interact with other students
 who were not assigned to the program and spillover effects
 may arise  thanks to
 the transmission
of knowledge among peers.
In social media, advertisements are shared by users
of the same network or the behavior of a single user might affect
other users. The effects of interference are typically referred to
as \emph{spillover effects}, in economics, or as \emph{peer
influence effects}, in social sciences. A unit's treatment might
affect other units' outcomes through a variety of mechanisms: the
spread of the treatment; through a change in its own outcome
which, in turn, affect the outcomes of other units; or through
other pathways involving intermediate variables. Regardless of the
mechanism through which it takes place, this dependence between
units' treatments and outcomes poses statistical challenges,
because potential outcomes for each unit must be indexed also by
the treatment received by other units.

When the target estimand is the average (individual) treatment effect, interference is seen as a nuisance and has to be taken into account, or dealt with in some way, to prevent bias \citep[e.g., see][]{Savje:2017}.
However, scientists and people making decisions about policies, interventions, products, and campaigns are also interested in the extent to which individuals are influenced by others, in order to leverage or prevent mechanisms of interference.
In settings with limited resources, beneficial spillover effects could be leveraged to increase the cost-effectiveness of an intervention. Similarly, marketers are interested in exploiting how specific individuals can have a large influence on their friends' decisions to promote the purchase of some products. On the contrary, when spillover effects are detrimental, researchers are interested in investigating the extent to which and how interference play a role, in order to preserve the effectiveness of the program.

In this paper, we consider the problem of estimating both the
causal effect of the individual treatment and the spillover effect, in
situations where interference occurs between units that are
related on a known network and the assignment mechanism of the treatment is not known.
 Our first aim is to formulate the problem under the potential outcome framework \citep{Rubin:1974,Rubin:1986} and to define causal estimands that are best suited for network data. We then provide identifying assumptions and discuss their plausibility in dependent data.
Our second aim is to quantify the bias of a naive estimator that neglects the presence of interference. 
Understanding the sources of bias is crucial in both the study design phase when information on connections is not easy to collect and in the final stage of drawing conclusions from the results.
Our third aim is to tackle the confounding problem of observational studies in networks 
by extending the propensity score approach to settings with interference from first order neighbors.
%by providing on the a clear definition of the propensity score under interference and showing its  balancing properties.
%We then reelaborate the propensity score approach under interference.
By revealing the balancing properties of this generalized propensity score we lay the basis for developing covariate adjustment methods for the estimation of treatment and spillover effects in observational network data. 
Finally, we rely on these results to propose a propensity score-based method that builds upon the literature on generalized propensity scores for continuous treatments \citep{Hirano:Imbens:2004}.

\subsection{Related work}

When assessing causal effects, the standard approach is to rule out the presence of interference.
The assumption of no interference is also called Individualistic Treatment Response (ITR)  \citep{Manski:2013}, or, combined with the assumption of no hidden versions of treatment, it is referred to as the Stable Unit Treatment Value Assumption (SUTVA) \citep{Rubin:1980}.
In many situations, however, no interference may not be plausible.
The risk of this assumption is demonstrated by \citet{Sobel:2006}. Examining randomized housing mobility studies in which households in poor areas are financed to relocate to better neighborhoods,
Sobel shows that ignoring interference can lead to entirely wrong conclusions about the effectiveness of the program. In fact, the observed mean difference between treated and control units can be decomposed into a total effect,i.e., the sum of the direct effect of receiving the treatment and the spillover effect from other units' treatment, for treated units and a spillover effect for control units.

%When the target estimand is the average (individual) treatment effect, interference is seen as a nuisance and has to be taken into account to prevent bias.
%However, scientists and people making decisions about policies, interventions, products, and campaigns are also interested in the extent to which individuals are influenced by others, in order to leverage or prevent mechanisms of interference.
%In settings with limited resources, beneficial spillover effects could be leveraged to increase the cost-effectiveness of an intervention. Similarly, marketers are interested in exploiting how specific individuals can have a large influence on their friends' decisions to promote the purchase of some products. On the contrary, when spillover effects are detrimental, researchers are interested in investigating the extent to which and how interference play a role, in order to preserve the effectiveness of the program.

In both cases, when spillover effects are seen as an inconvenience for the estimation of the treatment effect or when they are the target estimands, different strategies to get unbiased estimates of either the treatment or the spillover effects have been proposed, in both the design and the estimation phases.
These strategies depend on the extension of the interference mechanism.
When individuals can be partitioned into groups, it is often plausible to assume that interference occurs within groups but not across groups. This assumption is referred to as \emph{partial interference} \citep{Sobel:2006}. Under this assumption, cluster randomized trials are sometimes designed to rule out the presence of interference between treated and untreated clusters. Furthermore, to estimate both the treatment and spillover effects, one possible design is a sequential two stage randomization, where in the first stage groups are randomized to different treatment allocation strategies and in the second stage individuals are randomized to treatment or control conditional on the strategy assigned to their group in the first stage \citep{Hudgens:Halloran:2008}.
Some more recent work has examined interference when units are connected along more complicated network structures. Design strategies involve rearranging assignment of treatment to subjects in a manner that, incorporating information on network connectivity, is able to make a trade-off between bias and variance
\citep{Toulis:Kao:2013, Ugander:2013, Eckles:2014}.

With regard to inference strategies, there are various suggestions
on how to conduct causal inference under interference for
randomized experiments. \citet{Rosenbaum:2007} uses randomization
inference to both test and estimate the total treatment effect,
that is, the sum of treatment and spillover effect, even in the
presence of interference. \citet{Bowers:2013} and
\citet{Aronow:2012} propose the use of randomization inference to
separately test for interference. Building on these works,
\citet{Athey:2015} provide a general framework that applies to a
much larger class of non-sharp null hypotheses.
\citet{Aronow:Samii:2013} perform inference using the
Horvitz-Thompson estimator, which requires that inclusion
probabilities for complex network sampling designs can be
computed.

The literature on spillover effects is also rapidly evolving for observational studies.
%has only recently started to develop
However, most of the recently proposed methods rely on the assumption of partial interference.
\citet{Hong:Raudenbush:2006}, who evaluate the effect of grade
retention of low-achieving children on test scores, use a
parametric multilevel approach to mimic a two-stage experiment and
base their analysis on the assumption that 
%any possible peer
%effect can be summarized by a scalar function of the vector of
%treatment assignments within a school. 
the extent to which each child is affected by the retention of other
children only depends on whether they are exposed to a low versus high proportion
of kindergartners in the school.
Another similar example is
found in \citet{Verbitsky-Savitz:Raudenbush:2012}, evaluating the
effects of community policing program on neighborhoods crime rates
using a three-level generalized hierarchical linear model.
Building on the work by \citet{Hudgens:Halloran:2008},
\citet{TchetgenTchetgen:VanderWeele:2012} proposed
inverse-probability-weighted estimators (IPW) for treatment
effects also in the presence of partial interference.
\citet{Perez-Heydrich:2014} showed the performance of these IPW
estimators using simulation studies and applied them to analyze an
individually-randomized, placebo-controlled trial of cholera
vaccination. 

On the other hand, work on causal inference for observational network data is still in its infancy.
\cite{Liu:2016} extend these IPW estimators in the
presence of general forms of interference on networks. They focus
on direct and indirect effects based on Bernoulli allocation
strategies similar to those defined by
\citet{Hudgens:Halloran:2008}. Although their estimators allows
for general forms of interference, their estimator and their asymptotic results are
proved under partial interference and clustered data. 
 \citet{VanderLaan:2014} and then \citet{Sofrygin:vanderLaan:2017} propose a TMLE estimator for
similar treatment and spillover effects and prove asymptotic
results under IID assumptions.
Finally \citet{Ogburn:2017} extend this TMLE estimator to allow
for dependence due to both contagion and homophily and derive
asymptotic results that allow the
number of ties per node to increase as the network grows.

\subsection{Contributions}
%In this paper, we consider the problem of estimating both the
%causal effect of the treatment and the spillover effect, in
%situations where interference occurs between units that are
%related on a known network and data are observational. 
We start by providing a general formalization of the problem of interference in networks under the potential
outcome framework. As in previous works \citep{VanderLaan:2014, Aronow:Samii:2013}, we replace SUTVA with an
assumption that limits the propagation of treatment to immediate
neighbors, ruling out the influence by neighbors' of neighbors,
and simplifies the mechanism of interference, in that only a
summary of the neighbors' treatment vector matters. This
assumption will reduce the number of possible potential outcomes,
which are indexed only by the treatment received by the unit, the
individual treatment, and a summary of the treatment received by
his neighbors, the neighborhood treatment. Under this framework,
we provide new causal estimands
for treatment and spillover effects. 
We first define causal estimands as average comparisons of potential outcomes under different values of the treatment of both the unit and his neighbors. We then marginalize these estimands over the observed distribution of the neighbors' treatment. When compared to the estimands in \citet{VanderLaan:2014} and in \citet{Liu:2016}, where the treatment vector is drawn from a hypothetical intervention, our estimands have a more descriptive nature.

We provide an
uncounfoundedness assumption that is similar to the conditional exchangeability in \citet{Liu:2016} and the randomization assumption in \citet{VanderLaan:2014}. However, our assumption is weaker because it only implies the independence of the treatments with the potential outcomes of each single unit. In addition we prove identification results and discuss the implications and the plausibility of this assumption in network settings.

Key contributions of this paper is the derivation of
bias formulas for the treatment effect when SUTVA is wrongly
assumed. Our results differ from the one provided by
\citet{Sobel:2006} in that we deal with observational studies and
we explicitly show the two key factors driving the bias: the
extent to which a unit's potential outcome is affected by the
treatment received by his neighbors and the residual association
between the individual and the neighborhood treatment after
conditioning for covariates. These results are fundamental for
applied researchers as they demystify the misconception that bias
resulting from ignoring interference solely depends on the level
of interference itself. A simulation study is used to validate the
analytical results in different scenarios, with different level of interference and different correlation between the individual and the neighborhood treatment. 

In the second part of
the paper, we
provide a clear definition of the propensity score under interference and elucidate its  balancing properties. 
Under interference on the immediate neighborhood, balance must be achieved across arms defined not only by the individual treatment but also by the neighborhood treatment. In this spirit, under the neighborhood interference, we first define the joint propensity score as the probability of assignment to a particular individual and neighborhood treatment given the observed covariates. We discuss the inclusion of different kinds of covariates representing individual characteristics, neighbors' characteristics but also network properties. We then prove the balancing properties of this joint propensity score and we show that unconfoundedness holds not only conditional on the joint propensity score but also conditional on the individual and neighborhood propensity scores.
Relying on this unconfoudedness assumption, 
we  propose a joint propensity
score-based estimator, based on subclassification on the
individual binary-treatment propensity score and parametric
adjustment for the neighborhood multivalued-treatment propensity
score. 
%\sout{This semi-parametric estimator follows directly from our
%formalization of the problem and overcomes issues about the
%variance of IPW estimators due to the problem of small weights,
%whose presence is even more relevant in such a multivariate
%setting.}

The remainder of this article is organized as follows. In Section
\ref{sec:notation}, we introduce notation and formalize the
problem by providing general definitions of estimands and
identifying assumptions. The bias for the treatment effect when we
wrongly assume no interference is derived in Section
\ref{sec:bias}. In Section \ref{sec:ps} we give a formal definition and show the balancing properties of the generalized propensity score under neighborhood interference. 
In Section \ref{sec:estimator}  we present the
joint propensity score-based estimator for main effects and
spillover effects. The simulation study, which validates the bias
formulas and assesses the performance of our estimator, is
described in Section \ref{sec:sim}. Finally, in Section
\ref{sec:concl} we discuss our findings and highlight potential
future work. In Appendix \ref{app:condeffects} we develop propensity score-based estimators for 
conditional effects, that is, main and spillover effects among units that are observed under specific values of  individual and neighborhood treatments.
In Appendix \ref{app:dgp} we detail the data generating model for
the simulation study, leveraging Add-Health network and covariate
data. 
%Additional results of the simulation, including point
Details of the proposed estimator, the specific models used for the simulation study, as well as a proposed approach on how to conduct statistical inference, are presented in Appendix
\ref{app:addres}. Proofs of the theorems and corollaries are reported in Appendix
\ref{app:proofs}.

 %%% %%% %%%
 %%% %%% %%%
 %%% %%% %%%

\section{Interference based on the exposure to neighborhood treatment}
\label{sec:notation}

\subsection{Notation}

Let us denote an undirected network $G$ as a pair $(\N,\E)$, where $\N$ is a set of nodes (units) with cardinality $N$, and $\E$ is a set of edges with generic element $(i,j)=(j,i)$, which represents the presence of a link between unit i and unit j.
This network $G$ is our population of interest.
%\sout{of which we may only observe a random sample of cardinality $M\ll N$.}

Let us define a partition of the set $\N$ around node $i$ as $(i, \N_i, \N_{-i})$, where the set $\N_i$ has cardinality $N_i$ and contains all nodes $j$ connected to $i$ by an edge in $\E$, i.e., the {\em neighboring} nodes of $i$ in $G$ (e.g. friends or neighbors), and the set $\N_{-i}$ contains all nodes other than $i$ that are not in $\N_i$. $\N_i$ is referred to as the \textit{neighborhood} of unit i, regardless of whether the presence of edges is defined by physical proximity, friendship or any other type of relationship. For consistency with the literature of social networks, the number of neighbors $N_i$ is referred to as \textit{degree} of unit i.
It is worth noting that our proposed formulation and methods can be easily extended to directed networks.
Moreover, here we assume connections to be fixed and known. This assumption is more plausible when the type of relationship underlying the mechanism of interference is a concrete and objective concept, and information on connections between units can be objectively gathered from available databases. An example of objective relationship is the one defined by geographic proximity, membership to the same group, collaboration or networking between firms or units, friendship on social media, etc. However, real world networks are often uncertain, as social interactions between individuals may be either unobservable, or measured with error. Here we assume that information on social ties is correctly measured. Extensions to include uncertainty in the social network are possible, but are beyond the scope of this paper.

Let now $Z_i\in\{0,1\}$ be a binary variable representing the treatment assignment to unit i and $Y_i\in\mathcal{Y}$ the observed outcome of unit $i$. $\vZ$ and $\vY$ are the corresponding vectors in the whole population $\mathcal{N}$.  For each unit $i$, the partition $(i, \N_i, \N_{-i})$ defines the following partitions of the treatment and outcome vectors: $(Z_i,\vZ_{\mathcal{N}_i},\vZ_{\mathcal{N}_{-i}})$ and $(Y_i,\vY_{N_i},\vY_{N_{-i}})$.
Finally, let $\vX_i\in\mathcal{X}$ be a vector of $K$ covariates for unit $i$. In social network data, $\vX_i$ can be decomposed into two subvectors: individual covariates, denoted with $\vX_i^{ind}\in\mathcal{X}^{ind}$, and neighborhood covariates, denoted with $\vX_i^{neigh}\in\mathcal{X}^{neigh}$. $\vX_i^{ind}$ denotes the set of individual-level characteristics (e.g., demographic factors, socio-economic factors, health status, ...) or contextual covariates (e.g., environment socio-economic factors, geographical factors, ...). Conversely, $\vX_i^{neigh}$ may include three types of neighborhood-level covariates: variables representing the structure of the neighborhood $\N_i$ (the degree $N_i$, the topology, etc.), network properties at node-level representing the position of unit's neighborhood in the graph (e.g., centrality, betweenness,the number of shared neighbors, ... ), and aggregational covariates as functions of  individual-level covariates in the neighborhood, i.e., $X_{ik}^{neigh}=h_{ik}(\vX^{ind}_{k\mathcal{N}_i})$., where $\vX^{ind}_{k\mathcal{N}_i}$ is the vector of the k individual characteristic in the neighborhood of unit i and $h_{ik}(\cdot)$ is a summarizing function.
%$h_i: (\mathcal{X}^{ind})^{N_i}\rightarrow \mathcal{H}_i$
%summarizing the matrix $\vX^{ind}_{\mathcal{N}_i}$ into a vector of dimension $|\mathcal{H}_i|<|\mathcal{X}^{ind}|\times N_i$.
For instance, if the covariate $X_{ik}^{ind}$ is sex of unit i then $X_{ik}^{neigh}$ can be the proportion of males among the neighbors of unit i. Similarly, if the covariate $X_{ik}^{ind}$ is income of unit i then $X_{ik}^{neigh}$ can be the average income among the neighbors of unit i. $h_{ik}(\cdot)$ could also be a function of comparison between the covariate If the covariate  $X_{ik}^{ind}$ of unit i and the same covariate for his neighbors, e.g. if $X_{ik}^{ind}$ is town of residence of unit i then $X_{ik}^{neigh}$ can be the number of friends living in the same town.
With the distinction between $\vX_i^{ind}$ and $\vX_i^{neigh}$ we want to emphasize that in network settings we need to include covariates representing the structure of the neighborhood or the type of neighbors, i.e., $\vX_i^{neigh}$, that are not usually taken into account in settings without interference or in settings with partial interference.

\subsection{Potential outcomes and neighborhood interference}

Here, we extend the potential outcomes notation to include the presence of (network) interference.
In general, the outcome observable at node $i$ is a function of the entire treatment assignment vector $\vZ$ and can be written as $Y_i(\vZ)$. As pointed out by \citet{Rubin:1986}, this potential outcome is well defined only if the following assumption holds:
\begin{assumption}[No Multiple Versions of Treatment (Consistency)]
\label{ass: consistency}
%\[Y_i(\vZ)=Y_i(\vZ') \!\!\!\!\! \tag*{$\forall \vZ, \vZ': \vZ=\vZ'$}\]
\[Y_i=Y_i(\vZ)\]
This says that the mechanism used to assign the treatments does not matter and assigning the treatments in a different way does not constitute a different treatment.
\end{assumption}
This assumption is the first component of a fundamental assumption usually made in the potential outcomes approach to causal inference: the stable unit treatment value assumption (SUTVA)  \citep{Rubin:1980, Rubin:1986}.
The second component of SUTVA is the more critical assumption of no interference between individuals \citep{Cox:1958}.
If we write the assignment vector $\vZ$ as $(Z_i,\vZ_{\mathcal{N}_i},\vZ_{\mathcal{N}_{-i}})$, this assumption states that $Y_i(Z_i,\vZ_{\mathcal{N}_i},\vZ_{\mathcal{N}_{-i}}) = Y_i(Z_i,\vZ'_{\mathcal{N}_i},\vZ'_{\mathcal{N}_{-i}})$, $\forall \,\, \vZ_{\mathcal{N}_i}, \vZ'_{\mathcal{N}_i},\vZ_{\mathcal{N}_{-i}},\vZ'_{\mathcal{N}_{-i}}$.
In contrast, in this paper we are interested in relaxing the no interference assumption and allow the existence of network interference.
Formally, we replace the no interference assumption by a new assumption of interference within the neighborhood.

\begin{assumption}[Neighborhood Interference]
\label{ass:SUTNVA}
%A function $g_i: \{0,1\}^{N_i}\rightarrow \mathcal{G}_i$ exists, such that,
Given a function $g_i: \{0,1\}^{N_i}\rightarrow \mathcal{G}_i$, $\forall i \in \N$,
$ \forall \, \vZ_{\mathcal{N}_{-i}},\vZ'_{\mathcal{N}_{-i}} $ and $\forall \, \vZ_{\mathcal{N}_i}, \vZ'_{\mathcal{N}_i}: g_i(\vZ_{\mathcal{N}_i})=g_i(\vZ'_{\mathcal{N}_i})$, the following equality holds:
\[ \qquad Y_i(Z_i,\vZ_{\mathcal{N}_i},\vZ_{\mathcal{N}_{-i}}) = Y_i(Z_i,\vZ'_{\mathcal{N}_i},\vZ'_{\mathcal{N}_{-i}})\]
\end{assumption}

Assumption \ref{ass: consistency} and Assumption \ref{ass:SUTNVA} together can be referred to as Stable Unit Treatment on Neighborhood Value Assumption (SUTNVA, pronounced {\em Sut-Ton-Va}).
Assumption \ref{ass:SUTNVA} rules out the dependence of the outcome of unit i, $Y_i$, from the treatment received by units outside his neighborhood, i.e., $\vZ_{\mathcal{N}_{-i}}$, but allows $Y_i$ to depend on the treatment received by his neighbors, i.e., $\vZ_{\mathcal{N}_i}$. Moreover, this dependence  is assumed to be through a specific function $g_i(\cdot)$. Let  us denote with $G_i$ the variable resulting from applying this function to the neighborhood treatment vector, i.e., $G_i=g_i(\vZ_{\mathcal{N}_i})$. $G_i$ can be, for instance, the simple number or the proportion of treated neighbors, i.e., $G_i=\sum_{j\in \N_i}Z_j$ or $G_i=\frac{\sum_{j\in \N_i}Z_j}{N_i}$, respectively. If we think that different neighbors can affect the outcome of unit i in a different way, then $G_i$ can also be a weighted sum of the treatment vector,  $G_i=\sum_{j\in \N_i}w_{ij}Z_j$, where the weights can be covariates or the elements of the adjacency matrix  in a weighted network. For example, the spillover effect of the treatment received by the best friends can be higher than the one of other friends. We represent this situation by making the potential outcome of unit i depend on a variable $G_i$, which results from giving more weight to the treatment of the best friends. The domain of $G_i$ will depend on how the function $g_i(\cdot)$ is defined. For example, if we consider the simple number of treated neighbors, then $\mathcal{G}=\{0,1,\ldots, N_i\}$.
Note that $G_i$ can be any function of treatment vector in the neighborhood $\vZ_{\mathcal{N}_i}$, even the identity function.
This formulation is similar to the `exposure mapping' introduced by
\citet{Aronow:Samii:2013} and the one in \citet{VanderLaan:2014}.
Here we assume the function $g_i(\cdot)$ to be known and well-specified. A discussion on consequences of misspecification of this function is beyond the scope of this paper and can be found in \citet{Aronow:Samii:2013}.

\subsection{Individual and neighborhood treatments}

Under Assumption \ref{ass:SUTNVA}, each unit is subject to
two treatments: the {\em individual treatment},  $Z_i$, and the
{\em neighborhood treatment}, $G_i$. We define as \textit{joint
treatment} the bivariate treatment to which each unit is exposed.
To fix ideas, think of a dichotomous treatment $Z_i$, such as
whether or not an individual received a vaccine, in epidemiology
applications, or a coupon, in marketing applications.
We say that node $i$ or, equivalently, experimental unit $i$, is \emph{assigned to treatment} if $Z_i=1$, and that it is \emph{assigned to control} (or to an alternative treatment) if $Z_i=0$.
We say that a node $i$ is {\em exposed to neighborhood treatment} $G_i=g$ if $g_i(\vZ_{\mathcal{N}_i})=g$.
The distinction between the two expressions referring to either assignment or exposure to a treatment is due to the different nature of the two treatments. In fact, while the individual treatment $Z_i$ can be assigned or self-selected depending on individual and neighborhood characteristics, the neighborhood treatment $G_i$ is the result of a mapping function that operates on the treatments received by the neighbors.

The assignment mechanism is then the probability distribution of
the joint treatment in the whole sample, given all covariates and
potential outcomes. Formally, the assignment mechanism can be
written as follows: $P(\vZ,\mathbf{G}|\vX, \{\vY(z,g), z=0,1; g\in
\mathcal{G}\})$, where $\vZ$ is the vector of the individual
treatments received by all the units in the sample,
$\mathbf{G}$ is the the vector of the neighborhood treatments to
which units are exposed, $\vX$ is the covariate matrix collecting all the vectors $\vX_i$ in the sample, and $\vY(z,g)$ is the collection of the potential outcomes $Y_i(z,g)$, under treatments z and g, for all units. Nevertheless, the vector of neighborhood
treatments $ \mathbf{G}$ is a function of $\mathbf{Z}$, where the
deterministic function that links the two treatment vectors
depends on the mechanism of interference, which  determines the
function $g_i(\cdot)$, and on the structure of the social network,
which determines the neighborhood of each unit and thus the
subvector $\vZ_{\mathcal{N}_i}$ to which the function $g_i(\cdot)$
is applied. As a consequence, given the social network and the
function $g_i(\cdot)$ for all units, the domain of the  joint
treatment $(\mathbf{Z}, \mathbf{G})$ has cardinality $2^N$ and it
is clearly a subset of $\{0,1\}^N \times \mathcal{G}$. The
assignment mechanism then reduces to
\begin{equation}
\label{eq: assmec}
P(\vZ,\mathbf{G}|\vX, \{\vY(z,g), z=0,1; g\in \mathcal{G}\})\!=\!
\begin{cases}
P(\vZ|\vX, \{\vY(z,g), z=0,1; g\in \mathcal{G}\}) \,\,\,\, \text{if} \,\,\, \mathbf{G}=\mathbf{g}(\vZ)\\
0 \quad \text{otherwise}
\end{cases}
\end{equation}
where $\mathbf{g}(\vZ)$ is the N-vector $[g_1(\vZ_{\mathcal{N}_1}), \ldots, g_N(\vZ_{\mathcal{N}_N})]$.
Expression \eqref{eq: assmec} for the assignment mechanism reflects the fact that the only intervention we can conceive is on the individual treatments $\vZ$,
whereas the neighborhood treatments $\mathbf{G}$ directly follow from that same intervention.

\subsection{Causal estimands: Main effects and spillover effects}
\label{sec:effects}

Under SUTNVA (Assumption \ref{ass:
consistency} and Assumption \ref{ass:SUTNVA}), potential outcomes
can be indexed only by the the individual treatment and the
neighborhood treatment, reducing to $Y_i(Z_i,G_i)$. The potential
outcome $Y_i(z,g)$ -- which is the simplified expression for
$Y_i(Z_i=z,G_i=g)$ -- represents the potential outcome of unit i
under individual treatment $Z_i=z$ and if a summary of the
treatment vector of his neighborhood, $\vZ_{\mathcal{N}_i}$,
through the function $g_i(\cdot)$, had value $g$. A potential
outcome $Y_i(z,g)$ is defined only for a subset of nodes where
$G_i$ can take on value $g$. We denote this subset by $V_g=\{i\! :g\in\mathcal{G}_i\}$, with cardinality $v_g$. For instance, in the
case where $G_i$ is the number of treated neighbors, $V_g$ is the
set of nodes with degree $N_i\geq g$, that is, with at least $g$
neighbors. 
It is worth noting that each unit can belong to different subsets $V_g$, depending on the cardinality of $\mathcal{G}_i$.

Additional assumptions are needed for the definition of potential outcomes for units with degree $N_i$ equal to zero 
%should be treated separately. 
In fact, for these kind of units the neighborhood treatment is in principle not defined
and therefore they do not belong to any subset $V_g$. 
%for them potential outcomes of the form $Y_i(z,g)$,with $g\neq0$ are not defined for these kind of units. 
We denote by $V_\emptyset$ the subsets of units without neighbors, i.e., $V_{\emptyset}=\{i\! :N_i=0\}$.
For these units we could define potential outcomes of the form $Y_i(z)$ and analyze them separetely. Otherwise, 
%potential outcomes $Y_i(z,0)$ could instead be defined with the assumption
we could make the assumption 
that units with no neighbors would exhibit outcomes as if they did
have neighbors and their neighborhood treatment were zero, i.e.,
$Y_i=Y_i(Z_i,0)$ if $N_i=0$. In this case $V_{\emptyset} \subset V_{0}$.

We take a perspective where the potential outcomes of the population $G$ of cardinality $N$ are fixed quantities and expectations are simple averages of these outcomes. This perspective is sometimes referred to as `super-population' perspective \citep{Imbens:Rubin:2015, Hernan:Robins:2017}.

With regard to the individual treatment, we first define the causal effect for a fixed level of the neighborhood treatment. Formally,
we define the (individual) \textit{treatment effect}, also called \textit{main effect}, by
\begin{equation}
\label{eq:taug}
\tau(g)= E\bigm[Y_i(Z_i=1,G_i=g) - Y_i(Z_i=0,G_i=g) | \, i \in V_g \bigm]
\end{equation}
that is, $\tau(g)$, with $g\in \mathcal{G}$, denotes the causal effect of the individual treatment when the neighborhood treatment is set to level $g$. Next, define the \textit{overall main effect} $\tau$ by the average effect of the individual treatment over the probability distribution of the neighborhood treatment, that is
\begin{equation}
\label{eq:tau}
\tau= \sum_{g \in \mathcal{G}}\tau(g) P(G_i=g)
%\tau= \sum_{g \in \mathcal{G}}\tau(g) P(G_i=g| \, i \in V_g )P(\in V_g)
\end{equation}
where $\mathcal{G}=\bigcup_i \mathcal{G}_i$ \footnote{ Under the
`super-population perspective' the expected value operator
$E[\cdot | i \in U]$, where $U$ is a subset of the
super-population, must be understood as $\frac{1}{|U|}\sum_{i \in
U} (\cdot)$. Similarly the $P(\cdot)$ operator equals
$\frac{1}{N}\sum_{i \in \mathcal{N}} I(\cdot)$
\citep{Imbens:Rubin:2015, Hernan:Robins:2017}. }. We now define
the causal effects of the neighborhood treatment, often referred
to as spillover effects or peer effects. We define the
\textit{spillover effect} of having the neighborhood treatment set
to level $g$ versus $0$, when the unit is under the individual
treatment $z$, by
\begin{equation}
\label{eq:deltagz}
\delta(g;z)= E\bigm[Y_i(Z_i=z,G_i=g) - Y_i(Z_i=z,G_i=0) | \, i \in V_g \bigm]
\end{equation}
Finally, define the \textit{overall spillover effect} $\Delta(z)$ by the average of the spillover effects $\delta(g;z)$ over the distribution of the neighborhood treatment, that is
\begin{equation}
\label{eq:Delta}
\Delta(z)= \sum_{g \in \mathcal{G}}\delta(g;z) P(G_i=g)
%\Delta(z)= \sum_{g \in \mathcal{G}}\delta(g;z) P(G_i=g| \, i \in V_g)P(i \in V_g)
\end{equation}

The main effects $\tau(g)$ in \eqref{eq:taug} and spillover effects $\delta(z;g)$ in \eqref{eq:deltagz} are average comparisons of potential outcomes under fixed values of the individual and neighborhood treatment. Differently, in the overall main and spillover effects in \eqref{eq:tau} and \eqref{eq:Delta}, the individual treatment is kept fixed while the neighborhood treatment is drawn from its observed distribution. The latter estimands 
are similar to the ones introduced in the few papers on causal inference in observational network data. However, previous work takes the average over hypothetical interventions, (e.g. general stochastic interventions in \citet{VanderLaan:2014} and its extensions or Bernoulli trial in \citet{Liu:2016})
On the contrary, our estimands in  \eqref{eq:tau} and \eqref{eq:Delta} replace the hypothetical intervention on the whole treatment vector with a hypothetical intervention that fixes the treatment of a unit i and leaves the treatment of its neighbors to the observed value. 
%For this have a more descriptive nature in that they allow one to 
Thanks to this definition, we can disentangle the total effect that we could observe on units that receive the treatment and are also exposed to the treatment. If we define \textit{total effect} as the following average comparison
\begin{equation}
\begin{aligned}
\label{eq:tot}
TE=&\sum_{g \in \mathcal{G}} E\bigm[Y_i(Z_i=1,G_i=g) - Y_i(Z_i=0,G_i=0) | \, i \in V_g \bigm] P(G_i=g)\\
%&=\sum_{g \in \mathcal{G}} E\bigm[Y_i(Z_i=1,G_i=g) - Y_i(Z_i=0,G_i=0) | \, i \in V_g \bigm] P(G_i=g)\\
%&+\sum_{g \in \mathcal{G}} E\bigm[Y_i(Z_i=1,G_i=g) - Y_i(Z_i=0,G_i=0) | \, i \in V_g \bigm] P(G_i=g)\\
%&=\tau + \Delta(0)
\end{aligned}
\end{equation}
it is straightforward to show that this is equal to the sum of the overall main and spillover effects:
\begin{equation*}
\begin{aligned}
%\label{eq:tot}
%TE=&\sum_{g \in \mathcal{G}} E\bigm[Y_i(Z_i=1,G_i=g) - Y_i(Z_i=0,G_i=0) | \, i \in V_g \bigm] P(G_i=g)\\
&=\sum_{g \in \mathcal{G}} E\bigm[Y_i(Z_i=1,G_i=g) - Y_i(Z_i=0,G_i=g) | \, i \in V_g \bigm] P(G_i=g)\\
&+\sum_{g \in \mathcal{G}} E\bigm[Y_i(Z_i=0,G_i=g) - Y_i(Z_i=0,G_i=0) | \, i \in V_g \bigm] P(G_i=g)\\
&=\tau + \Delta(0)
\end{aligned}
\end{equation*}
% It is worth noting that by averaging over the observed distribution of the neighborhood treatment, we do not rule out any observed mechanism of peer influence in the treatment receipt. 

%\footnote{\cngL{It is worth noting that even if we cannot conceive an intervention that collects all these i-specific interventions, causal estimands are well-defined.}}

Both the main and spillover effects $\tau(g)$ and $\delta(z;g)$ are based on the comparison between the marginal mean of two different potential outcomes. Formally, we denote with $\mu(z,g)$ the following quantity:
\[
 \mu(z,g)=E\big[Y_i(z,g)| \, i \in V_g\big] \qquad \forall z\in \{0,1\}, g\in\mathcal{G}
\]
that is, the marginal mean of the potential outcome $Y_i(z,g)$ in the subset $V_g$ of units where this potential outcome is well-defined. We can view
$ \mu(z,g)$ as an average dose-response function (ADRF) depending on the dose of two different treatments, i.e., the individual treatment, which is binary, and the neighborhood treatment, which is a discrete variable.

%\cngL{From now on we will for simplicity we will remove the conditioning on $V_g$ from the expected values and probabilities. We could think of a function $g(\cdot)$ such that $V_g$ coincides with the whole population $\mathcal{N}$ for each value $g$.}

\subsection{Unconfoundedness of the joint treatment}
\label{sec:unconf}

Because the causal effects of interest depend on the comparison between two quantities $\mu(z,g)$ with different values of the joint treatment, identification results can focus on the identification of the ADRF $\mu(z,g)$.
In the presence of interference, the typical unconfoundedness assumption for identification of causal effects under SUTVA must be restated. In particular, under SUTVNA the unconfoundedness
assumption must be defined for both the individual and neighborhood treatment.
\begin{assumption}[Unconfoundedness of Individual and Neighborhood Treatment]
\label{ass: Totunconf}
\[
Y_i(z,g) \ind Z_i,G_i \mid  \vX_i \qquad \forall z\in \{0,1\}, g
\in \mathcal{G}_i, \forall i.
\]
This assumption states that the individual and neighborhood treatments are independent of the potential outcomes of unit $i$, conditional on the vector of covariates $\vX_i$.
\end{assumption}

\begin{theorem}[Identification of ADRF]
\label{theo: identification}
Under Assumption \ref{ass: consistency} (No Multiple Versions of Treatment), Assumption \ref{ass:SUTNVA} (Neighborhood Interference) and Assumption \ref{ass: Totunconf} (Unconfoundedness), we have
%\sout{an unbiased estimator of $\mu(z,g)$ can be obtained by an unbiased estimator of the conditional outcome mean, that is,}
 \begin{equation}
 \label{eq:identification}
 E\big[Y_i(z,g)| \, i \in V_g\big]=\sum_{\vx \in \mathcal{X }}E\bigm[Y_i| Z_i=z, G_i=g, \vX_i=\vx, i \in V_g]P(\vX_i=\vx \, | \, i \in V_g)
% \overline{Y}^{obs}_{z, g}=\sum_{\vx \in \mathcal{X }}E\bigm[Y_i| Z_i=z, G_i=g, \vX_i=\vx, i \in V_g]P(\vX_i=\vx \, | \, i \in V_g)
 \end{equation}
%\sout{which can be estimated from the observed data.}
\end{theorem}
\textit{Proof}. See Appendix \ref{app:proofs}.

For convenience we denote the conditional outcome mean in the second term of Equation \eqref{eq:identification} by $\overline{Y}^{obs}_{z, g}$.
Theorem \ref{theo: identification} implies that the average potential outcome $Y_i(z,g)$ among the subset of units $V_g$ is equal to the weighted average of the observed outcomes of units with $Z_i=z$ and $G_i=g$ and with the same values of covariates, i.e. $\overline{Y}^{obs}_{z, g}$.
If the population at hand $\N$ is actually the population of interest we can easily compute the latter average. Therefore, we can obtain an unbiased estimate of the dose-response function $ \mu(z,g)$, for all values of $z$ and $g$, and consequently all the causal effects of interest, namely the main effects and the spillover effects. 
%On the other hand, if the population at hand is a sample, of size $M \ll N$, of the population of interest $\N$, 
In case we only have a sample, of size $M \ll N$, of the population of interest $\N$,
$\overline{Y}^{obs}_{z, g}$ must be estimated form the sample. In this case,  Theorem \ref{theo: identification} implies that an unbiased estimator of $\mu(z,g)$ can be obtained by an unbiased estimator of the conditional outcome mean $\overline{Y}^{obs}_{z, g}$.

Assumption \ref{ass: Totunconf} does not specify the entire
assignment mechanism $p(\mathbf{Z}, \mathbf{G} |
\{\mathbf{Y}(z,g),  z\in {0,1},g \in \mathcal{G} \}, \mathbf{X})$,
and thus has no implications on the independence between the
treatment assignments or between the potential outcomes of
different units. Therefore,  Assumption \ref{ass: Totunconf} may
hold irrespective of the independence between the treatment
assignments or between the potential outcomes of different units.

In this regard, it is worth discussing the plausibility of
Assumption \ref{ass: Totunconf} in network settings, because certain dependences may
cause the assumption to fail.
Assuming the unconfoundedness of the joint treatment means
assuming that the vector $\vX_i$ contains all the potential
confounders of the relationship between the joint treatment and
the potential outcomes for a specific unit i. The plausibility of
this assumption depends on how the vector $\vX_i$ is defined in
relation to the assignment mechanism.

Assumption \ref{ass: Totunconf} rules out the presence of latent variables (not included in $\vX_i$) that affect both the individual treatment and/or the neighborhood treatment and the potential outcome of a specific unit i. This has several implications.
For one, in principle the assumption does not rule out the presence of homophily, that is tendency of individuals who share similar characteristics to form ties. In fact, homophily does not violate the unconfoudedness assumption in the cases where characteristics driving the homophily mechanism i) are included in $\vX_i$, ii) even if unobserved they do not affect the outcomes iii) they correspond to  treament, that is people who share the same treatment/exposure variable tend to form ties. The only situation where homophily is a threat to identification is when variables underlying the network formation process  are not included in $\vX_i$ and affect the outcome.

In addition Assumption \ref{ass: Totunconf}  has implications on the correlation of units' potential outcomes.
Consider the case where $Y_i(z,g)$ are correlated with the vector
of potential outcomes in his neighborhood, i.e., $\vY_{\N_i}(z,g)$,
and depend on the vector of covariates $\vX_i$: this does not pose
in general identification threats, unless $\vY_{\N_i}(z,g)$ is associated with 
$G_i$ after conditioning on $\mathbf{X}_i$. For
example, 
%if the neighborhood treatment $G_i$ depends  factors that
%are not included in $\mathbf{X}_i$,  such as 
second wave
covariates, may be associated
with both $G_i$ and $Y_{\N_i}(z,g)$ and therefore their omission may induce association between 
$G_i$ and $Y_{i}(z,g)$ trough the dependence between $Y_{i}(z,g)$ and $Y_{\N_i}(z,g)$.
 In this case, Assumption \ref{ass:
Totunconf} would be violated. 
The dependence of $G_i$ from second
wave covariates may however be diluted in the application of two
summarizing function, i.e. $h_i(\cdot)$ and $g_i(\cdot)$, as well
as in the link function that relates the probability of the
individual treatment and the neighborhood covariates. 
Therefore,
in many circumstances, we believe that, depending on the choice of
$\mathbf{X}_i$, Assumption \ref{ass: Totunconf} may plausibly
hold, at least approximately, with dependent outcomes.
If we instead consider the case where potential outcomes of neighboring units are independent,
the dependence of  $\vY_{\N_i}(z,g)$ with second wave covariates do not obviously pose identification issues.

\subsection{Conditional main and spillover effects}

In Section \ref{sec:effects} and \ref{sec:unconf} we have defined marginal causal effects and provided identification results for the marginal mean of potential outcomes, the ADRF.  Given the bivariate and multi-valued nature of the joint treatment, the estimation of these marginal quantities poses some challenges. In fact, for most of the units in the sample we do not observe neither of the two potential outcomes involved in the effect of interest. We will see in Section \ref{sec:estimator} one possible estimation approach.
Nevertheless, we could focus on a specific set of units who exhibits specific values of either the individual or the neighborhood treatment and for whom the observed outcome corresponds to one of the two potential outcomes of interest. For instance, for the estimation of the effect of the individual treatment we might focus on units for whom the neighborhood treatment is zero, i.e., $G_i=0$. Causal effects defined on such a subpopulation of units require weaker identifying conditions and can be estimated using methods for binary treatments.
Definitions of these conditional causal effects, together with identification results and possible estimation methods, are reported in Appendix \ref{app:condeffects}.

 %%% %%% %%%
 %%% %%% %%%
 %%% %%% %%%

\section{Bias when SUTVA is wrongly assumed}
\label{sec:bias}

\subsection{Naive Estimator}
Under SUTVA, potential outcomes can be indexed only by the individual treatment, i.e., $Y_i(z)$,
regardless of the treatment received by other units. Therefore, in this case, the average (individual) treatment effect can be defined as
\begin{equation}
\tau_{sutva}=E\bigm[Y_i(Z_i=1) - Y_i(Z_i=0) \bigm]
\end{equation}
In observational studies, several covariate-adjusted estimators for this quantity have been proposed. All these estimators are designed to consistently estimate the quantity
\begin{equation}
\tau^{obs}_{X^{\star}}=\sum_{\vx \in \mathcal{X }^{\star}}E\bigm[Y_i| Z_i=1, \vX^{\star}_i=\vx \, ]- E[Y_i| Z_i=0, \vX^{\star}_i=\vx\bigm]P(\vX^{\star}_i=\vx)
\end{equation}
where $\vX^{\star}_i\in \mathcal{X }^{\star}$ is the subset of covariates used for the adjustment methods. If interference is ruled out, we can imagine that we would not take neighborhood covariates into account and we would only adjust for individual covariates, eventually including contextual covariates, i.e., $\vX^{\star}_i=\vX^{ind}_i$. Thus, under SUTVA, if individual covariates form a sufficient set for unconfoundedness, that is, $Y_i(z) \ind Z_i | \vX^{\star} \,\, \forall z=0,1$, then $\tau_{sutva}=\tau^{obs}_{X^{\star}}$, and hence an unbiased estimator of $\tau^{obs}_{X^{\star}}$ would provide an unbiased estimate of the average treatment effect $\tau_{sutva}$ (e.g., \citealp{Rosenbaum:2010, Imbens:Rubin:2015}).

However, in the presence of interference, potential outcomes of
the form $Y_i(z)$ are not well-defined and, thus, these estimators
would clearly not estimate the quantity $\tau_{sutva}$. Moreover,
in general they would not even estimate main effects $\tau(g)$ or $\tau$ , given that
estimators of $\tau_{x}^{obs}$ compare units belonging to the two
treatment arms defined by the individual treatment $Z_i$,
regardless of the neighborhood treatment $G_i$.

\subsection{Outline of Bias Results}
Here, we derive results for the bias for the overall main effect $\tau$
of a na\"ive approach that neglects interference. Bias is expressed as
the difference between $\tau$ and the quantity $\tau_{x}^{obs}$.
When $\tau_{x}^{obs}$ is estimated from a sample,
then the difference between $\tau$ and the quantity $\tau_{x}^{obs}$ represents
the bias for $\tau$ of an unbiased estimator of  $\tau_{x}^{obs}$, that we would
na\"ively use if SUTVA is assumed, 
%that we would have when SUTVA is wrongly assumed and a
%covariate-adjusted estimator of the quantity $\tau_{x}^{obs}$ is
%na\"ively used. 

We must distinguish between two cases depending
on whether or not Assumption \ref{ass: Totunconf} holds
conditioning only on the subset of covariates $\vX^{\star}$, that
is, $Y_i(z,g) \ind Z_i,G_i \mid  \vX^{\star}_i, \forall z\in
\{0,1\}, g \in \mathcal{G}_i$.

Section \ref{sec:bias1} is concerned with bias results under the unconfoudedness assumption:
\begin{itemize}
\item Theorem \ref{theo: biasA} provides an expression for the the quantity $\tau_{x}^{obs}$ in terms of potential outcomes of the form $Y_i(z,g)$
\item Corollary \ref{cor: bias1} shows that  if $Z_i$ and $G_i$ are conditionally independent an unbiased estimator of $\tau_{X^{\star}}^{obs}$ is unbiased for the overall main effect $\tau$
\item Corollary \ref{cor: bias2} provides expressions for the different between $\tau_{x}^{obs}$ and   $\tau$ when $Z_i$ and $G_i$ are conditionally dependent and highlights the two main sources of bias 
\end{itemize}

Section \ref{sec:bias2} is concerned with bias results when the unconfoudedness assumption does not hold conditional on a set of covariates $\vX^{\star}$:
\begin{itemize}
\item Theorem \ref{theo: biasB} provides an expression for the the difference between $\tau_{x}^{obs}$ and $\tau$, combining the bias due to interference and the bias due to unmeasured confounders
\item Corollary \ref{cor: bias3} simplifies the expression of the bias when $Z_i$ and $G_i$ are conditionally independent 
\item Corollary \ref{cor: bias4} shows that when SUTVA does hold the difference between $\tau_{x}^{obs}$ and $\tau$ is only due to unmeasured confounders and is the same as in the previous case where SUTVA does not hold but $Z_i$ and $G_i$ are conditionally independent
\end{itemize}

\subsection{Bias of Naive Estimator When Unconfoudedness Holds}
\label{sec:bias1}
\begin{subtheorem}{theorem}
\label{theo: bias}
\begin{theorem}\label{theo: biasA}
Let $G=(\N,\E)$ be a known social network and let $\mathcal{N}_i$ be the neighborhood of unit $i$ as defined by the presence of edges. Let $Z_i\in \{0,1\}$ be a binary treatment assigned to unit $i$  and let $G_i$ be a deterministic function of the subset of the treatment vector $\vZ$ in the neighborhood $\mathcal{N}_i$, that is, $G_i=g_i(\vZ_{\mathcal{N}_i})$, with $g_i: \{0,1\}^{N_i}\rightarrow \mathcal{G}_i$. If
\begin{enumerate}
\item Assumption \ref{ass: consistency} holds
\item Assumption \ref{ass:SUTNVA} holds, given function $g_i(\cdot)$ for each unit $i\in\mathcal{N}$
\item Assumption \ref{ass: Totunconf} holds conditional on $\vX^{\star}_i$, i.e., $ Y_i(z,g)\ind Z_i,G_i|\vX^{\star}_i, \forall z\in \{0,1\}, g \in \mathcal{G}_i$
\end{enumerate}
then the following equality holds
\vspace{-0.5em}
\begin{equation}
\label{eq: tauobs}
\begin{aligned}
\tau_{X^{\star}}^{obs}=&\sum_{\vx \in \mathcal{X^{\star} }}\bigg(\sum_{g\in \mathcal{G}}E[Y_i(1,g)| \vX^{\star}_i=\vx ,i \in V_g]P(G_i=g|Z_i=1,\vX^{\star}_i=\vx)\\
&\quad- E[Y_i(0,g)| \vX^{\star}_i=\vx , i \in V_g]P(G_i=g|Z_i=0,\vX^{\star}_i=\vx)\bigg) P(\vX^{\star}_i=\vx)\\
 \end{aligned}
\end{equation}
\end{theorem}
\textit{Proof}. See Appendix \ref{app:proofs}.

\begin{cor}
\label{cor: bias1}
Under the three conditions of Theorem \ref{theo: biasA} and the additional condition
\begin{enumerate}
 \setcounter{enumi}{3}
 \item $Z_i$ and $G_i$ are independent conditional on $\vX^{\star}_i$, i.e., $Z_i \ind G_i| \vX^{\star}_i$
\end{enumerate}
the following equality holds:
\[\tau_{X^{\star}}^{obs}=\tau\]
Therefore, an unbiased estimator of $\tau_{X^{\star}}^{obs}$ is unbiased for the overall main effect $\tau$, even in the presence of interference, if $Z_i$ and $G_i$ are conditionally independent.
\end{cor}
\textit{Proof}. See Appendix \ref{app:proofs}.

\begin{cor}
\label{cor: bias2}
Under the three conditions of Theorem \ref{theo: biasA} and if  $Z_i \nind G_i| \vX^{\star}_i$, an unbiased estimator of $\tau_{X^{\star}}^{obs}$ would be biased for the overall main effect $\tau$, with bias given by
\beq
\label{eq: biasgen}
\begin{aligned}
\tau_{X^{\star}}^{obs}-\tau&=\sum_{\vx \in \mathcal{X }^{\star}}\sum_{g\in \mathcal{G}}\bigg(E[Y_i|Z_i=1, G_i=g, \vX^{\star}_i=\vx , i \in V_g]-E[Y_i|Z_i=1, G_i=g', \vX^{\star}_i=\vx , i \in V_g]\bigg)\\
& \qquad \qquad\qquad \bigg( P(G_i=g|Z_i=1,\vX^{\star}_i=\vx)-P(G_i=g|\vX^{\star}_i=\vx)\bigg)P(\vX^{\star}_i=\vx)\\
&\quad -\sum_{\vx \in \mathcal{X }^{\star}}\sum_{g\in \mathcal{G}}\bigg(E[Y_i|Z_i=0, G_i=g, \vX^{\star}_i=\vx , i \in V_g ]-E[Y_i|Z_i=0, G_i=g', \vX^{\star}_i=\vx , i \in V_g]\bigg)\\
& \qquad \qquad\qquad\bigg( P(G_i=g|Z_i=0,\vX^{\star}_i=\vx)-P(G_i=g|\vX^{\star}_i=\vx)\bigg)P(\vX^{\star}_i=\vx)
\end{aligned}
\eeq
\noindent If the spillover effect of the neighborhood treatment $G_i$ at level $g$ vs level $g'$ does not depend on the individual treatment $Z_i$, the bias formula is reduced to
\beq
\label{eq:bias}
\begin{aligned}
&=\sum_{\vx \in \mathcal{X }^{\star}}\sum_{g\in \mathcal{G}}\bigg(E[Y_i|Z_i=z, G_i=g,\vX^{\star}_i=\vx , i \in V_g]-E[Y_i|Z_i=z, G_i=g',\vX^{\star}_i=\vx , i \in V_g]\bigg)\\
& \qquad \qquad\qquad \bigg( P(G_i=g|Z_i=1,\vX^{\star}_i=\vx)-P(G_i=g|Z_i=0,\vX^{\star}_i=\vx)\bigg)P(\vX^{\star}_i=\vx)
\end{aligned}
\eeq
irrespective of the value of $z \in \{0,1\}$.
\end{cor}
\textit{Proof}. See Appendix \ref{app:proofs}.

Theorem \ref{theo: biasA} concerns the bias of the estimation approach when we wrongly assume away the presence of interference, but we are able to adjust for a set of covariates $\vX^{\star}$ that satisfy the uncondoudedness assumption (Assumption \ref{ass: Totunconf}).
The core result of the theorem is the expression for $\tau_{X^{\star}}^{obs}$ in \eqref{eq: tauobs}, which shows that $\tau_{X^{\star}}^{obs}$ is actually comparing different types of potential outcomes with different values of the neighborhood treatment. Then, Corollary \ref{cor: bias1} states that if the individual and neighborhood treatments are independent conditional on $\vX^{\star}_i$, then using covariates-adjusted estimation methods that assume SUTVA would yield unbiased estimates for the overall main effect $\tau$, even if SUTVA does not hold. On the contrary, as stated in Corollary \ref{cor: bias2}, a residual correlation between $Z_i$ and $G_i$, after conditioning on  $\vX^{\star}_i$, would result in bias.
The formula presented in \eqref{eq:bias} shows that the bias depends on two factors: the level of interference and the residual association between the individual treatment $Z_i$ and the neighborhood treatment $G_i$ after conditioning for $\vX^{\star}_i$.
There can be several reasons for such an association. For instance, the individual treatment $Z_i$ and the neighborhood treatment $G_i$ can be linked through neighborhood covariates that are not included in $\vX^{\star}_i$. Moreover, there can be peer influence in the treatment uptake. Such situation is plausible in most realistic applications where a unit's choice to take the treatment might depend also on other units's choices. Another cause of an association between $Z_i$ and $G_i$ can be the presence of homophily, that is, similar characteristics underlying the neighborhood structure and driving the assignment mechanism.

\subsection{Bias of Naive Estimator When Unconfoudedness Does Not hold}
\label{sec:bias2}
\begin{theorem}\label{theo: biasB}
 Under Assumptions \ref{ass: consistency} and \ref{ass:SUTNVA} , if Assumption \ref{ass: Totunconf} does not hold conditional on $\vX^{\star}_i$, i.e., $Y_i(z,g)\nind Z_i,G_i|\vX^{\star}_i$, but holds conditional on $\vX^{\star}_i$ and an additional vector of covariates $\mathbf{U}_i\in \mathcal{U}$, i.e., $Y_i(z,g)\ind Z_i,G_i|\vX^{\star}_i, \mathbf{U}_i$,
 an unbiased estimator of $\tau_{X^{\star}}^{obs}$ would be biased for the overall main effect $\tau$, with bias $\tau_{X^{\star}}^{obs}-\tau$ given by
\begin{align*}
\label{eq:biasgu}
%\begin{aligned}
&=\sum_{\vx \in \mathcal{X^{\star} }}\sum_{g\in \mathcal{G}}\sum_{u\in \mathcal{U}} \bigg(E[Y_i|Z_i=1, G_i=g, \mathbf{U}_i=\mathbf{u},\vX^{\star}_i=\vx , i \in V_g ]\\
&\hspace{3.3cm}-E[Y_i|Z_i=1, G_i=g' ,\mathbf{U}_i=\mathbf{u}',\vX^{\star}_i=\vx , i \in V_g ]\bigg)\\
& \qquad \qquad\qquad \qquad\qquad \bigg( P(\mathbf{U}_i=\mathbf{u}|Z_i=1,G_i=g, \vX^{\star}_i=\vx)P(G_i=g|Z_i=1,\vX^{\star}_i=\vx)\\
&\qquad \qquad\qquad\qquad\qquad -P(\mathbf{U}_i=\mathbf{u}|G_i=g, \vX^{\star}_i=\vx)P(G_i=g|\vX^{\star}_i=\vx)\bigg)P(\vX^{\star}_i=\vx)\\
&\quad -\sum_{\vx \in \mathcal{X^{\star} }}\sum_{g\in \mathcal{G}}\sum_{u\in \mathcal{U}} \bigg(E[Y_i|Z_i=0, G_i=g, \mathbf{U}_i=\mathbf{u},\vX^{\star}_i=\vx , i \in V_g ]\stepcounter{equation}\tag{\theequation}\\
&\hspace{3.5cm}-E[Y_i|Z_i=0, G_i=g', \mathbf{U}_i=\mathbf{u}',\vX^{\star}_i=\vx , i \in V_g ]\bigg)\\
& \qquad \qquad\qquad \qquad\qquad\bigg( P(\mathbf{U}_i=\mathbf{u}|Z_i=0,G_i=g, \vX^{\star}_i=\vx)P(G_i=g|Z_i=0,\vX^{\star}_i=\vx)\\
&\qquad \qquad\qquad\qquad\qquad -P(\mathbf{U}_i=\mathbf{u}|G_i=g, \vX^{\star}_i=\vx)P(G_i=g|\vX^{\star}_i=\vx)\bigg)P(\vX^{\star}_i=\vx)
%\end{aligned}
\end{align*}
\noindent If both the spillover effect of the neighborhood treatment $G_i$ at level $g$ vs level $g'$ and the effect of the unmeasured confounder $\mathbf{U}_i$ at level $\mathbf{u}$ vs level $\mathbf{u}'$ do not depend on the individual treatment $Z_i$, the bias formula reduces to
\begin{equation}
\begin{aligned}
&=\sum_{\vx \in \mathcal{X^{\star} }}\sum_{g\in \mathcal{G}}\sum_{u\in \mathcal{U}} \bigg(E[Y_i|Z_i=z, G_i=g, \mathbf{U}_i=\mathbf{u},\vX^{\star}_i=\vx , i \in V_g ]\\
&\hspace{3.3cm}-E[Y_i|Z_i=z, G_i=g', \mathbf{U}_i=\mathbf{u}',\vX^{\star}_i=\vx , i \in V_g ]\bigg)\\
& \qquad \qquad\qquad\qquad \bigg( P(\mathbf{U}_i=\mathbf{u}|Z_i=1,G_i=g, \vX^{\star}_i=\vx)P(G_i=g|Z_i=1,\vX^{\star}_i=\vx) \\& \qquad \qquad\qquad\qquad\quad -P(\mathbf{U}_i=\mathbf{u}|Z_i=0,G_i=g, \vX^{\star}_i=\vx)P(G_i=g|Z_i=0,\vX^{\star}_i=\vx)\bigg)P(\vX_i=\vx)
\end{aligned}
\end{equation}
irrespective of the value of $z \in \{0,1\}$.
\end{theorem}
\textit{Proof}. See Appendix \ref{app:proofs}.

\begin{cor}
\label{cor: bias3}
Under the following conditions
\begin{enumerate}
\item Assumption \ref{ass: consistency} holds
\item Assumption \ref{ass:SUTNVA} holds given the function $g_i(\cdot)$ for each unit $i\in\mathcal{N}$
\item Assumption \ref{ass: Totunconf} holds conditional on $\vX^{\star}_i$ and $\mathbf{U}_i$, i.e., $ Y_i(z,g)\ind Z_i,G_i|\vX^{\star}_i, \mathbf{U}_i, \forall z\in \{0,1\}, g \in \mathcal{G}_i$
 \item $Z_i$ and $G_i$ are independent conditional on $\vX^{\star}_i$, i.e., $Z_i \ind G_i| \vX^{\star}_i$
\end{enumerate}
an unbiased estimator of $\tau_{X^{\star}}^{obs}$ would be biased for the overall main effect $\tau$, with bias $\tau_{X^{\star}}^{obs}-\tau$ given only by the unmeasured confounder $\mathbf{U}_i$:
\begin{equation}
\label{eq: biasU}
\begin{aligned}
&=\sum_{\vx \in \mathcal{X^{\star} }}\sum_{u\in \mathcal{U}} \bigg(E[Y_i|Z_i=z, \mathbf{U}_i=\mathbf{u},\vX^{\star}_i=\vx \, ]-E[Y_i|Z_i=z, \mathbf{U}_i=\mathbf{u}',\vX^{\star}_i=\vx \, ]\bigg)\\
& \qquad \qquad\qquad\qquad \bigg( P(\mathbf{U}_i=\mathbf{u}|Z_i=1, \vX^{\star}_i=\vx) -P(\mathbf{U}_i=\mathbf{u}|Z_i=0,\vX^{\star}_i=\vx)\bigg)P(\vX_i=\vx)
\end{aligned}
\end{equation}
where $z \in \{0,1\}$
\end{cor}
\textit{Proof}. See Appendix \ref{app:proofs}.
\begin{cor}
\label{cor: bias4}
Under the following conditions
\begin{enumerate}
\item SUTVA holds
\item Assumption \ref{ass: Totunconf} holds conditional on $\vX^{\star}_i$ and $\mathbf{U}_i$, i.e., $ Y_i(z,g)\ind Z_i,G_i|\vX^{\star}_i, \mathbf{U}_i, \forall z\in \{0,1\}, g \in \mathcal{G}_i$
\end{enumerate}
an unbiased estimator of $\tau_{X^{\star}}^{obs}$ would be biased for the overall main effect $\tau$, an unbiased estimator of $\tau_{X^{\star}}^{obs}$ would be biased for the overall main effect $\tau$, with bias $\tau_{X^{\star}}^{obs}-\tau$ given only by the unmeasured confounder $\mathbf{U}_i$ as in Equation \eqref{eq: biasU}.
\end{cor}
\textit{Proof}. See Appendix \ref{app:proofs}.

Theorem \ref{theo: biasB} states that, if in the estimation of the treatment effect we wrongly assume SUTVA and adjust for a set of covariates $\vX^{\star}_i$ that does not suffice for unconfoundedness to hold, the bias due to interference is combined with the bias due to unmeasured confounders $\mathbf{U}_i=\vX_i\backslash \vX^{\star}_i$. However, if either the individual treatment $Z_i$ and the neighborhood treatment $G_i$ are independent given $\vX^{\star}_i$ or there is no interference between units, then covariate-adjusted estimators would be biased only because of unmeasured confounders.
The vector of unmeasured confounders $\mathbf{U}:i$ can include neighborhood covariates $\vX^{neigh}_i$ that might affect the individual treatment $Z_i$ directly or through the neighborhood treatment $G_i$.
Typically, when SUTVA is assumed these kind of covariates are not taken into account in the estimation procedure. Theorem \ref{theo: biasB} shows that we should pay careful attention to the problem of dependence between units, even if proper interference can be ruled out.
\end{subtheorem}

 %%% %%% %%%
 %%% %%% %%%
 %%% %%% %%%

\section{Definition and Properties of Generalized Propensity Score Under Neighborhood Interference}
\label{sec:ps}
%\cmntL{New section on Propensity Score}

In this section we contribute to the literature by extending
the definition of propensity score under neighborhood
interference. Under Assumption \ref{ass:SUTNVA}, in fact, each
unit is exposed to a  bivariate treatment; therefore the
definition of the propensity score must be generalized to be the
joint probability of an individual with certain observed
characteristics of being assigned to an individual treatment and
being exposed to a neighborhood treatment. We show that the new
propensity score has balancing properties that are similar to the
propensity score under SUTVA. 

%\section{Propensity Score-Based Estimator for Main Effects and Spillover Effects}
%\label{sec:estimator}

%Relying on the identification result of Theorem \ref{theo: identification},
%we could obtain an unbiased estimator of $\mu(z,g)$ using an unbiased estimator of the conditional mean $\overline{Y}^{obs}_{z, g}$. For example,
%under simple random sampling, this quantity could be estimated by taking the mean of the observed outcomes within cells defined by covariates (stratification) \citep{Imbens:Rubin:2015}. Nevertheless, the presence of continuous covariates or a large number of covariates poses some challenges in the estimation of $\overline{Y}^{obs}_{z, g}$. Here we propose a propensity score-based estimator of the average dose-response function $\mu(z,g)$, that allows to estimate marginal main and spillover effects $\tau(g)$ and $\delta(z;g)$.

\subsection{Joint Propensity Scores}
%\cmntL{Should we introduce the ps as an average over those with same $\vX_i$ (which includes neighborhood covariates) first and then introduce the neighborhood individualistic assumption which defines the ps as the unit-level probability?}
%
%The propensity score \citep{Rosenbaum:Rubin:1983, Rosenbaum:2010, Imbens:Rubin:2015} is defined as the average assignment probability for units with the same value of covariates.  Given an unconfounded and individualistic assignment, the propensity score reduces to the unit-level assignment probability.
%
%\textcolor{red}{x FABRI: Riguarda la definizione di super-population propensity score nel libro. Lo definisce cosi:
%`$e(x) = Pr(Wi = 1|Xi = x)$ where the probability is
%taken both over the assignment mechanism and over the random sampling.
%Note that with our definition of superpopulations the assignment mechanism is automatically
%individualistic (of course, given $\phi$).' Non capisco.}
%
%\footnote{Note that here the individualistic property, which limits the dependence of the treatment assignment for unit $i$ on the outcomes and covariates of other units, is plausible given our formalization of the joint treatment and the way the vector of covariates $\vX_i$ is defined.}
%%Because our formalization involves a bivariate treatment,
%Under the neighborhood interference (Assumption \ref{ass:SUTNVA}),
We define as the \textit{joint propensity score}, denoted by
$\psi(z; g; x)$, as the joint probability distribution of  the
individual treatment and the neighborhood treatment given the
observed covariates:
\begin{equation}
\label{eq:jps}
\begin{aligned}
\psi(z; g; x)&=P(Z_i=z, G_i=g| \vX_i=\vx)\\
\end{aligned}
\end{equation}
%We refer to $\psi(z; g; x)$ as the \textit{joint propensity score}.
$\psi(z; g; x)$ is the probability for unit i of being
exposed to treatment z and  neighborhood treatment g, given his
observed individual and neighborhood characteristics $\vx$.

The joint propensity score does not necessarily correspond
to the unit-level assignment probability, which is  in general
expressed as $Pr(Z_i=z, G_i=g| \vX,  \{\vY(z,g), z=0,1; g\in
\mathcal{G}\})$ (unit-level version of assignment mechanism in
\eqref{eq: assmec}). However, if the unit-level assignment
probability of being exposed to treatment z and neighborhood
treatment g only depends unit-level variables, i.e., $Pr(Z_i=z,
G_i=g| \vX, \{\vY(z,g), z=0,1; g\in \mathcal{G}\})=Pr(Z_i=z,
G_i=g| \vX_i, \{Y_i(z,g), z=0,1; g\in \mathcal{G}_i\})$,
\footnote{This property is similar to the individualistic
property of the assignment mechanism in \citet{Imbens:Rubin:2015}.
However, here it is defined on the extended assignment mechanism
defined on both the individual and the neighborhood treatment and
the vector of covariates $\vX_i$ include neighbors'
characteristics. } and unconfoundedness holds given $\vX_i$
(Assumption \ref{ass: Totunconf}), then the unit-level assignment
probability coincides with the joint propensity score.

Given the definition of the joint propensity score in \eqref{eq:jps}, we can prove the following two properties.

\begin{prop}[Balancing Property]
\label{prop:balance}
The joint propensity score is a balancing score, that is, 
\[P(Z_i=z, G_i=g| \vX_i, \psi(z; g; \vX_i))=P(Z_i=z, G_i=g|\psi(z; g; \vX_i))\].
\end{prop}
\textit{Proof}. See Appendix \ref{app:proofs}.

Proposition \ref{prop:balance} implies that if a group of
units have the same  $\psi(z; g; \vx)$, then the
distribution of covariates $\vX$ for the group is the same 
between the arm with $Z_i=z$ and $G_i=g$ and all the other arms.

\begin{prop}[Conditional Unconfoundedness of $Z_i$ and $G_i$ given the joint propensity score]
\label{prop:PSZuncon}
If Assumption \ref{ass: Totunconf} holds given $\vX_i$, then
\[
Y_i(z,g) \ind Z_i, G_i | \psi(z; g; \vX_i) \qquad \forall z\in \{0,1\}, g \in \mathcal{G}_i
\]
\end{prop}
\textit{Proof}. See Appendix \ref{app:proofs}.

Proposition \ref{prop:PSZuncon} states that if
unconfoundedness holds conditional on covariates $\vX_i$ than the
distribution of the potential outcome $Y_i(z,g)$ is independent of
the individual and the neighborhood treatment of units with the
same value of the joint propensity score $\psi(z; g; \vX_i)$ for
the corresponding values z and g. This is a crucial result in that
it allows imputing missing potential outcomes $Y_i(z,g)$ across
arms for units with the same propensity of being exposed to
treatment z and neighborhood treatment g. Therefore, it is
sufficient to adjust the joint propensity score to account for
confounding bias in the estimation of both main and spillover
effects.

%\sout{As a consequence, we can get an unbiased estimator of $\mu(z,g)$ by adjusting for the joint propensity score, that is,
%using an unbiased estimator of the quantity
%%\[
%%E\bigm[E[Y_i| Z_i=z, G_i=g, \psi(z; g; \vX_i)]|Z_i=z, G_i=g\bigm]
%%\]
%where the average is taken over the empirical distribution of the joint propensity score in the population.}
%\sout{Note that the unbiasedness of the estimator is with respect to the sampling scheme.
%Our proposed estimation strategy in Section \ref{sec:est.strategy} is appropriate for simple random
%sampling of units or clusters of separated neighborhoods.} \cmntL{Moved below}

\subsection{Individual Propensity Score and Neighborhood Propensity Score}
%\cmntL{Moved below}
%\sout{Because the joint treatment is bivariate and the neighborhood treatment $G_i$ can potentially take on many different values, depending on the function $g_i(\cdot)$, we can neither stratify nor match on this joint propensity score }\citep{Imbens:2000}. \sout{To solve this issue, we propose a semi-parametric approach, which exploits the following factorization of the joint propensity score:}

We can consider the following factorization of the joint propensity score:
\begin{equation}
\label{eq:fact}
\begin{aligned}
\psi(z; g; x)&=P(Z_i=z, G_i=g| \vX_i=\vx)\\
&=P(G_i=g| Z_i=z, \vX^g_i=\vx^g)P(Z_i=z|\vX^z_i=\vx^z)
\end{aligned}
\end{equation}
where $\vX^g_i\in \mathcal{X}^g\subset  \mathcal{X}$ is the subset
of covariates affecting the neighborhood treatment, and
$\vX^z_i\in \mathcal{X}^z\subset  \mathcal{X}$ is the subset of
covariates affecting the individual treatment. In principle
vectors $\vX^z_i$ and $\vX^g_i$ could be different. In fact,
individual characteristics  collected in $\vX_i^{ind}$ should be
included in $\vX^z_i$, but the type of neighboring units and the
neighborhood structure, i.e., $\vX_i^{neigh}$, could also affect
the probability of individual treatment for unit i. Also, the
probability of a unit's neighbors receiving treatments
$\vZ_{\N_i}$ summarized in g will depend on neighborhood
characteristics in $\vX_i^{neigh}$, on neighborhood structure that
is likely to affect the mapping of $\vZ_{\N_i}$ into g, but might
as well depend on individual characteristics of unit i in
$\vX_i^{ind}$.  We denote the probability of having the
neighborhood treatment at level $g$ conditional on a specific
value $z$ of the individual treatment and on the vector of
covariates $\vX^g_i$, i.e., $P(G_i=g| Z_i=z, \vX^g_i=\vx^g)$, with
$\lambda(g; z; x^g)$ and refer to it as \textit{neighborhood
propensity score}. Similarly, we denote the probability of having
the individual treatment at level $z$ conditional on covariates
$\vX^z_i$, i.e., $P(Z_i=z|\vX^z_i=\vx^z)$, with $\phi(z;x^z)$ and
we refer to it as the \textit{individual propensity score}. Given
this factorization (Eq. \ref{eq:fact}), the unconfoundedness
assumption holds conditioning on the two types of propensity
scores separately.
\begin{prop}[Conditional Unconfoundedness of $Z_i$ and $G_i$ given the individual propensity score and the neighborhood propensity score]
\label{prop:2PSuncon}
If Assumption \ref{ass: Totunconf} holds given $\vX_i$, then
\[
Y_i(z,g) \ind Z_i, G_i | \lambda(g; z; \vX_i^g), \phi(1;\vX_i^z) \qquad \forall z\in \{0,1\}, g \in \mathcal{G}_i
\]
\end{prop}
\textit{Proof}. See Appendix \ref{app:proofs}.

As Proposition \ref{prop:PSZuncon}, Proposition \ref{prop:2PSuncon} is a key result in that it allows deriving adjustment methods that separately adjust for the individual and the neighborhood treatment.

\section{Propensity Score-Based Estimator for Main Effects and Spillover Effects}
\label{sec:estimator}

%\cmntL{New section on estimator. }
%Note that the unbiasedness of the estimator is with respect to the sampling scheme. Our proposed estimation strategy in Section \ref{sec:est.strategy} is appropriate for simple random sampling of units or clusters of separated neighborhoods.

Here we propose an estimator that relies on results of the
previous section and follows directly from the formalization of
the individual and neighborhood treatment. Consider a sample of
the population $G$. The unbiasedness of the proposed estimator
relies on a random sampling mechanism that preserves the
connections among units. Different sampling schemes are possible.
Here we consider two sampling schemes. 
A possible method is cluster sampling where disjoint
clusters are randomly sampled and but networks in each cluster are not
necessarily fully connected. School sampling with friendship networks within school is a good example.
Another option is an egocentric sampling method  where randomly selected units, called `ego', are the units of analysis, but, via interview, they are asked to nominate a list of persons (`alters') with whom they have a specific type of relationship%a survey is used to collect information
%on their ties and this is stored as attributes of the egos. 
\citep{Kolaczyk:2009, Perri:2018}

\subsection{Individual and Neighborhood Propensity Score Estimator}

%\cmntL{Moved here these 6 paragraphs}

Based on the identification result of Theorem \ref{theo: identification},
we could obtain an unbiased estimator of $\mu(z,g)$ using an unbiased estimator of the conditional mean $\overline{Y}^{obs}_{z, g}$. For example,
%\sout{under simple random sampling,} 
this quantity could be estimated by taking the mean of the observed outcomes within cells defined by covariates (stratification) \citep{Imbens:Rubin:2015}. Nevertheless, the presence of continuous covariates or a large number of covariates poses some challenges in the estimation of $\overline{Y}^{obs}_{z, g}$. Relying on results of the previous section, we propose a propensity score-based estimator of the average dose-response function $\mu(z,g)$, that allows estimating marginal main and spillover effects $\tau(g)$ and $\delta(z;g)$.

If Assumption \ref{ass: Totunconf} of unconfoundedness holds given $\vX_i$, then we can define the joint propensity score as
the probability distribution of the individual and neighborhood treatments given covariates $\vX_i$.
\footnote{Note that we included in the joint propensity score the set of covariates that is sufficient to satisfy the unconfoundedness assumption. Oftentimes, the probability of unit i being exposed to a certain neighborhood treatment does depend on non-neighboring characteristics, given that the probability of his neighbors being assigned to treatment might depend on their neighbors' covariates. In this case the joint propensity score will not coincide with the unit-level assignment mechanism.}
As a consequence of Proposition \ref{prop:PSZuncon}, we can get an unbiased estimator of $\mu(z,g)$ by adjusting for the joint propensity score $\psi(z; g; \vX_i)$, that is, using an unbiased estimator of the quantity
\[
E\big[E[Y_i| Z_i=z, G_i=g, \psi(z; g; \vX_i)]|Z_i=z, G_i=g\big]
\]
where the outer expectation is taken over the empirical distribution of the joint propensity score in the population.
However, because the joint treatment is bivariate and the neighborhood treatment $G_i$ can potentially take on many different values, depending on the function $g_i(\cdot)$, it is not easy to stratify or match on this joint propensity score \citep{Imbens:2000}. To solve this issue, we propose a semi-parametric approach, which exploits
the factorization of the joint propensity score in \eqref{eq:fact}.

According to Proposition \ref{prop:2PSuncon}, we can adjust for the joint propensity score by adjusting separately for both the individual propensity score $\phi(1;x^z)$ and for the neighborhood propensity score $\lambda(g; z; x^g)$,that is, using an unbiased estimator of the quantity
\[
E\big[E[Y_i| Z_i=z, G_i=g, \phi(1;\vX_i^z), \lambda(g; z; \vX_i^g)]|Z_i=z, G_i=g\big]
\]

The need for propensity score-adjustment has a different meaning for the two types of propensity scores. When estimating spillover effects of the form $\delta(z,g)$, adjustment for the neighborhood propensity score $\lambda(g;z,\vX_i^g)$ is required
in order to correct for the covariate imbalance across the two arms with the values of the neighborhood treatment that we are comparing, i.e., $G_i=g$ and $G_i=0$.
Conversely, adjustment for the individual propensity score $\phi(1;\vX_i^z)$ is required when we are estimating marginal spillover effects. In fact, the ADRF $\mu(z,g)$ for each value of $z\in \{0,1\}$ is estimated using data of the corresponding individual treatment arm. Conditioning on $\phi(1;\vX_i^z)$ allows generalizing the result to the opposite arm.
When estimating main effects this argument is reversed.

Moreover, the two propensity scores are of different nature. The individual treatment is binary and thus we can adjust for the individual propensity score using the standard propensity score-based adjustment methods for binary treatment. On the contrary, the neighborhood treatment has a discrete domain $\mathcal{G}_i$ with cardinality depending on the function $g_i$. $\lambda(g; z; x^g)$ can be seen as the \textit{generalized propensity score} (GPS) defined by \citet{Hirano:Imbens:2004} for continuous treatments. To adjust for the neighborhood propensity score, we can then use a similar model-based approach.
To reduce the bias due to a possible model-misspecification, we propose the use of a subclassification on the individual propensity score $\phi(1;x^z)$ and, within subclasses that are approximately homogenous in $\phi(1;x^z)$, a model-based approach for the neighborhood propensity score . 
%-- where there is sufficient balance across individual treatment groups -- , .

\subsection{Estimating Procedure: Subclassification and GPS}
\label{sec:est.strategy}

Here we describe the details of the estimating procedure (\textit{Subclassification and GPS}).
Note that the proposed estimation strategy is particularly appropriate when the domain of the neighborhood treatment $\mathcal{G}_i \subset \mathcal{R}$.

\begin{enumerate}[label=\arabic*)]
\item We derive a subclassification on the individual propensity score $\phi(1;\vX^z_i)$ as follows:
\begin{enumerate}[label=\alph*)]
\item We estimate $\phi(1;\vX^z_i)$ with a logistic regression for $Z_i$ conditional on covariates $\vX^z_i$;
\item We predict $\phi(1;\vX^z_i)$ for each unit;
\item We identify J subclasses $B_j$, with $j=1, \ldots, J$, defined by similar values of $\phi(1;\vX^z_i)$ and where there is sufficient balance between individual treatment groups, i.e., $\vX_i^z\ind Z_i| i\in B_j$.\end{enumerate}
\item Within each subclass $B_j$, we repeat the following steps to estimate $\mu_j(z,g)=E\big[Y_i(z,g)|\,  i\in B^g_j\big]$, where $B_j^g=V_g\cap B_j$:
\begin{enumerate}[label=\alph*)]
\item We estimate the parameters of a model for the neighborhood propensity score $ \lambda(g; z; x^g)$: $\lambda(z; g; \vX^g_i)= Pr(G_i=g| Z_i=z, \vX^g_i)=f^{G}(g, z, \vX^g_i)$
%For example, if $G_i$ is defined as the number of treated units in the neighborhood $\mathcal{N}_i$, then we can assume a binomial distribution for $G_i$ and model the probability of being treated using a logistic regression weighted by the unit's degree $N_i$;
\item We use the observed data ($Y_i, Z_i, G_i, \vX^g_i$) and $\widehat{\Lambda}=\lambda(G_i; Z_i; \vX^g_i)$ to estimate the parameters of a model $Y_i(z,g)\mid \lambda(z; g; \vX^g_i) \sim f^Y(z, g, \lambda(g; z; \vX^g_i))$;
\item  For a particular level of the joint treatment $(Z_i=z, G_i=g)$, for each unit $i \in B^g_j$ we predict the neighborhood propensity score evaluated at that level of the treatment, i.e., $\lambda(g; z; \vX^g_i)$, and use it to predict the potential outcome $Y_i(z,g)$.
\item To estimate the dose-response function $\mu_j(z,g)$ we average the potential outcomes over $\lambda(z; g; \vX^g_i)$
\[\widehat{\mu}_j(z,g)=\frac{\sum_{i\in B^g_j} \widehat{Y}_i(z,g)}{|B^g_j|}\]
\end{enumerate}
\item We derive the average dose-response function as follows:
\begin{equation*}
\widehat{\mu}(z,g)=\sum_{j=1}^J \widehat{\mu}_j(z,g) \pi^g_j
\end{equation*}
where $\pi^g_j=\frac{|B^g_j|}{v_g}$.
\end{enumerate}

%Standard errors and confidence intervals can be derived using
%bootstrap methods. We propose the use of resampling with
%replacement at the unit-level. The choice of this  resampling
%strategy is consistent with the sampling scheme that samples units
%at random and collects treatment and covariates information on
%their neighbors. However, standard errors derived with such
%bootstrapping procedure are not valid for more general sampling
%schemes. Alternative resampling methods should be used with
%different sampling schemes, such as those which sample separated
%clusters of data, where there are no links between clusters, but
%individuals within clusters are organized in networks and
%interfere.
%
%%In fact, since we treat the neighborhood treatment $G_i$ as if it were a second treatment assigned to unit $i$, we are allowed to resample individuals, regardless of their neighbors. However, standard errors derived with such a bootstrapping procedure are valid only if there is no correlation between neighbors' outcomes. If this is not the case, we would have to use a block bootstrap method  that attempts to replicate the correlation by resampling  blocks of data.
%%
%%Given the complex structure of social networks, where neighborhoods are not well separated and different units may share few neighbors, we could use graph partitioning methods as sampling procedures. Investigating these alternative sampling procedures, however, goes beyond the scope of this paper.

It is worth noting that there are three key differences between our GPS approach and the one in \citet{Hirano:Imbens:2004}: i) it is performed within each subclass defined by $\phi(1;x^z)$, ii) both the GPS and the outcome model will include the individual treatment $Z_i=z$, iii) $G_i$ is a discrete treatment whose domain $\mathcal{G}_i$ depends on the function $g_i(\cdot)$, and iv) the dose-response function $\mu(z,g)$ is not always defined for all units.

The problem of statistical inference with units connected in network is not straightforward, given the correlation structure of the data. 
In Appendix \ref{app:addres} we propose a bootstrapping method with resampling
at different levels depending on the sampling scheme.

%%% %%% %%%
%%% %%% %%%
%%% %%% %%%

\section{Realistic Simulation Study leveraging Add Health data}
\label{sec:sim}

We use a simulation study to illustrate how the proposed methods may be applied. Our aim is twofold:  i) to validate the analytical derivation of the bias for the main effect when interference is wrongly ruled out; ii) to show the performance of the proposed estimators in a realistic sample.
We use friendship network data collected through the National Longitudinal Study of Adolescent Health (Add Health). We limited our analysis to 29 schools for a total of 16410 students.
Let us assume that we are interested in estimating the effect of health coverage on flu infection.
Such effect could, in principle, be estimated from the real data. However, to show the performance of the estimation procedures in different scenarios, both the treatment and the outcome are generated using different generating processes. Let the individual treatment variable $Z_i$ denote whether student $i$ was covered or not by some health insurance, and let $Y_i$ denote the number of days student $i$ missed school because of illness in one given year. The assumption of no-interference might be questionable here, because we can think that health insurance leads to better health for the covered student, which in turn can reduce the chance of spreading infectious diseases, such as flu, to the student's friends. Therefore, even if we are only interested in estimating the main effect of the individual health insurance, not taking interference into account might result in biased estimates. For simplicity, let us consider two individual covariates:  $\mathtt{race}_i$, indicating student $i$'s race (1 if white and 0 if other), and $\mathtt{grade}_i$, a discrete variable indicating student $i$'s grade. Let $\vX^{ind}_i=(\mathtt{race}_i, \mathtt{grade}_i)$ and $\vX^{neig}_i=(\frac{\sum_{k\in\mathcal{N}_i}\mathtt{race}_k}{N_i},  \frac{\sum_{k\in\mathcal{N}_i}\mathtt{grade}_k}{N_i}, N_i)$.

The simulation study considers four scenarios of dependence between $Z_i$ and $G_i$. In all scenarios but the third, $G_i$ is the proportion of friends with health insurance among the first five best friends. In the third scenario $G_i$ is the number of `treated' friends among all friends.
\begin{enumerate}[leftmargin=*, label=Scenario \arabic*:]
\item $Z_i$ is generated depending on individual race and grade. Hence, $Z_i$ and $G_i$ are independent conditional on $\vX^{ind}_i$.
\item $Z_i$ is generated depending on individual race and grade, and on friends' race and grade. Hence, $Z_i$ and $G_i$ are dependent if we condition only on $\vX^{ind}_i$, but independent conditional on $\vX^{neig}_i$.
\item $Z_i$ is generated depending on individual race and grade, and on the student's degree. Hence, $Z_i$ and $G_i$ are dependent if we condition only on $\vX^{ind}_i$, but independent conditional on $\vX^{neig}_i$.
\item $Z_i$ is generated depending on individual race and grade and on $G_i$. Hence $Z_i$ and $G_i$ are directly correlated and are not independent even if we condition on $\vX^{ind}_i$ or $\vX^{neig}_i$. (Here data are generated using an iterative procedure).
\end{enumerate}
Details of the association between $Z_i$, $G_i$ and the covariates in the four scenarios are in Appendix \ref{apx:scenarios}.
The outcome models are described in Appendix \ref{apx:outcome}. It is worth noting that the generating models are slightly different in the two sets of simulations where the focus is either on main effects or on spillover effects. In the first set of simulations, the outcome distribution, reported in Equation \eqref{eq:outcome1} in Appendix  \ref{apx:outcome1}, only depends on individual covariates and does not depend on neighborhood characteristics. Therefore, unconfoundedness (Assumption \ref{ass: Totunconf}) holds conditional on $\vX^{ind}_i$. This is to show that, even if the outcome does not depend on neighborhood covariates that do affect $Z_i$, a bias can still occur due to the induced correlation between $G_i$ and $Z_i$ and interference. In the second set of simulations with a focus on spillover effects, the outcome model, reported in Equation \eqref{eq:outcome2} in Appendix  \ref{apx:outcome2}, depends on both individual and neighborhood covariates, and has a more complicated structure with additional interaction terms.
For each scenario, we consider different levels of interference by changing the value of the parameter $\delta$ in Equations \eqref{eq:outcome1} and \eqref{eq:outcome2}.

\begin{table}
\caption{Computed Bias for $\tau$.\label{tab:bias}}
\centering
\begin{tabular}{cc|ccc}
\multicolumn{2}{c}{Scenario}&& &\\
 & interference &bias($\emptyset$)& bias($\vX^{ind}_i$)&bias($\vX^{z}_i$)\\ \hline
 \multirow{3}{4cm}{\centering 1\\($Z_i\ind G_i|\vX_i^{ind}$)}&low& -5.977 &-0.045&-0.045\\
 &medium&   -6.200& -0.072&-0.072\\
 &high&  -6.323&-0.090&-0.090 \\[0.3cm]
 \multirow{3}{4cm}{\centering 2\\($Z_i\ind G_i|\vX_i^{ind},\vX_i^{neigh} $)}&low&-5.749  &-1.636&-0.034 \\
 &medium& -6.498  &-2.618&-0.054 \\
 &high&-6.998   &-3.273&-0.068\\[0.3cm]
 \multirow{3}{5cm}{\centering 3\\($Z_i\ind G_i|\vX_i^{ind},\mathtt{degree}_i$)}&low&   -4.158&-1.247&-0.047 \\
 &medium&-4.792  & -2.079& -0.075\\
 &high& -5.744   &  -3.327&-0.095\\[0.3cm]
 \multirow{3}{5cm}{\centering 4\\($Z_i\nind G_i|\vX_i^{ind},\vX_i^{neigh}$)}&low&  -9.504 &-1.414& -1.415\\
 &medium&   -11.681  & -2.263&-2.263\\
 &high&   -13.132 &-2.829&-2.825
\end{tabular}
\end{table}

\subsection{Main Effect: Bias of na\"ive estimators and GPS-based estimator}

The focus of this first simulation study is on the estimation of the overall main effect $\tau$ (Eq. \ref{eq:tau}).
Our aim here is to show in different scenarios the bias resulting from neglecting the presence of interference and using typical estimators of the treatment effect and compare it to the analytical results presented in Section \ref{sec:bias}.
%Our aim is twofold: 1) to show the bias resulting from neglecting the presence of interference and compare it to the analytical results, 2) to show the performance of the proposed dose-response estimator for the estimation of the marginal main effect.
%As shown in Section \ref{sec:cme}, conditional main effects $\tau_g(g)$ can be estimated using common covariate-adjusted estimators on the subset of units with $G_i=g$. We think it is not necessary to illustrate the performance of these estimators, which has been extensively assessed in the literature.

Table \ref{tab:bias} shows the bias computed using formulas in Theorem \ref{theo: bias}. In particular, for each scenario, we derived the bias resulting from neglecting interference and from adjusting for different sets of covariates: $\vX^{\star}_i=\{\emptyset, \vX^{ind}_i, \vX^{z}_i\}$. Given that unconfoundedness holds conditional on $\vX^{ind}_i$, for the approach that does not adjust for any covariates, i.e., $\vX^{\star}_i=\emptyset$, the bias is due to both unmeasured confounders and interference. Equation \eqref{eq:biasgu} has been used for the computation, since in the outcome model (Equation \eqref{eq:outcome1} of Appendix \ref{apx:outcome}) there is an interaction between the individual treatment $Z_i$ and $\texttt{race}_i$ (an unmeasured confounder in this case). When we do adjust for either the individual covariates $\vX^{ind}_i$ or all covariates $\vX^{z}_i$, the bias is derived using Equation \eqref{eq:bias}. In fact, in these approaches the potential confounders $ \vX^{ind}_i$ are included in the adjustment set and, thus, we do not have the bias due to unmeasured confounders. However, when the  the adjustment set is not sufficient to rule out the dependence between $Z_i$ and $G_i$, the presence of interference does produce bias.

We then run 500 replications of each scenario, applying the following estimators \citep{Imbens:Rubin:2015} of the overall main effect:

\begin{itemize}
\item A simple difference in means estimator (Unadjusted Estimator) comparing treated and untreated units;
\item An ordinary least squares estimator that regresses the outcome on individual treatment $Z_i$, adjusting only for individual covariates $\vX_i^{ind}$;
\item An estimator based on a subclassification on the individual propensity score $\hat{\phi}(1,\vX_i^{ind})$, which is estimated using only individual covariates $\vX_i^{ind}$;
\item An ordinary least squares estimator that regresses the outcome on individual treatment $Z_i$, adjusting for individual and neighborhood covariates, i.e, $\vX_i^{ind}$;
\item An estimator based on a subclassification on the individual propensity score $\hat{\phi}(1,\vX_i^{z})$, which is estimated using individual and neighborhood covariates, i.e, $\vX_i^{z}$;
\item The estimator proposed in Section \ref{sec:estimator} (Sublcassification \& GPS), which is based on subclassification on the individual propensity score and model-based adjustment for the neighborhood propensity score.
\end{itemize}

\begin{table}
\caption{Estimation of main effect $\tau$.\label{tab:main}}
%\centering
\hspace{-2.25cm}
{\small
\begin{tabular}{@{\!\!\!\!\!\!}c@{\!\!\!}c|c@{\!\,\,\,}c@{\!\,\,\,}c@{\!\,\,\,}c@{\!\,\,\,}c@{\!\,\,\,}c@{\!\,\,\,}c@{\!\,\,\,}c@{\!\,\,\,}c@{\!\,\,\,}c@{\!\,\,\,}c@{\!\,\,\,}c}
\multicolumn{2}{c|}{\multirow{2}{3cm}{Scenario}}&\multicolumn{2}{c}{\multirow{2}{2cm}{Unadjusted}}&
\multicolumn{2}{c}{\multirow{2}{1.8cm}{Regression}}&\multicolumn{2}{c}{\multirow{2}{1.5cm}{Subclass}}&\multicolumn{2}{c}{\multirow{2}{1.8cm}{Regression}}&\multicolumn{2}{c}{\multirow{2}{1.5cm}{Subclass}}&\multicolumn{2}{c}{Subclass} \\
\multicolumn{2}{c|}{}&\multicolumn{2}{c}{}&\multicolumn{2}{c}{}&\multicolumn{2}{c}{}&\multicolumn{2}{c}{}&\multicolumn{2}{c}{}&\multicolumn{2}{c}{$\hat{\phi}(1,\vX_i^{z})$}\\
\multicolumn{2}{c|}{}&\multicolumn{2}{c}{}&
\multicolumn{2}{c}{ $\sim  Z_i,\vX_i^{ind}$}&\multicolumn{2}{c}{ $\hat{\phi}(1,\vX_i^{ind})$}&\multicolumn{2}{c}{$\sim  Z_i,\vX_i^{z}$}&\multicolumn{2}{c}{$\hat{\phi}(1;\vX_i^{z})$}&\multicolumn{2}{c}{\& GPS} \\
\cmidrule(lr){3-4}\cmidrule(lr){5-6}\cmidrule(lr){7-8}\cmidrule(lr){9-10}\cmidrule(lr){11-12}\cmidrule(lr){13-14}
&$\delta$&Bias& RMSE&Bias& RMSE &Bias& RMSE&Bias& RMSE&Bias& RMSE&Bias& RMSE\\
\hline
\multirow{3}{4cm}{\centering 1\\($Z_i\ind G_i|\vX_i^{ind}$)}&low& -5.104  & 5.104&-3.429 & 3.431&-0.016&0.130&-3.254&3.256&0.006&0.074&-0.002&0.064\\
&medium& -5.278 &5.278 &-3.529 &3.532&-0.021&0.149&-3.256&3.258&0.007&0.090&0.002&0.067\\
&high& -5.396  & 5.396& -3.598&3.601&-0.021&0.167&-3.258&3.261&0.015&0.108&0.003&0.067\\[0.3cm]
\multirow{3}{4cm}{\centering 2\\($Z_i\ind G_i|\vX_i^{ind},\vX_i^{neigh} $)}&low& -5.136  &5.136 & -3.477& 3.477& -1.652&1.668&-2.156&2.157&-0.066&0.331&-0.000&0.036\\
&medium&  -5.884 &5.884 &-4.521 &4.521&-2.620&2.631&-2.318&2.319&-0.097&0.511&0.003&0.039\\
&high&-6.384   &6.384 &-5.222&5.223& -3.272&3.281&-2.430&2.432&-0.123&0.635&-0.002&0.036\\[0.3cm]
\multirow{3}{5cm}{\centering 3\\($Z_i\ind G_i|\vX_i^{ind},\mathtt{degree}_i$)}&low&-4.106   &4.107 &-2.079 &2.089&-1.146&1.149&-1.033&1.049&0.078&0.553&-0.002&0.230\\
&medium&-4.725   &4.726 &-2.886 &2.893&-1.911&1.913&-1.041&1.056&0.082&0.556&0.019&0.233\\
&high& -5.655  &5.656 & -4.098&4.103 &-3.058&3.059&-1.054&1.068&0.084&0.559&0.012&0.234\\[0.3cm]
\multirow{3}{5cm}{\centering 4\\($Z_i\nind G_i|\vX_i^{ind},\vX_i^{neigh}$)}&low&  -8.670 & 8.670& -7.104&7.104&-1.443&1.445&-7.076&7.076&-1.738&1.740&0.000&0.105\\
&medium& -10.761  &10.761 &-8.861 &8.862&-2.309&2.309&-8.824&8.823&-2.784&2.785& -0.003&0.095\\
&high& -12.154  & 12.154&-10.030&10.030&-2.883&2.884&-9.985 &9.985&-3.477&3.478&0.010&0.090\\
\end{tabular}}
\end{table}

Table \ref{tab:main} reports the mean bias and root mean squared error (RMSE) of all these estimators in all scenarios.
The first five estimators ignore the presence of interference, resulting in a bias that is proportional to the level of interference and the level of association between individual treatment and neighborhood treatment. In Scenario 1, where $Z_i$ and $G_i$ are independent conditional on individual covariates, all subclassification-based estimators are essentially unbiased, for all levels of interference.
Regression-based estimators are affected by a bias due to model-misspecification, in the range of $-3.50$, when adjusting for individual covariates, and $-3.25$, when adjusting for individual and neighborhood covariates. Finally, in the unadjusted difference in means estimators we have the usual bias due to unmeasured confounders.
In scenarios where the association between the $Z_i$ and $G_i$ is due to neighborhood covariates (Scenario 2 and Scenario 3),  for the unadjusted estimator and the regression estimator that only adjust for individual covariates, an additional bias due to interference is combined with the aforementioned bias due to unmeasured confounders and model-misspecification, respectively. The unadjusted estimator and the two subclassification-based estimators, adjusting for individual covariates or all covariates, show a bias that is comparable to the corresponding analytical bias reported in Table \ref{tab:bias}.
The two estimators --regression and subclassification-based -- that adjust for both individual and neighborhood covariates are able to reduce the bias due to interference. The model-misspecification bias for the regression estimator is still present, whereas the semiparametric sublcassifcation-based estimator is able to cancel out the bias.
In Scenario 4, where there is a direct correlation between $Z_i$ and $G_i$, an adjustment for neighborhood covariates cannot remove this correlation. Therefore, all regression-based and sublcassifcation-based estimators are affected by the bias due to interference.
Only an estimator that explicitly takes interference into account
is able to eliminate this kind of bias. The estimator proposed in this paper (Sublcassification \& GPS), which adjusts for both individual and neighborhood propensity scores, shows no bias in all scenarios and for all levels of interference, thanks to Assumption \ref{ass: Totunconf} holding. In fact, relying on Theorem \ref{theo: identification} and Proposition \ref{prop:2PSuncon}, an unbiased estimator of the quantity $ E_{\Lambda, \Phi}\bigm[E[Y_i| Z_i=z, G_i=g, \phi(1; \vX^z_i), \lambda(g;z,\vX^g_i)]|Z_i=z, G_i=g\bigm]$ is unbiased for the ADRF $\mu(z,g)$.

\subsection{Spillover Effects}

\begin{table}
\caption{Estimation of $\Delta(0)$}
%\centering
\hspace{-1cm}
\begin{tabular}{cc|c@{\!\,\,\,}cc@{\!\,\,\,\,\,}cc@{\!\,\,\,}cc@{\!\,\,\,}c}
\multicolumn{2}{c|}{\multirow{2}{2cm}{Scenario}}&\multicolumn{2}{c}{Unadjusted}&
\multicolumn{2}{c}{\multirow{2}{1.5cm}{Subclass}}&\multicolumn{2}{c}{\multirow{3}{1cm}{GPS}}&\multicolumn{2}{c}{Subclass}\\
\multicolumn{2}{c|}{}&\multicolumn{2}{c}{Regression}&\multicolumn{2}{c}{}&\multicolumn{2}{c}{}&\multicolumn{2}{c}{$\hat{\phi}(1,\vX_i^{z})$}\\
\multicolumn{2}{c|}{}&\multicolumn{2}{c}{$\sim Z_i, G_i$}&
\multicolumn{2}{c}{ $\hat{\phi}(1,\vX_i^{z})$}&\multicolumn{2}{c}{}&\multicolumn{2}{c}{\& GPS}\\
\cmidrule(lr){3-4}\cmidrule(lr){5-6}\cmidrule(lr){7-8}\cmidrule(lr){9-10}
&interference&Bias& RMSE &Bias& RMSE&Bias& RMSE&Bias& RMSE\\
\hline
\multirow{3}{4cm}{\centering 1\\($Z_i\ind G_i|\vX_i^{ind}$)}&low& 1.576  &1.581 &   2.871&2.901&-1.457&1.465& 0.003&0.091\\
&medium& 1.568 & 1.574&2.838&2.872&-1.462&1.472&0.005&0.114\\
&high& 1.571  &  1.580&2.878&2.910&-1.467&1.479&0.000&0.148\\[0.3cm]
\multirow{3}{4cm}{\centering 2\\($Z_i\ind G_i|\vX_i^{ind},\vX_i^{neigh} $)}&low& 3.503  & 3.506&  4.132& 5.259& -0.574&0.590& 0.033&0.098\\
&medium&  3.506 &3.510 & 4.037&4.933& -0.579&0.599&0.041& 0.118\\
&high&  3.485  & 3.489&4.204&5.232&-0.592&0.613 &0.034&0.129\\[0.3cm]
\multirow{3}{5cm}{\centering 3\\($Z_i\ind G_i|\vX_i^{ind},\mathtt{degree}_i$)}&low& 5.445   & 5.446&7.005&7.048&0.380&0.399&-0.035&0.091\\
&medium&   5.455 & 5.456&  7.009&7.050&0.381&0.401&-0.033&0.091\\
&high&  5.441 &5.442 & 7.054&7.096&0.381& 0.400&-0.033&0.099\\[0.3cm]
\multirow{3}{5cm}{\centering 3\\($Z_i\nind G_i|\vX_i^{z},\vX_i^{neigh}$)}&low& 3.002   & 3.002&2.577&2.584&-1.201&1.202&0.071&0.111\\
&medium&   3.002 & 3.002& 2.556&2.563&-1.202&1.203&0.068&0.111\\
&high&  3.005  & 3.005&2.548&2.555&-1.200&1.201& 0.066&0.111\\
\end{tabular}
\label{tab:delta0}
\end{table}%

The second simulation study focuses on the estimation of the marginal spillover effect $\Delta(0)$ and $\Delta(1)$.
Here we consider estimators that do take interference into account and are explicitly designed for the estimation of spillover effects. In observational studies, the presence of confounders $\vX_i$ requires the use of covariate-adjusted methods. Given unconfoundedness (Assumption \ref{ass: Totunconf}), according to Proposition \ref{prop:PSZuncon} and Proposition \ref{prop:2PSuncon},
we might adjust for either the joint treatment propensity score
or separately for the individual propensity score and the neighborhood propensity score, as in our proposed estimator.
Failing to adjust for either of the two propensity score would result in biased estimates.

We compare the performance of the following estimators, which differ for the way this covariate-adjustment is handled:

\begin{itemize}
\item An unadjusted regression-based estimator, that regresses the outcome on both the individual treatment $Z_i$ and the neighborhood treatment $G_i$ -- with an interaction term --, without adjusting for covariates;
%\item An ordinary least squares estimator that regresses the outcome on both the individual treatment $Z_i$ and the neighborhood treatment $G_i$ --with an interaction term --, adjusting only for individual and neighborhood covariates;
\item An estimator based on a subclassification on the individual propensity score ${\phi}(1,\vX_i^{z})$, which is estimated using both individual and neighborhood covariates;
\item A model-based estimator that regresses the outcome on both the individual treatment $Z_i$ and the neighborhood treatment $G_i$ -- with an interaction term --, adjusting for the neighborhood propensity score $\lambda(g;z,\vX^g_i)$ (GPS approach, \citealp{Hirano:Imbens:2004});
\item The estimator proposed in Section \ref{sec:estimator} (Sublcassification \& GPS), which is based on subclassification on the individual propensity score and model-based adjustment for the neighborhood propensity score.
\end{itemize}

\begin{table}
\caption{Estimation of $\Delta(1)$}
%\centering
\hspace{-1cm}
\begin{tabular}{cc|c@{\!\,\,\,}cc@{\!\,\,\,}cc@{\!\,\,\,\,\,}cc@{\!\,\,\,}cc@{\!\,\,\,}c}
\multicolumn{2}{c|}{\multirow{2}{2cm}{Scenario}}&\multicolumn{2}{c}{Unadjusted}&
\multicolumn{2}{c}{\multirow{2}{1.5cm}{Subclass}}&\multicolumn{2}{c}{\multirow{3}{1cm}{GPS}}&\multicolumn{2}{c}{Subclass}\\
\multicolumn{2}{c|}{}&\multicolumn{2}{c}{Regression}&\multicolumn{2}{c}{}&\multicolumn{2}{c}{}&\multicolumn{2}{c}{$\hat{\phi}(1,\vX_i^{z})$}\\
\multicolumn{2}{c|}{}&\multicolumn{2}{c}{$\sim Z_i, G_i$}&
\multicolumn{2}{c}{ $\hat{\phi}(1,\vX_i^{z})$}&\multicolumn{2}{c}{}&\multicolumn{2}{c}{\& GPS}\\
\cmidrule(lr){3-4}\cmidrule(lr){5-6}\cmidrule(lr){7-8}\cmidrule(lr){9-10}
&interference&Bias& RMSE &Bias& RMSE&Bias& RMSE&Bias& RMSE\\
\hline
\multirow{3}{4cm}{\centering 1\\($Z_i\ind G_i|\vX_i^{ind}$)}&low&3.195   &3.196 & 2.870&2.878&0.432&0.441&0.003&0.055\\
&medium&  3.197&3.198 &2.863&2.871&0.437&0.449&0.002&0.083\\
&high&  3.198  &3.200 &2.869&2.877&0.434&0.452&0.004&0.103\\[0.3cm]
\multirow{3}{4cm}{\centering 2\\($Z_i\ind G_i|\vX_i^{ind},\vX_i^{neigh} $)}&low&2.483   &2.485 &2.842 &2.862&0.386&0.393&0.001&0.059\\
&medium&   2.481 & 2.485&2.885&2.907&0.385&0.394&-0.000& 0.072\\
&high&2.473  & 2.477&2.882&2.919&0.379 &0.391&-0.005&0.088\\[0.3cm]
\multirow{3}{5cm}{\centering 3\\($Z_i\ind G_i|\vX_i^{ind},\mathtt{degree}_i$)}&low& 3.668  & 3.669&5.921&5.951&0.559&0.562&-0.038&0.045\\
&medium&  3.668 & 3.668&5.944& 5.975&0.558&0.561&-0.040&0.046\\
&high&3.669   & 3.669&5.943&5.982&0.558&0.561&-0.038&0.048\\[0.3cm]
\multirow{3}{5cm}{\centering 3\\($Z_i\nind G_i|\vX_i^{z},\vX_i^{neigh}$)}&low&-3.280   & 3.280&-5.182&5.182&0.367&0.370&-0.019&0.056\\
&medium& -3.277  &-3.277 &-5.182&5.182&0.368&0.371&-0.021&0.057\\
&high&   -3.281 &3.281 &-5.178&5.178&0.364&0.367& -0.026&0.059\\
\end{tabular}
\label{tab:delta1}
\end{table}%

In Tables \ref{tab:delta0} and  \ref{tab:delta1} we report the bias and root mean squared error (RMSE) of all estimators for spillover effects $\Delta(0)$ and $\Delta(1)$, respectively (Eq. \ref{eq:Delta}). In observational studies affected by interference, when SUTNVA assumption holds, estimators of spillover effects, relying on the correct function $g_i(\cdot)$, are unbiased as long as they properly adjust for confounding covariates. The unadjusted estimator and the two estimators that only adjust for one of the two propensity scores show a significant bias. The estimator that adjusts for the neighborhood propensity score $\lambda(g;z,\vX^g_i)$ performs better than the one based only on the individual propensity score ${\phi}(1,\vX_i^{z})$, given the different nature of adjustment. Our proposed estimator based on both propensity scores performs well in all scenarios.

\begin{figure}[t!]
\centering
\makebox{\includegraphics[scale=0.9,clip, trim=0.1cm 0.5cm 0.1cm 1.5cm]{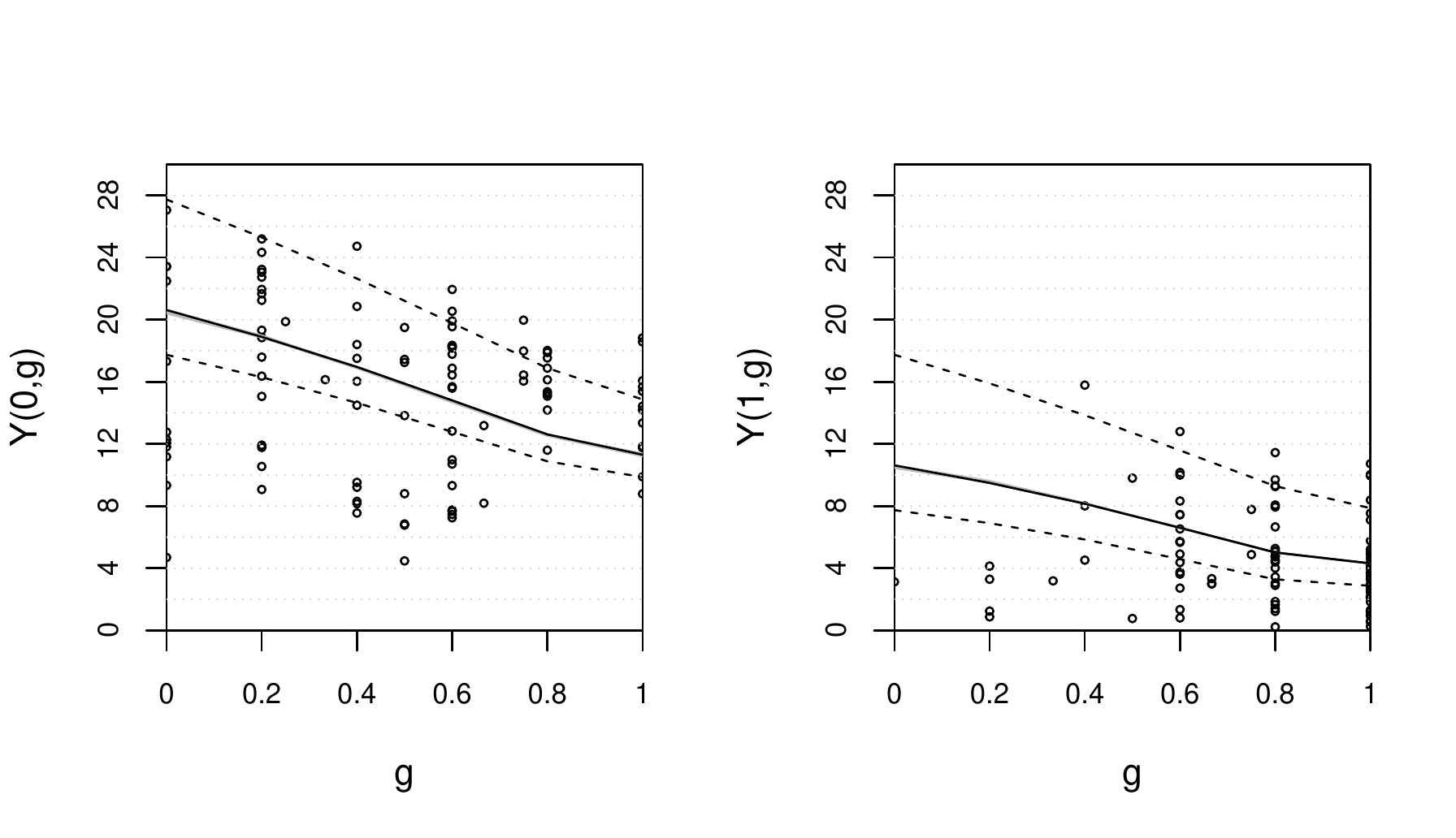}}
\caption{Estimated Dose-Response Functions $\mu(0,g)$ and $\mu(1,g)$.}
\label{fig:Yave}
\end{figure}

\begin{figure}[t!]
\centering
\makebox{\includegraphics[scale=0.9,clip, trim=0.1cm 0.5cm 0.1cm 1.5cm]{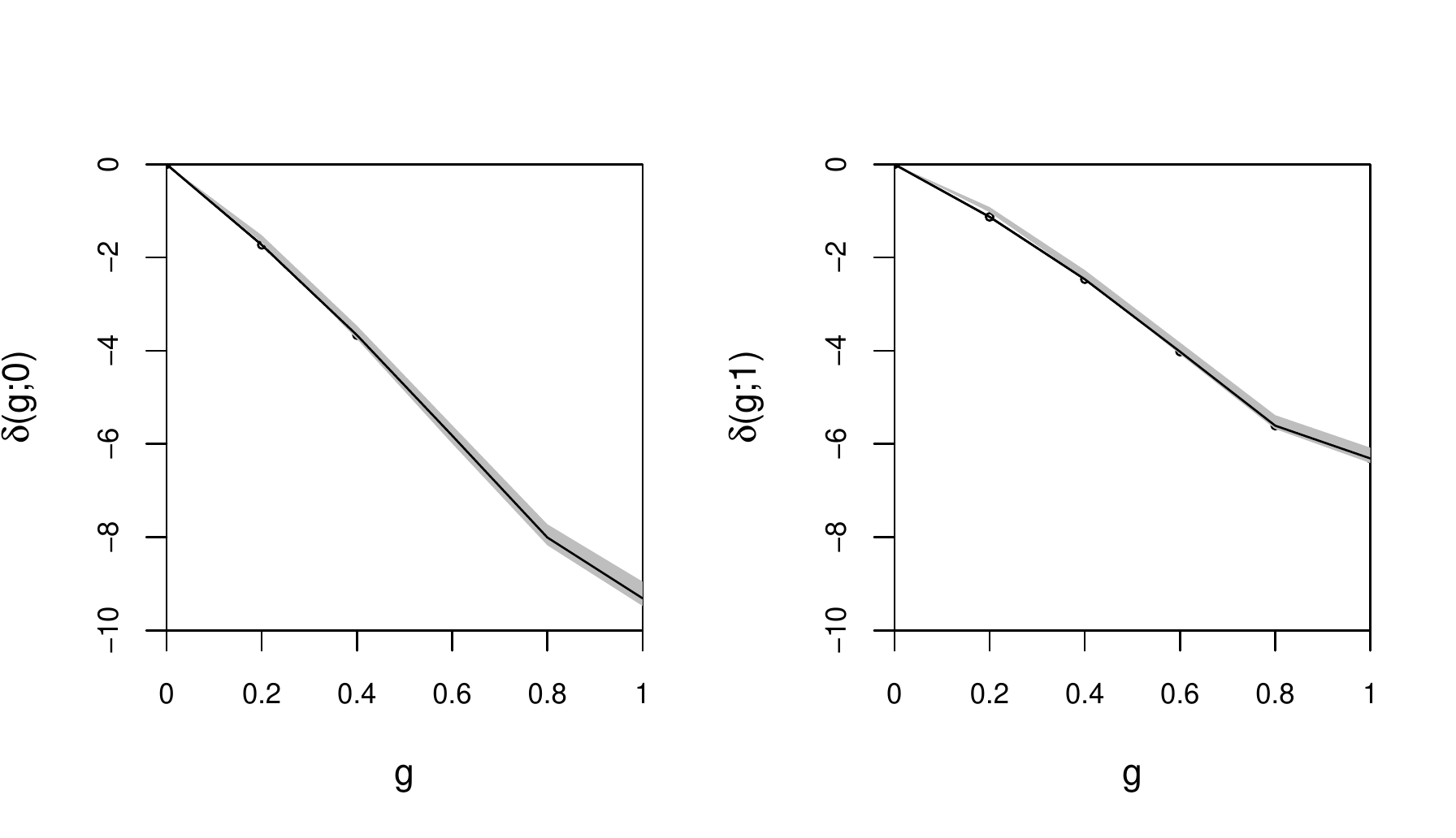}}
\caption{Estimated Spillover Effects $\delta(g;z)$.}
\label{fig:deltag}
\end{figure}

Figure \ref{fig:Yave} depicts the scatterplot of the observed outcomes and the estimated ADRFs $\mu(0,g)$ and $\mu(1,g)$ against all possible values of the neighborhood treatment $g\in \{0,0.2,0.4,0.6,0.8,1\}$. These plots are derived for Scenario 2, although all scenarios show similar results (Scenario 3 has a different domain for $g$ in the x-axis).
Solid lines represent the estimated marginal ADRFs, whereas dashed lines represent $\mu(0,g)$ (left) and $\mu(1,g)$ (right) estimated in two subclasses defined by the individual propensity score: the top and bottom lines correspond to a low and high value of ${\phi}(1,\vX_i^{z})$, respectively. The nonlinear trend is a result of the dependence of the outcome from the neighborhood propensity score in both the generation and the estimation model. $95\%$ confidence intervals are obtained using bootstrap methods and are illustrated by the gray bands. Here the small standard deviation is due to the large sample size. As mentioned, our dose-response estimator allows estimating both main and spillover effects, which can be easily computed from the functions $\mu(0,g)$ and $\mu(1,g)$. In particular, spillover effects $\delta(0,g)$, shown in Figure \ref{fig:deltag} (left), can be computed by subtracting $\mu(0,0)$ from $\mu(0,g)$ for each value of $g$.

\begin{figure}[t!]
\centering
\makebox{\includegraphics[scale=0.8,clip, trim=0.1cm 0.5cm 0.1cm 1.5cm]{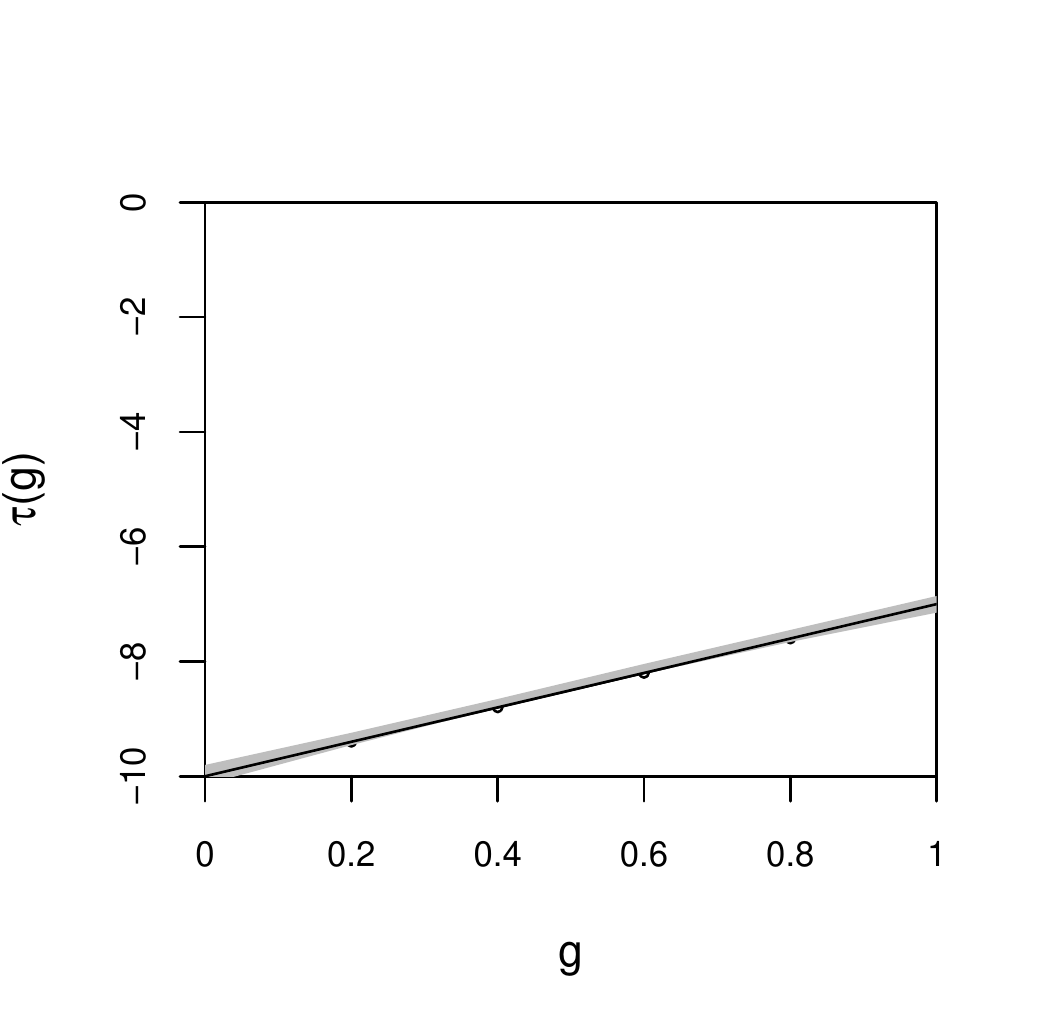}}
\caption{Estimated Main Effects $\tau(g)$.}
\label{fig:taug}
\end{figure}
Similarly, the difference between $\mu(1,g)$ and $\mu(1,0)$ leads to the estimation of spillover effects $\delta(1,g)$, shown in Figure \ref{fig:deltag} (right).
We can see that health insurance coverage among friends has an effect on a student's health. In particular, a greater proportion of insured friends result in a reduction of the average number of school days that a student would miss because of illness. The effect is larger for those who are not insured themselves ($\delta(0,g)>\delta(1,g), \,\, \forall g$). Figure \ref{fig:taug} depicts estimated main effects $\tau(g)$, which can be obtained from the difference between $\mu(1,g)$ and $\mu(0,g)$ for each value of $g$. The individual effect of being covered by some health insurance is reduced for those with a higher coverage among friends.
More estimation and inference results are reported in Appendix \ref{app:addres}.

%% NOTE
%GPS estimator without subclassfying on IPS is wrong because we are not trying to estimate conditiona but marginal effects. Simulations, not shown here, have shown GPS without bias if there is no interaction between G and ips and between G and Z.
%
%\textit{DUBBIO: Qui in realtà non servirebbe mostrare i diversi scenari, nè i diversi livelli di interferenza. Forse potrebbe avere senso mostrare solo il quarto scenario con diversi livelli di correlazione tra G e Z...e vedere che succede.}
%
%\textit{TO DO: Plots and Bootstrap. }

%%% %%% %%%
%%% %%% %%%
%%% %%% %%%

\section{Concluding Remarks}
\label{sec:concl}

The paper contains several contributions. We have introduced a
relaxed version of SUTVA, named SUTNVA, that simplifies in a
meaningful way the mechanism of interference in networks. Under
SUTNVA, potential outcomes are defined as a function of a
bivariate treatment variable, $(Z_i,G_i)$, which consists of the
individual treatment and the neighborhood treatment, which is a
function of the vector of treatments received by the unit's
neighbors. The framework presented in this paper is based on a
compound assignment mechanism for the bivariate individual and
neighborhood treatment, which is seen as a joint treatment to
which each unit is assigned.
Within this framework, we have introduced new causal estimands for the treatment and spillover effects , which disentangle the total effect of the individual treatment receipt and the exposure to 
the neighbors' treatments.
We have then laid out an unconfoundedness
assumption for the joint treatment, discussing its implications and plausibility
in networks.

We have  derived explicit bias formulas under several
scenarios for the treatment effects when SUTVA is wrongly assumed.
In this case, there can be a combination of a bias due to not
adjusting for neighborhood confounders and a bias due to the
presence of interference. The latter is proportional to the level
of interference and to the association between the individual and
the neighborhood treatments. Note that this result does not depend
on the choice of the function defining the neighborhood treatment.
An important result is that, when the set of covariates in the
adjustment set suffices for the independence of the two
treatments, the bias due to interference is canceled out; it
demystifies the misconception that bias resulting from neglecting
interference solely depends on the level of interference itself.
Regardless of the level of interference, the treatment effect
estimated using a naive approach would not be biased if the
treatments received by units of the same neighborhood are
conditionally independent. Oftentimes, this is likely to be true.
This means that we would only need to worry when there are reasons
to believe that there is a direct dependence among treatments in
the form of peer influence or that we have not included important
neighborhood covariates that are part of the assignment mechanism.
The implications of our derivations are significant for researchers in
both experimental and  observational studies. In experiments, our
results help to understand and predict the risk of different
design strategies. In observational studies, researchers will have
a better sense of whether they should trust their conclusions and
will make them think more carefully of ways to reduce the
probability of bias due to interference. For example, they may
collect covariates that are not necessarily confounders but are
predictive of both the individual treatment and the neighborhood
treatment. These results may  be used to develop a formal
sensitivity analysis which is part of our future research agenda.

%The correctness of these formulas was confirmed via simulations.

If we do take interference into account, proper covariate-adjustment methods are needed under the new version of unconfoundedness. Formal theory for designing and analyzing observational studies with units organized in a network is still in its infancy.
We have extendend the theory on propensity score as a balancing score. The proven balancing and unconfoudedness properties of both the joint propensity score and the individual and neighborhood propensity scores can 
open a whole new field of research. It could give rise to different propensity score-based estimators that remove bias by adjusting for either the joint propensity score or for the individual and neighborhood propensity scores separately. The easier way to adjust for the propensity score of a multivariate treatment is through a Horvitz-Thompson inverse probability-weighted estimator. This idea has already been used in the interference literature, however only in experimental settings, or in observational studies but under partial interference (e.g. \citet{TchetgenTchetgen:VanderWeele:2012, Aronow:Samii:2013, Liu:2016}). A clear use of Horvitz-Thompson inverse probability-weighted estimator in observational network data is still lacking.
%(it is in fact not clear at all how Liu et al. 2016 IPW estimator can be applied with a network, i.e., when units are not clustered in noninterfering groups: their estimator is rather general, but there are no indications of how to estimate weights with network data. 
We have instead focused on another adjustment method based on Hirano and Imbens (2004) model-based generalized propensity score approach. 

We have proposed a semi-parametric propensity-score based estimator
of both treatment and spillover effects.
We define a joint propensity score that balances individual and neighborhood covariates across units under different levels of such a bivariate treatment. If we assume interference to operate only through a function of the vector of friends' treatments, then spillover effects can be seen as a dose-response function of a multivalued treatment. We factorize the bivariate-treatment propensity score and estimate the dose response function by combining subclassification on the individual binary-treatment propensity score and parametric adjustment for the neighborhood multivalued-treatment propensity score.
The reason for the focus on this type of estimator stems from its flexibility. In fact, although it  heavily relies on the correctness of the outcome model, extensions to more complex nonlinear models is straightforward. In addition, thanks to its imputation approach that uses the estimated model to impute the specific potential outcomes of interest, it can be used to estimate different types of causal estimands.

The first simulation study, focused on the main effect, was used to i) assess the performance of the proposed estimator in the estimation of the main effect, ii) show the bias resulting from neglecting interference and using typical parametric and semi-parametric estimators of the treatment effect, ii) shed light on the different sources of bias through different scenarios iii) validate the nonparametric bias formulas. The performance of our semi-parametric propensity-score based estimator was also assessed in the second simulation study for the estimation of the spillover effects. These simulations were also useful to emphasize the need for the adjustment for the joint propensity score in different scenarios.

As with all papers, this also has some limitations. The causal estimands
are defined conditional on the observed social network that we assume
fully known and fixed. The estimator we propose is consistent with the way
we formulated the problem of interference and, for this reason, it is unbiased for the estimation
of treatment and spillover effects.  This is confirmed by our simulation study, where the estimator is consistent with the generating model. However, because the estimating procedure is based on the subclassification on the individual propensity score and a model-based approach for the neighborhood propensity score, in real studies it might be affected by a bias due to a residual imbalance in the subclasses or to model-misspecification.
%as with linear regression, also our estimator may be subject to bias due to model misspecification.
%In real studies, ...The only bias of our estimator, if any, would originate from the estimation of this quantity. The estimating procedure, based on the subclassification on the individual propensity score and a model-based approach for the neighborhood propensity score, might be indeed affected by a bias due to a residual imbalance in the subclasses or to model-misspecification. . e' coerente
Due to the complexity of the problem, we expect the true model
to be non linear, and, thus, we prefer our semi-parametric estimator over linear regression
on individual and neighborhood treatments.
We are working on making our semi-parametric estimator less model-dependent.

The problem of statistical inference with units connected in network is very tricky given the correlation structure of the data. 
\citet{Aronow:Samii:2013} derive finite-sample variance for an IPW estimator in experiments. \citet{Liu:2016} rely on the assumption of partial interference, whereas \citet{VanderLaan:2014} and its extensions rely on independence assumptions. To the best of our knowledge, this problem is still an open research area. 
In Appendix \ref{app:addres} we propose a bootstrapping procedure with resampling
at the unit-level or at the cluster-level.
These resampling technique quantify uncertainty due to the two sampling schemes, egocentric or cluster sampling, respectively.
% that samples units at random and collects information on their neighbors and its properties do not depend on the structural correlation between outcomes of different units. 
%This approach is also valid when units are independent given the observed covariates. 
Under alternative sampling schemes, different resampling methods to derive
standard errors should  be further investigated.
%Such approach is justified by the formalization of our framework. However, in complex correlation structures, the issue of alternative resampling methods to derive
%standard errors should  be further investigated.

Directions for future research include using Bayesian semiparametric
approaches to inference, which could potentially overcome the problems of
quantifying uncertainty, developing novel sensitivity analysis techniques
to evaluate how departures from the unconfoundedness and SUTNVA affect
treatment effect estimates, developing new methodological tools to account
for network uncertainty when true interactions between individuals are
either not observed or measured with error.

\renewcommand{\refname}{\textbf{References}}
%\bibliographystyle{plainnat}

%%% %%% %%%
%%% %%% %%%
%%% %%% %%%

\clearpage
\appendix

%%% %%% %%%
%%% %%% %%%
%%% %%% %%%

\setcounter{page}{1}
\addcontentsline{toc}{part}{\appendixname}
\section*{Online Appendix}
\section{Propensity Score-Based Estimator for Conditional Main and Spillover Effects}
\label{app:condeffects}
\subsection{Conditional Main Effects }
\label{sec:cme}

%Under SUTNVA the outcomes reduce to $Y_i(Z_i,G_i)$, that is, the potential outcome of unit i under treatment $Z_i$ and if $G_i$ were a summary of the treatment vector of his neighborhood, $\vZ_{N_i}$, through the function $g_i(\cdot)$.
%A potential outcome $Y_i(z,g)$ is defined only for a subset of nodes $V_g=\{i\in\N: g\in\mathcal{G}_i\}$, with cardinality $v_g$. For instance, in the case of number of treated neighbors $V_g$ is the set of nodes with at least $g$ neighbors.
%We can now define the \textit{average treatment effect} or \textit{main effect} as
%\begin{equation}
%\tau(g)= \frac{1}{v_g} \sum_{i \in V_g} \bigm( Y_i(Z_i=1,G_i=g) - Y_i(Z_i=0,G_i=g) \bigm)
%\end{equation}
%

In one experiment, we can observe one of the two potential outcomes $Y_i(1,g)$ and $Y_i(0,g)$ defining the main effect $\tau(g)$ only for units who exhibit a neighborhood treatment $G_i=g$. For all the other units in $V_g$ these potential outcomes are potentially observable but none of them is actually observed. For this reason let us define a \textit{conditional main effect}, which is the average effect of the individual treatment only for those with $G_i=g$:
\begin{equation}
\label{eq:taugg}
\tau_g(g)= E \bigm[ Y_i(Z_i=1,G_i=g) - Y_i(Z_i=0,G_i=g) | G_i=g \bigm]
\end{equation}

If we condition on a particular value of $G_i$, then we only have one treatment, which is the individual treatment $Z_i$.
In observational studies, this treatment depends on a set o covariates. To be able to get an unbiased estimate of $\tau_g(g)$, we need to make the following assumption:

\begin{assumption}[Conditional Unconfoundedness of the Individual Treatment]
\label{ass:uncon1}
\[
Y_i(z,g) \ind Z_i \mid G_i=g, \vX^{z|g}_i \qquad \forall z\in\{0,1\}, g\in\mathcal{G}_i
\]
\end{assumption}
\noindent
Assumption \ref{ass:uncon1} states that units with neighborhood treatment $G_i=g$ and with the same value of the vector $\vX^{z|g}_i$ have the same probability of being treated, which does not depend on the potential outcomes $Y_i(z,g)$.
$ \vX^{z|g}_i$ is a sub-vector of $ \vX_i$ containing all those covariates that would affect the decision of taking the treatment for unit i, given that the neighborhood treatment is $g$. It is worth noting that Assumption \ref{ass: Totunconf} given $\vX_i$ implies Assumption \ref{ass:uncon1}. The proof just follows from the fact that joint independence implies conditional independence and that by definition of $ \vX^{z|g}_i$ we have that $P(Z_i=1| G_i, \vX_i)=P(Z_i=1|G_i, \vX^{z|g}_i)$.
Under Assumption \ref{ass:uncon1} an unbiased estimator of \eqref{eq:taugg} can be obtained by any unbiased estimator of the quantity
\[\tau^{obs}_{g, x}=\sum_{\vx \in \mathcal{X }}E\bigm[Y_i| Z_i=1, G_i=g, \vX_i=\vx \, ]- E[Y_i| Z_i=0, G_i=g, \vX_i=\vx\bigm]P(\vX_i=\vx)\]

\subsubsection{Conditional Propensity Scores of the Individual Treatment}
\label{app:condind}

If the vector $\vX^{z|g}_i$ has large dimension or includes continuous covariates, then adjustment for these covariates cannot be performed by stratification. As \citet{Rosenbaum:Rubin:1983} proposed for the no-interference case, adjustment for covariates can be achieved through the use of propensity scores. Let us define the \textit{conditional individual propensity score} as the probability of being treated, given a value of the neighborhood treatment and the vector of covariates:
\begin{equation}
e(x^{z|g};g)=P(Z_i=1| G_i=g, \vX^{z|g}_i=\vx)
\end{equation}
The balancing property and the property of unconfoundedness given the conditional individual propensity score can be formalized as follows.

\begin{prop}[Balancing Property of the Conditional Individual Propensity Score]
The propensity score of the individual treatment is a balancing score:
\[Z_i \ind  \vX^{z|g}_i \mid G_i=g, e(\vX^{z|g}_i ;g)\]
\end{prop}
\textit{Proof}. Refer to \citet{Rosenbaum:Rubin:1983}.

%\begin{proof}
%We have to show that $P(Z_i=1|G_i=g, \vX^z_i,  e(\vX^z_i ;g))=P(Z_i=1|G_i=g, e(\vX^z_i ;g))$. First consider the left hand side:
%\[P(Z_i=1|G_i=g, \vX^z_i,  e(\vX^z_i ;g))=P(Z_i=1|G_i=g, \vX^z_i)=e(\vX^z_i ;g)\]
%where the first equality follows because the propensity score is a function of $\vX^z_i$ and the second is by the definition of the propensity score. Second, consider the right hand side.
%\begin{equation*}
%\begin{aligned}
%P(Z_i=1|G_i=g, e(\vX^z_i ;g))&=E_{\vX^z}[P(Z_i=1|G_i=g, \vX^z_i, e(\vX^z_i ;g))|e(\vX^z_i ;g)]\\
%&=E_{\vX^z}[P(Z_i=1|G_i=g, \vX^z_i)|e(\vX^z_i ;g)]\\&=E_{\vX^z}[e(\vX^z_i ;g)|e(\vX^z_i ;g)]=e(\vX^z_i ;g)
%\end{aligned}
%\end{equation*}

%By the definition of probability and iterated expectations,
%\end{proof}

\begin{prop}[Conditional Unconfoundedness of $Z_i$ given the Conditional Individual Propensity Score]
\label{prop:cPSZuncon}
If assumption \ref{ass:uncon1} holds
\[Y_i(z,g) \ind Z_i \mid G_i=g, e(\vX^{z|g}_i ;g)\]
\end{prop}

\noindent \textit{Proof}. Refer to \citet{Rosenbaum:Rubin:1983}.

%\begin{proof}
%We have to show that $P(Z_i=1|G_i=g, \vY(z,g),  e(\vX^z_i ;g))=P(Z_i=1|G_i=g, e(\vX^z_i ;g))$.The proof proceeds by showing that both the left and the right hand sides of this equation are equal to the propensity score itself and, hence, they are also equal to each other. In doing so we will make use of the assumption of unconfoundedness \ref{ass:uncon1}.
%Notice that in the proof of the Balancing Property we have already shown
%that the right hand side of the equation is equal to the propensity score, i.e., $P(Z_i=1|G_i=g, e(\vX^z_i ;g))=e(\vX^z_i ;g)$. Now to prove that $P(Z_i=1|G_i=g, \vY(z,g),  e(\vX^z_i ;g))=e(\vX^z_i ;g)$ we use the law of iterated equations:
%\begin{equation*}
%\begin{aligned}
%P(Z_i=1|G_i=g, \vY(z,g),  e(\vX^z_i ;g))&=E_{\vX^z}[P(Z_i=1|G_i=g, \vX^z_i, \vY(z,g), e(\vX^z_i ;g))|\vY(z,g), e(\vX^z_i ;g)]\\
%&=E_{\vX^z}[P(Z_i=1|G_i=g, \vY(z,g),\vX^z_i)|\vY(z,g), e(\vX^z_i ;g)]\\&=E_{\vX^z}[P(Z_i=1|G_i=g, \vX^z_i)|\vY(z,g), e(\vX^z_i ;g)]\\&=E_{\vX^z}[e(\vX^z_i ;g)|e(\vX^z_i ;g)]=e(\vX^z_i ;g)
%\end{aligned}
%\end{equation*}
%where the second equality holds because $e(\vX^z_i ;g)$ is a function of $\vX^z_i$ and the third follows from assumption \ref{ass:uncon1}.
%,
%\end{proof}

\subsubsection{Estimator based on Propensity Scores}

One possible estimator based on the propensity score $e(\vX^{z|g}_i ;g)$ is a subclassification estimator \citep{Imbens:Rubin:2015}, which stratifies the subpopulation where $G_i=g$ into J strata $B_j, j \in\{1, \ldots, J\}$, defined by similar values of the propensity score $e(\vX^{z|g}_i ;g)$, i.e., $B_j=\{i\in V_g: G_i=g, b_{j-1}<e(\vX^{z|g}_i ;g)\leq b_j\}$. Within each stratum the conditional main effect can be estimated non-parametrically by
\begin{equation}
\widehat{\tau}^j_g(g)=\frac{\sum_{i\in B_j} Y_i\mathds{1}(Z_i=1)}{\sum_{i\in B_j}\mathds{1}(Z_i=1)}- \frac{\sum_{i\in B_j} Y_i\mathds{1}(Z_i=0)}{\sum_{i\in B_j}\mathds{1}(Z_i=0)}
\end{equation}
Finally the conditional main effect $\tau_g(g)$ can be estimated by the following average:
\begin{equation}
\widehat{\tau}_g(g)=\sum_{j=1}^J \widehat{\tau}^j_g(g) \pi_j
\end{equation}
where $\pi_j=\frac{|B_j|}{\sum_{i\in V_g} 1(G_i=g)}$

\subsection{Conditional Spillover Effects }
\label{app:condspil}

Let us now focus on spillover effects for units that are in the control group. We define as \textit{conditional spillover effect}, denoted with $\delta_0(g;0)$, the following quanity:
\begin{equation}
\delta_0(g;0)= E \bigm[ Y_i(Z_i=0,G_i=g) - Y_i(Z_i=0,G_i=0) | \, i \in V_g, Z_i=0 \bigm]
\end{equation}
In this case, even units with $Z_i=0$ can exhibit an outcome that is not one of the two potential outcomes defining the effect. This is because $G_i$ is not a binary treatment.
Nevertheless,  if we are only interested in estimating $\delta_0(g;0)$ for a specific value $g$,
we can ease the problem by defining a new variable
\[
T_i(g)=
\begin{cases}
1 \qquad \text{if} \quad G_i=g\\
0 \qquad\text{if} \quad G_i=0\\
\text{NA} \quad \text{if} \quad G_i\neq g,0
\end{cases}
\]
We can then focus on the subpopulation with $Z_i=0$ and $T_i\in \{0,1\}$ and proceed with a standard estimation procedure for the effect of a binary treatment binary $T_i$. The identifying unconfoudedness assumption can be expressed as follows.
\begin{assumption}[Conditional Unconfoundedness of the Neighborhood Treatment]
\label{ass: Tunconf}
\[Y_i(0,g) \ind T_i(g) \mid Z_i=0, \vX^{g|0}_i, G_i=\{g,0\}\]
\end{assumption}
\noindent In principle, $\vX^{g|0}_i$ can be a subvector of $\vX_i$. This is the set of covariates that makes the probability of having a specific value of the neighborhood treatment for units under control independent of the potential outcome $Y_i(0,g)$. As with Assumption \ref{ass:uncon1} in Section \ref{sec:cme}, Assumption \ref{ass: Totunconf} does imply Assumption \ref{ass: Tunconf}.

\subsubsection{Conditional Propensity Scores of the Neighborhood Treatment}

To use a propensity score-based estimator, we can define the \textit{conditional neighborhood propensity score} as follows:
\begin{equation}
r_{g,0}(x)=P(T_i=1| Z_i=0, \vX^{g|0}_i=\vx, G_i=\{g,0\})
\end{equation}
satisfying the following properties.

\begin{prop}[Balancing Property of the Conditional Neighborhood Propensity Score]
The conditional neighborhood propensity score is a balancing score:
\[T_i \ind  \vX^g_i \mid Z_i=0, G_i=\{0,1\}, r_{g,0}(\vX^g_i)\]
\end{prop}
\textit{Proof}. Refer to \citet{Rosenbaum:Rubin:1983}.

\begin{prop}[Conditional Unconfoundedness of $Z_i$ given the Conditional Neighborhood Propensity Score]
\label{prop:cPSGuncon}
If assumption \ref{ass: Tunconf} holds
\[Y_i(z,g) \ind T_i(g) \mid Z_i=0, r_{g,0}(\vX^{g|0}_i), G_i=\{g,0\}\]
\end{prop}
\textit{Proof}. Refer to \citet{Rosenbaum:Rubin:1983}.

One possible estimator exploiting this result can be a subclassification-based estimator, as in Section \ref{app:condind}.

As shown in Sections \ref{sec:cme} and \ref{app:condspil}, according to Propositions \ref{prop:cPSZuncon} and \ref{prop:cPSGuncon},
conditional effects can be estimated using common propensity score-based estimators for binary treatments on a subset of units. We do not illustrate the performance of these estimators, which has been extensively assessed in the literature (e.g., \citealp{Imbens:Rubin:2015}).

%%% %%% %%%
%%% %%% %%%
%%% %%% %%%

\section{Data Generating Model for Simulations}
\label{app:dgp}

%The simulation study considers four scenarios of dependence between $Z_i$ and $G_i$. In all scenarios but the third, $G_i$ is the proportion of friends with health insurance among the first five best friends. In the third scenario $G_i$ is the number of treated friends among all friends.
%
%\begin{enumerate}[leftmargin=*, label=Scenario \arabic*:]
%\item $Z_i$ is generated depending on individual race and grade. Hence, $Z_i$ and $G_i$ are independent conditional on $\vX^{ind}_i$.
%\item $Z_i$ is generated depending on individual race and grade, and on friends' race and grade. Hence, $Z_i$ and $G_i$ are dependent if we condition only on $\vX^{ind}_i$, but independent conditional on $\vX^{neig}_i$.
%\item $Z_i$ is generated depending on individual race and grade, and on the student's degree. Hence, $Z_i$ and $G_i$ are dependent if we condition only on $\vX^{ind}_i$, but independent conditional on $\vX^{neig}_i$.
%\item $Z_i$ is generated depending on individual race and grade and on $G_i$. Hence $Z_i$ and $G_i$ are directly correlated and are not independent even if we condition on $\vX^{ind}_i$ or $\vX^{neig}_i$. (Here data are generated using an iterative procedure).
%
%\end{enumerate}

 As introduced in Section \ref{sec:sim}, the simulation study exploits the friendship network collected in the Add Health Data. This dataset also provided information on two students' characteristics, namely, race and grade. In our simulation we have: $\mathtt{race}_i\in\{0,1\}$ and $\mathtt{race}_i\in\{6,7,...,12\}$. On the contrary, the treatment ($Z_i, G_i$) and the outcome $Y_i$ variables were generated using different generating processes. Here we describe the data generating model for the four scenarios.

\subsection{Correlation Structure of the Joint Treatment and Covariate Balance}
\label{apx:scenarios}

\paragraph{Scenario 1.}
In the fist scenario, the individual treatment $Z_i$ is generated depending on individual race and grade, with the following propensity score
\begin{equation}
\texttt{logit}(P(Z_i=1))=-18+ 2\mathtt{grade}_i+ 3\mathtt{race}_i
\end{equation}
and the neighborhood treatment $G_i$ is defined as the proportion of treated friends among the first five best friends. This generating procedure leads to a specific correlation structure between neighborhood and individual treatments and covariates. Here we describe the distribution of these variables in one simulated dataset. The following table shows the number of treated and untreated students and the average proportion of treated friends in the whole population, for treated and for untreated units. The proportion of treated students of 0.743 is similar to the actual proportion of students with health insurance in the real dataset.
\[
\begin{array}{l|rrr}
                 & All    & Z_i=1 & Z_i=0\\
                 \hline
              N  & 16410 &12188 & 4222\\
\overline{Z} &0.743 &1 &0\\
\overline{G}&0.662 &0.677&0.617 \\
\hline
\end{array}
\]

Table \ref{tab:sce1_covZ} describes the balance of covariate distributions across individual treatment arms. The imbalance of individual covariates follows from the propensity score, whereas for neighborhood covariates the small difference is presumably due to the presence of homophily, that is, students are more likely to have friends with similar characteristics.
\begin{table}
\caption{Covariate balance across individual treatment arms \label{tab:sce1_covZ}}
\centering
\begin{tabular}{lrrrr}
Variables   &   $\bar{X}_T$ &   $\bar{X}_C$ &   Stand.  Diff.\\
\hline
 Grade &    10.097  &   7.531   &   1.873   \\
 Race   &   0.754   &   0.364       &0.905      \\
 Friends' Grade &   9.186   &   8.949&  0.227   \\
 Friends' Race  &   0.588   &   0.541   &   0.144   \\
 Degree &   7.892   &   8.053   &   -0.037  \\[5pt]
 $G_i$  &   0.694   &   0.633 & 0.230   \\
\end{tabular}
\end{table}
For the same reason, the proportion of treated friends depends on individual covariates, other than neighborhood covariates as follows from the individual propensity score depending on individual covariates. The covariate balance across units with different neighborhood treatment is described by Table \ref{tab:sce1_covGdic}, which shows the distribution of covariates in two arms defined by a dichotomized neighborhood treatment, and by Table \ref{tab:sce1_covGdic}, which reports the coefficients of a weighted logistic regression.
\begin{table}
\caption{Covariate balance across dichotomized neighborhood treatment arms \label{tab:sce1_covGdic}}
\centering
\begin{tabular}{lrrrr}
Variables   &   $\bar{X}_{G_i\geq0.5}$  &   $\bar{X}_{G_i<0.5}$ &   Stand.  Diff.\\
\hline
 Grade &    9.546   &   9.070   &   0.270   \\
 Race   &   0.651 & 0.661       &-0.020     \\
 Friends' Grade &   9.451   &   8.027&  1.584   \\
 Friends' Race  &   0.636   &   0.373   &   0.847   \\
 Degree &   7.783   &   8.442   &   -0.153  \\[5pt]
 $Z_i$  &   0.764   &   0.671 & 0.219   \\
\end{tabular}
\end{table}
In this scenario, the assignment mechanism only based on individual covariates, should result in a zero correlation between  the neighborhood treatment and the individual treatment after conditioning for $\vX^{ind}_i$.
The small association seen in Table \ref{tab:sce1_covZ} and \ref{tab:sce1_covGdic}, is perhaps due to the homophily mechanism, as we can see in Table \ref{tab:sce1_covGcoef}, where the association is zero after conditioning for covariates.

\begin{table}
\caption{Coefficients of logistic regression of neighborhood treatment \label{tab:sce1_covGcoef}}
\centering
\begin{tabular}{lrrrr}
Variables   &   Estimate    &   SE\\
\hline
 Grade &    0.011   &   0.007       \\
 Race   &   -0.021  &   0.020           \\
 Friends' Grade &   0.92    &   0.010   \\
 Friends' Race  &   1.401   &   0.028       \\
 Degree &   -0.002  &   0.002       \\[5pt]
 $Z_i$  &   0.017   &   0.028   \\
\end{tabular}
\end{table}

\begin{figure}
\centering
\includegraphics[scale=0.45]{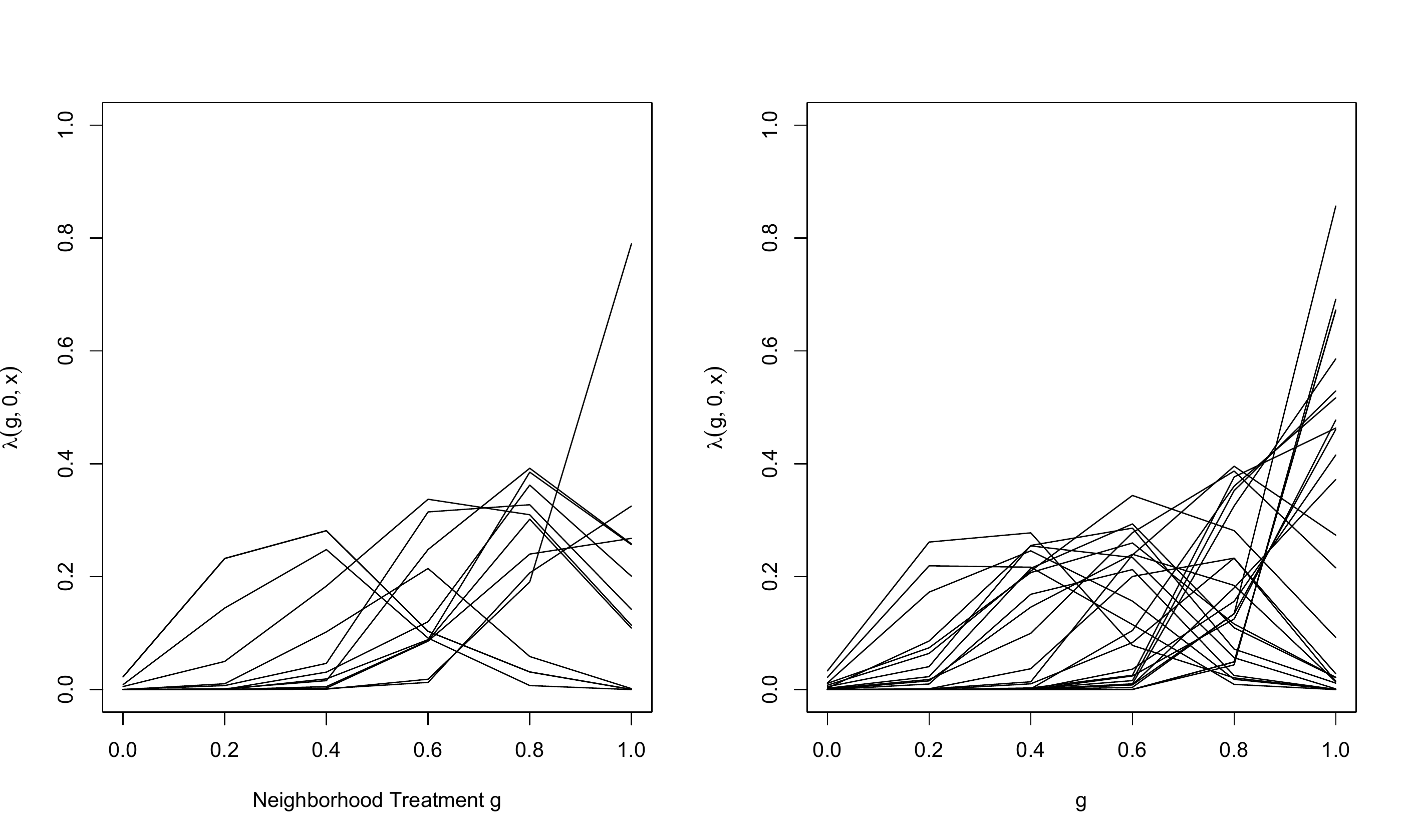}
\caption{Neighborhood propensity score across race: 0 (left), 1 (right).}
\label{fig:gps.race}
\end{figure}

\begin{figure}
\centering
\includegraphics[scale=0.8]{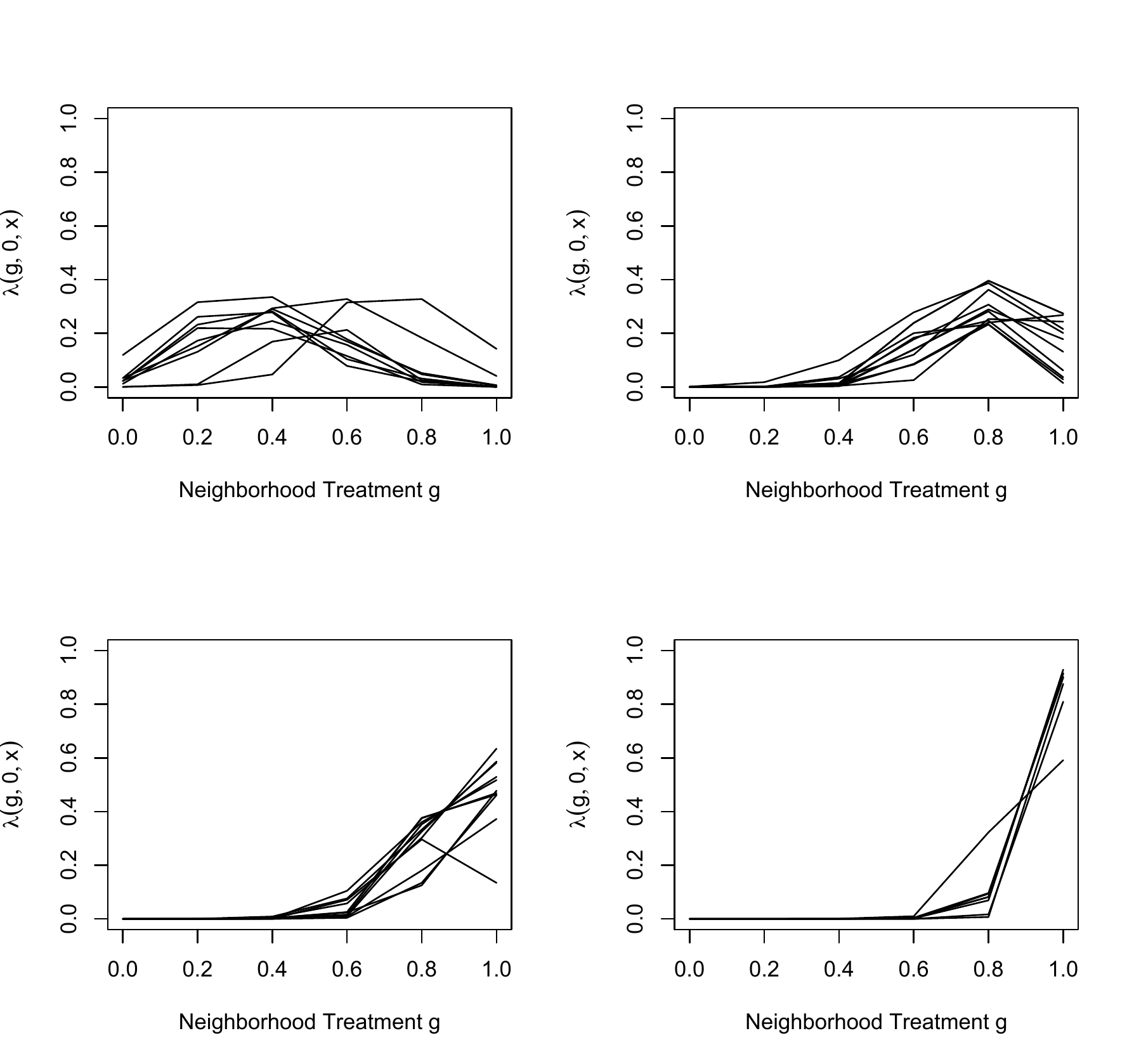}
\caption{Neighborhood propensity score across grade: 7 ( top left), 9 (top right), 10 (bottom left), 12 (bottom right).}
\label{fig:gps.grade}
\end{figure}

\begin{figure}
\centering
\includegraphics[scale=0.8]{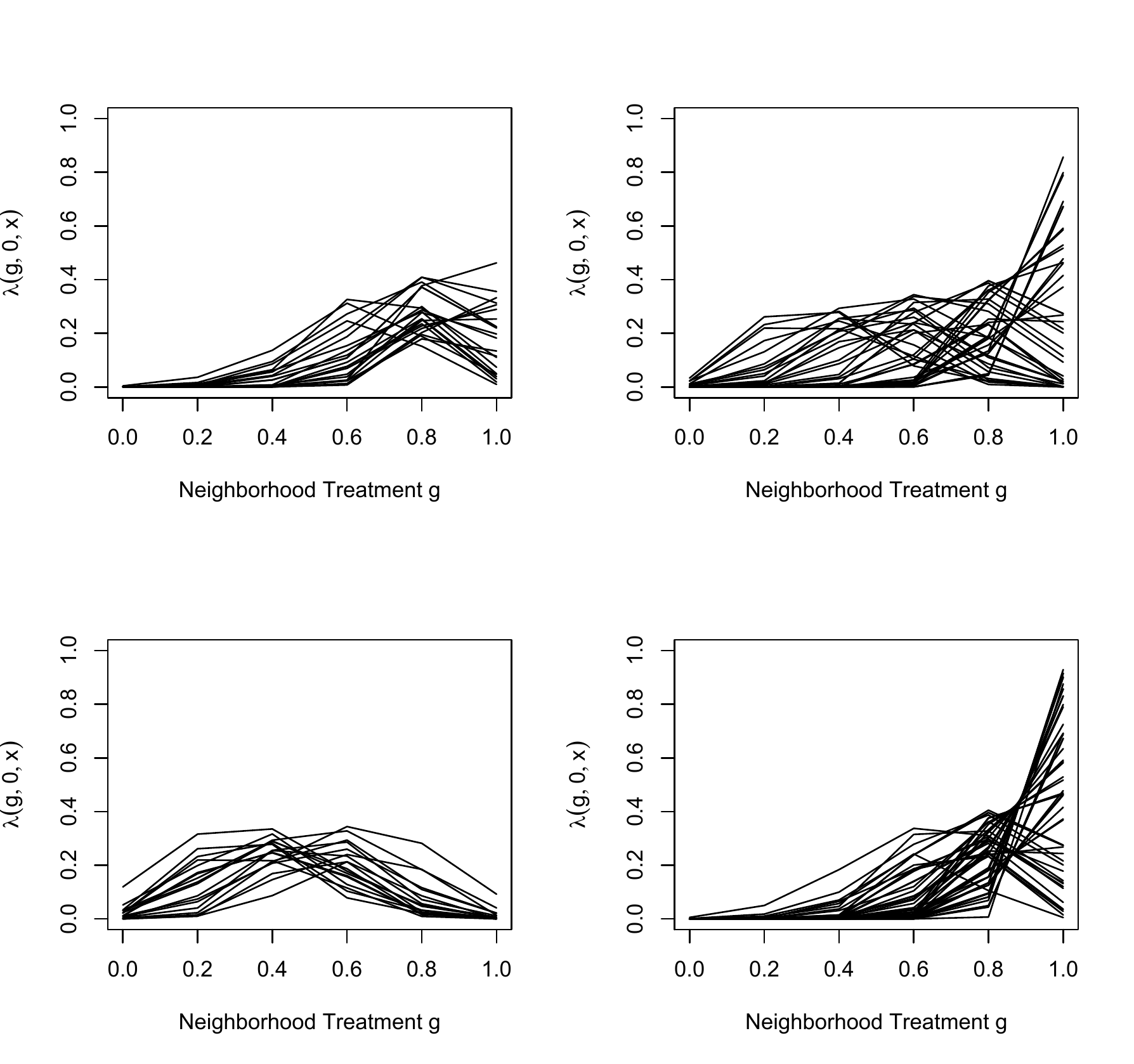}
\caption{Neighborhood propensity score across friends' race (top left: prevalence of non-white friends, top right: prevalence of white friends) and grade (bottom left: prevalence of friends in lower grades, bottom right: prevalence of friends in higher grades).}
\label{fig:gps.friends}
\end{figure}

\paragraph{Scenario 2.}
The individual treatment $Z_i$ is generated depending on individual race and grade, and on friends' race and grade with the following propensity score
\begin{equation}
\texttt{logit}(P(Z_i=1))=-47+ 2\mathtt{grade}_i+ 4\mathtt{race}_i+ 3\mathtt{friends.grade}_i+ 5\mathtt{friends.race}_i
\end{equation}
 The neighborhood treatment $G_i$ is the proportion of treated friends among the first five best friends. This assignment mechanism leads to the following distribution of $Z_i$ and $G_i$
\[
\begin{array}{l|rrr}
                 & All    & Z_i=1 & Z_i=0\\
                 \hline
              N  & 16410 &12064 & 4346\\
\overline{Z}&0.735 &1 &0\\
\overline{G}&0.626&0.686&0.459 \\
%\hbox{Outcome}   &&& \\
\hline
\end{array}
\]
and covariate balance across individual treatment arms, shown in Table \ref{tab:sce2_covZ}.
\begin{table}
\caption{Covariate balance across individual treatment arms \label{tab:sce2_covZ}}
\centering
\begin{tabular}{lrrrr}
Variables   &   $\bar{X}_T$ &   $\bar{X}_C$ &   Stand.  Diff.\\
\hline
 Grade &    10.024  &   7.650   &   1.660   \\
 Race   &   0.752   &   0.355       &0.919  \\
 Friends' Grade &   9.286   &   8.637&  0.656   \\
 Friends' Race  &   0.632   &   0.405   &   0.730   \\
 Degree &   7.906 & 8.018   &   -0.026  \\[5pt]
 $G_i$  &   0.687   &   0.505 & 0.687   \\
\end{tabular}
\end{table}
As in Scenario 1, the neighborhood treatment $G_i$ will depends on neighborhood covariates, as seen in Table \ref{tab:sce2_covGdic} and Table \ref{tab:sce2_covGcoef}.
Hence, $Z_i$ and $G_i$ are dependent if we condition only on $\vX^{ind}_i$, but independent conditional on $\vX^{neig}_i$. In Table \ref{tab:sce2_covGcoef} we report the coefficients of a weighted logistic regression of the neighborhood treatment on covariates. We used this estimated model to predict the neighborhood propensity score $\lambda(g; 0, \vX^g_i)$ for each unit. As an illustration, here we report the difference in the predicted $\lambda(g; 0, \vX^g_i)$ across units with different race (Figure \ref{fig:gps.race}), grade (Figure \ref{fig:gps.grade}), friend's race and grade (Figure \ref{fig:gps.friends}). The dependence of the the neighborhood treatment from friends' grade is presumably due to homophily, since students are likely to have friends in the same class.

\begin{table}
\caption{Covariate balance across dichotomized neighborhood treatment arms \label{tab:sce2_covGdic}}
\centering
\begin{tabular}{lrrrr}
Variables   &   $\bar{X}_{G_i\geq0.5}$  &   $\bar{X}_{G_i<0.5}$ &   Stand.  Diff.\\
\hline
 Grade &    9.561   &   9.120   &   0.300   \\
 Race   &   0.656 & 0.649       &0.013  \\
 Friends' Grade &   9.512   &   8.128&  1.585   \\
 Friends' Race  &   0.632   &   0.432   &   0.630   \\
 Degree &   7.803   &   8.272   &   -0.109  \\[5pt]
 $Z_i$  &   0.820   &   0.393 & 0.627   \\
\end{tabular}
\end{table}

\begin{table}
\caption{Coefficients of logistic regression of neighborhood treatment \label{tab:sce2_covGcoef}}
\centering
\begin{tabular}{lrrrr}
Variables   &   Estimate    &   SE\\
\hline
 Grade &    0.016   &   0.006       \\
 Race   &   0.025   &   0.020           \\
 Friends' Grade &   0.935   &   0.010   \\
 Friends' Race  &   0.978   &   0.029       \\
 Degree &   0.000   &   0.002       \\[5pt]
 $Z_i$  &   0.044   &   0.028   \\
\end{tabular}
\end{table}


% NOTA.
% Sembra che questo modello che non e' reale non catturi la correlazione tra Z e G

\paragraph{Scenario 3.}
This scenario was generated with the purpose of making $Z_i$ and $G_i$ associated trough a unit's degree. In fact, $Z_i$ is generated depending on individual covariates, and on the student's degree, using
\begin{equation}
\texttt{logit}(P(Z_i=1))=-49+ 3\mathtt{grade}_i+ 4\mathtt{race}_i+ 4N_i
\end{equation}
$G_i$ is the number of treated friends among all friends and, thus, it depends on the degree.
From the following distribution we can see that the average number of treated friends differs in the two individual treatment arms.
\[
\begin{array}{l|rrr}
                 & All    & Z_i=1 & Z_i=0\\
                 \hline
              N  & 16410 &11865 & 4544\\
\overline{Z}&0.723 &1 &0\\
\overline{G} &5.209&6.119&2.833 \\
\hline
\end{array}
\]
Covariate balance across treatments are reported in Tables \ref{tab:sce3_covZ}, \ref{tab:sce3_covGdic} and \ref{tab:sce3_covGcoef}.
\begin{table}
\caption{Covariate balance across individual treatment arms \label{tab:sce3_covZ}}
\centering
\begin{tabular}{lrrrr}
Variables   &   $\bar{X}_T$ &   $\bar{X}_C$ &   Stand.  Diff.\\
\hline
 Grade &    9.694   &   8.628   &   0.653   \\
 Race   &   0.692   &   0.532       &0.346  \\
 Friends' Grade &   9.631   &   8.806&  0.566   \\
 Friends' Race  &   0.678   &   0.586   &   0.268   \\
 Degree &   9.417 & 3.264   &   1.597   \\[5pt]
 $G_i$  &    6.786  &   2.323 & 1.440   \\
\end{tabular}
\end{table}
\begin{table}
\caption{Covariate balance across dichotomized neighborhood treatment arms \label{tab:sce3_covGdic}}
\centering
\begin{tabular}{lrrrr}
Variables   &   $\bar{X}_{G_i\geq0.5}$  &   $\bar{X}_{G_i<0.5}$ &   Stand.  Diff.\\
\hline
 Grade &     9.408  &    9.464  &   -0.033  \\
 Race   &   0.666 & 0.642       &0.050  \\
 Friends' Grade &   9.512   &   8.128&  1.585   \\
 Friends' Race  &   0.632   &   0.432   &   0.630   \\
 Degree &   11.284  &   4.87    &   1.845   \\[5pt]
 $Z_i$  &   0.989   &    0.548 &    4.422   \\
\end{tabular}
\end{table}
\begin{table}
\caption{Coefficients of logistic regression of neighborhood treatment \label{tab:sce3_covGcoef}}
\centering
\begin{tabular}{lrrrr}
Variables   &   Estimate    &   SE\\
\hline
 Grade &    -0.004  &    0.007      \\
 Race   &    0.012  &   0.018           \\
 Friends' Grade &   0.014   &   0.008   \\
 Friends' Race  &   0.084   &   0.024       \\
 Degree &   -0.002  &   0.002       \\[5pt]
 $Z_i$  &   0.033   &   0.024   \\
\end{tabular}
\end{table}

%Naturalmente G come proporzione non dipenderebbe da degree, cioe' solo  il codominio. Per fare dipendere anche G da degree in modo da avere dipendenza tra G e Z, definisco G come il numero dei trattati e non solo la proporzione, tra tutti gli amici.

\paragraph{Scenario 4.}
In this scenario the generating procedure is such that the individual and neighborhood treatments are directly correlated. We used an iterative procedure that, starting form initial values, it assigns the individual treatment depending on individual covariates and on the neighborhood treatment, using the following model
\begin{equation}
\texttt{logit}(P(Z_i=1))=-20+ 2\mathtt{grade}_i+ 3\mathtt{race}_i+ 4G_i
\end{equation}
At each iteration, the neighborhood treatment $G_i$ is simply computed as the proportion of treated friends among the first 5 best friends.
\[
\begin{array}{l|rrr}
                 & All    & Z_i=1 & Z_i=0\\
                 \hline
              N  & 16410 &11773 & 4637\\
\overline{Z} &0.717 &1 &0\\
\overline{G} &0.719&0.915&0.223 \\
\hline
\end{array}
\]
Covariate balance across treatments are reported in Tables \ref{tab:sce4_covZ}, \ref{tab:sce4_covGdic} and \ref{tab:sce4_covGcoef}. Here we can see that there is still a residual dependence between $Z_i$ and $G_i$ even after conditioning for all types of covariates.
\begin{table}
\caption{Covariate balance across individual treatment arms \label{tab:sce4_covZ}}
\centering
\begin{tabular}{lrrrr}
Variables   &   $\bar{X}_T$ &   $\bar{X}_C$ &   Stand.  Diff.\\
\hline
 Grade &    10.204  &   7.431   &   2.197   \\
 Race   &   0.744   &   0.417       &0.750  \\
 Friends' Grade &    9.200  &   8.928&  0.261   \\
 Friends' Race  &   0.591   &   0.536   &   0.170   \\
 Degree &   7.939 & 7.919   &   0.004   \\[5pt]
 $G_i$  &    0.930  &   0.233 & 4.125   \\
\end{tabular}
\end{table}
\begin{table}
\caption{Covariate balance across dichotomized neighborhood treatment arms \label{tab:sce4_covGdic}}
\centering
\begin{tabular}{lrrrr}
Variables   &   $\bar{X}_{G_i\geq0.5}$  &   $\bar{X}_{G_i<0.5}$ &   Stand.  Diff.\\
\hline
 Grade &     10.085 &    7.455  &   1.778   \\
 Race   &   0.716 & 0.463       &0.560  \\
 Friends' Grade &   9.205   &   8.880&   0.313  \\
 Friends' Race  &   0.589   &   0.534   &   0.169   \\
 Degree &   7.948   &    7.887  &   0.014   \\[5pt]
 $Z_i$  &   0.925   &    0.105 &    3.122   \\
\end{tabular}
\end{table}

\begin{table}
\caption{Coefficients of logistic regression of neighborhood treatment \label{tab:sce4_covGcoef}}
\centering
\begin{tabular}{lrrrr}
Variables   &   Estimate    &   SE\\
\hline
 Grade &     0.802  &    0.013      \\
 Race   &    0.534  &   0.027           \\
 Friends' Grade &   0.176   &   0.039   \\
 Friends' Race  &   0.106   &   0.012       \\
 Degree &   0.026   &   0.003       \\[5pt]
 $Z_i$  &   1.934   &   0.034   \\
\end{tabular}
\end{table}

\subsection{Outcome Models}
\label{apx:outcome}

\subsubsection{Outcome Models for Main Effects Simulations}
\label{apx:outcome1}

In the first simulation study for the estimation of main effects, the potential outcomes
have the following distribution:
\begin{equation}
 \label{eq:outcome1}
\begin{aligned}
Y_i(z,g)| \vX_i^{ind}\sim &\mathcal{N}\big(\mu(z, g, \vX_i^{ind}), 1\big)\\
\mu(z, g,\vX_i^{ind})=15-7\mathbf{I}(\phi(1;\vX_i^{ind})\geq0.7)&-15z+ 3z\mathbf{I}(\phi(1;\vX_i^{ind})\geq0.7)
+\delta g
\end{aligned}
\end{equation}
with
$\vX^{ind}_i=[\mathtt{race}_i+\mathtt{grade}_i]$ and $\vX^{g}_i=[\mathtt{race}_i,\mathtt{grade}_i,\mathtt{friends.race}_i,\mathtt{friends.grade}_i, \mathtt{degree}_i ]$ and $\delta \in (-5,-8,-10)$ (or $\delta \in (-0.3,-0.5,-0.8)$ for Scenario 3), corresponding to a low, medium and high level of interference.
According to this model, the main effects are
\begin{equation}
 \label{eq:main}
\begin{aligned}
\tau(g)=-15 + 3\mathbf{I}(\phi(1;\vX_i^{ind})\geq0.7)
\quad \forall g\in \mathcal{G}\quad \Longrightarrow \quad \tau=-15+ 3\mathbf{I}(\phi(1;\vX_i^{ind})\geq0.7)
\\
\end{aligned}
\end{equation}
and spillover effects are
\begin{equation}
\begin{aligned}
\delta(g;z)=\delta g  \quad \Longrightarrow \quad \Delta(z)=\delta E[G_i]\quad \forall z=0,1
\end{aligned}
\end{equation}

\subsubsection{Outcome Models for Spillover Effects Simulations}
\label{apx:outcome2}

In the second simulation study for the estimation of spillover effects, the potential outcomes
have the following distribution:
\begin{equation}
 \label{eq:outcome2}
\begin{aligned}
&\qquad Y_i(z,g)| \vX_i^{ind},\vX_i^{g} \sim \mathcal{N}\big(\mu(z, g, \vX_i^{ind},\vX_i^{g}), 1\big)\\
\mu(z, g, \vX_i^{ind},\vX_i^{g})=&15+\mathtt{friends.grade}_i+7\mathtt{friends.race}_i-10\mathbf{I}(\phi(1;\vX_i^{ind})\geq0.7)-10z\\
&+\delta g -10\lambda(g;z,\vX_i^{g})+5g\mathbf{I}(\phi(1;\vX_i^{ind})\geq0.7)+3zg
\end{aligned}
\end{equation}
with
$\vX^{ind}_i=[\mathtt{race}_i+\mathtt{grade}_i]$ and $\vX^{g}_i=[\mathtt{race}_i,\mathtt{grade}_i,\mathtt{friends.race}_i,\mathtt{friends.grade}_i, \mathtt{degree}_i ]$ and $\delta \in (-5,-8,-10)$ (or $\delta \in (-0.3,-0.5,-0.8)$ for Scenario 3), corresponding to a low, medium and high level of interference.
In Scenario 3, where $G_i$ is the number of treated friends, the coefficients of the last two terms of Equation \eqref{eq:outcome2} are scaled by $1/10$.
According to this model, the main effects are
\begin{equation}
\begin{aligned}
\tau(g)=-10 +3g \quad \Longrightarrow \quad \tau=-10 +3E[G_i]\\
\end{aligned}
\end{equation}
and spillover effects are
\begin{equation}
 \label{eq:spill2}
\begin{aligned}
&\delta(g;z)=\delta g  5gE[\mathbf{I}(\phi(1;\vX_i^{ind})\geq0.7)]-10\lambda(g;z,\vX_i^{g})+3zg \\
&\Longrightarrow \, \Delta(z)=\delta E[G_i]+5E[G_i]E[\mathbf{I}(\phi(1;\vX_i^{ind})\geq0.7)]-10E[\lambda(G_i;z,\vX_i^{g})]+3zE[G_i]
\end{aligned}
\end{equation}

%\cmntL{I moved this section here}

\subsection{True Main and Spillover Effects}

In Table \ref{tab:scen} we present a description of all scenarios in terms of correlation between $Z_i$ and $G_i$, the true values of the overall main effect $\tau$ (according to the outcome model in Equation  \eqref{eq:outcome1}), and the overall spillover effects $\Delta(0)$ and $\Delta(1)$ (according to the outcome model in Equation  \eqref{eq:outcome2}). True values are computed as the mean of the values estimated in each simulation data according to equations \eqref{eq:main} and  \eqref{eq:spill2}, respectively.
%We also report the bias for $\tau$ that would result from na\"ively estimating the main effect of the individual treatment, without taking interference into account and only adjusting for individual covariates. This bias is computed using Equation \eqref{eq:bias} and according to the outcome model in Equation \eqref{eq:outcome1}.
%Within each combination of individual covariates $ \vX^{ind}_i= \vx^{ind}$, the first part of Equation \eqref{eq:bias} , i.e., $E[Y^{obs}_i|Z_i=z, G_i=g, \vX^{ind}_i=\vx^{ind}]-E[Y^{obs}_i|Z_i=z, G_i=g', \vX^{ind}_i=\vx^{ind}]$, represents the level of interference and, when the potential outcomes follow the distribution in Equation \eqref{eq:outcome1}, it is measured by $\delta g$. The second part of Equation \eqref{eq:bias}, i.e., $ P(G_i=g|Z_i=1,\vX^{ind}_i=\vx^{ind})-P(G_i=g|Z_i=0,\vX^{ind}_i=\vx^{ind})$, depends on residual dependence between conditional on the individual covariates.
We also report, for each scenario, a measure of partial correlation between $Z_i$ and $G_i$, conditional on the individual covariates:
%Instead of the value of this quantity for each $g\in\mathcal{G}$, we report the following measure of partial correlation:
\[
\rho_{ZG|X^{ind}}=\frac{R^2_{G|Z,X^{ind}}-R^2_{G|X^{ind}} }{1-R^2_{G|X^{ind}}}
\]
It is worth noting that the bias for the overall main effect $\tau$, when $\vX^{\star}_i=\vX^{ind}_i$, is not directly proportional to this measure of association, but to the quantity $ P(G_i=g|Z_i=1,\vX^{ind}_i=\vx)-P(G_i=g|Z_i=0,\vX^{ind}_i=\vx)$, for each value $g\in \mathcal{G}$, as seen in Equation \eqref{eq:bias}.
\begin{table}
\caption{Simulation Scenarios and True Effects}
\centering
\begin{tabular}{cc|c|ccc}
\multicolumn{2}{c|}{Scenario}&& &&\\
& interference &$\rho_{ZG|X^{ind}}$& $\tau$&$\Delta(0)$&$\Delta(1)$\\
\hline
\multirow{3}{4cm}{\centering 1\\($Z_i\ind G_i|\vX_i^{ind}$)}&low& 0.002  &-13.107 &-1.961&-0.513\\
&medium&   0.002& -13.107 &-3.421&-1.974\\
&high&  0.002 &-13.107  &-4.384&-2.937\\[0.3cm]
\multirow{3}{4cm}{\centering 2\\($Z_i\ind G_i|\vX_i^{ind},\vX_i^{neigh} $)}&low&0.166   & -13.109 &-2.542&-0.905\\
&medium& 0.166  &-13.109  &-4.223&-2.585\\
&high&0.166   &-13.109 &-5.336& -3.699\\[0.3cm]
\multirow{3}{5cm}{\centering 3\\($Z_i\ind G_i|\vX_i^{ind},\mathtt{degree}_i$)}&low&   0.236&-13.177  &-0.997&0.605\\
&medium&0.236   & -13.177 &-2.065&-0.462\\
&high& 0.236  & -13.177&-3.667&-2.064\\[0.3cm]
\multirow{3}{5cm}{\centering 4\\($Z_i\nind G_i|\vX_i^{z},\vX_i^{neigh}$)}&low&  0.403 &-13.109  &-1.789&-1.239\\
&medium&   0.403 &-13.109 &-3.861&-3.311\\
&high&   0.403 &-13.109 &-5.243&-4.693\\
\end{tabular}
\label{tab:scen}
\end{table}%
%%% %%% %%%
%%% %%% %%%
%%% %%% %%%

\section{Subclassification and Generalized Propensity Score Estimator}
\label{app:addres}

\begin{enumerate}[label=\arabic*)]
\item We derive a subclassification on the individual propensity score $\phi(1;\vX^z_i)$ as follows:
\begin{enumerate}[label=\alph*)]
\item We estimate $\phi(1;\vX^z_i)$ with a logistic regression for $Z_i$ conditional on covariates $\vX^z_i$;
\item We predict $\phi(1;\vX^z_i)$ for each unit;
\item We identify J subclasses $B_j$, with $j=1, \ldots, J$, defined by similar values of $\phi(1;\vX^z_i)$ and where there is sufficient balance between individual treatment groups, i.e., $\vX_i^z\ind Z_i| i\in B_j$.\end{enumerate}
\item Within each subclass $B_j$, we repeat the following steps to estimate $\mu_j(z,g)=E\big[Y_i(z,g)|\,  i\in B^g_j\big]$, where $B_j^g=V_g\cap B_j$:
\begin{enumerate}[label=\alph*)]
\item We estimate the parameters of a model for the neighborhood propensity score $ \lambda(g; z; x^g)$: $\lambda(z; g; \vX^g_i)= Pr(G_i=g| Z_i=z, \vX^g_i)=f^{G}(g, z, \vX^g_i)$
%For example, if $G_i$ is defined as the number of treated units in the neighborhood $\mathcal{N}_i$, then we can assume a binomial distribution for $G_i$ and model the probability of being treated using a logistic regression weighted by the unit's degree $N_i$;
\item We use the observed data ($Y_i, Z_i, G_i, \vX^g_i$) and $\widehat{\Lambda}=\lambda(G_i; Z_i; \vX^g_i)$ to estimate the parameters of a model $Y_i(z,g)\mid \lambda(z; g; \vX^g_i) \sim f^Y(z, g, \lambda(g; z; \vX^g_i))$;
\item  For a particular level of the joint treatment $(Z_i=z, G_i=g)$, for each unit $i \in B^g_j$ we predict the neighborhood propensity score evaluated at that level of the treatment, i.e., $\lambda(g; z; \vX^g_i)$, and use it to predict the potential outcome $Y_i(z,g)$.
\item To estimate the dose-response function $\mu_j(z,g)$ we average the potential outcomes over $\lambda(z; g; \vX^g_i)$
\[\widehat{\mu}_j(z,g)=\frac{\sum_{i\in B^g_j} \widehat{Y}_i(z,g)}{|B^g_j|}\]
\end{enumerate}
\item We derive the average dose-response function as follows:
\begin{equation*}
\widehat{\mu}(z,g)=\sum_{j=1}^J \widehat{\mu}_j(z,g) \pi^g_j
\end{equation*}
where $\pi^g_j=\frac{|B^g_j|}{v_g}$.
\end{enumerate}

\subsection{Individual Treatment, Neighborhood Treatment and Outcome Models}

%\cmntL{NEW SECTION}

Here we detail the estimation procedure used in the simulation study. We provide details on the models used for the individual propensity score, the neighborhood propensity score and the outcome.

\begin{enumerate}[label=\arabic*)]
\item We derive a subclassification on the individual propensity score $\phi(1;\vX^z_i)$ as follows:
\begin{enumerate}[label=\alph*)]
\item We first assume a Bernoulli distribution for the individual treatment, 
\begin{equation}
\begin{aligned}
Z_i \sim Ber(\phi(1;\vX^z_i))
\end{aligned}
\end{equation}
with 
\begin{equation}
\begin{aligned}
\textrm{logit}(\phi(1;\vX^z_i))= \alpha_0+ \boldsymbol{\alpha_x^T}\vX_i^z
\end{aligned}
\end{equation}
Parameters $\boldsymbol{\alpha}=[\alpha_0, \boldsymbol{\alpha_x^T}]$ are estimated using a logistic regression. 
\item  Given estimated parameters $\widehat{\boldsymbol{\alpha}}$, for each unit we compute the individual propensity score $\phi(1;\vX^z_i)$.
\item We identify J subclasses $B_j$, with $j=1, \ldots, J$, defined by similar values of $\phi(1;\vX^z_i)$ and where there is sufficient balance between individual treatment groups, i.e., $\vX_i^z\ind Z_i| i\in B_j$. In the simulation study we choose 5 subclasses based on quintiles. For recommendations on the choice of the number of subclasses and the boundaries see \citet{Imbens:Rubin:2015}
\end{enumerate}
\item Within each subclass $B_j$, we repeat the following steps to estimate $\mu_j(z,g)=E\big[Y_i(z,g)|\,  i\in B^g_j\big]$, where $B_j^g=V_g\cap B_j$:
\begin{enumerate}[label=\alph*)]
\item We posit a model for the neighborhood treatment assuming a distribution that depends on its specific definition. 
In scenarios 1,2, and 4, where $G_i$ is defined as  the proportion of treated neighbors, we posit the following model:
\begin{equation}
G_i N_i\sim Bin(N_i, \pi(z, \vX_i^g)) 
\end{equation}
with 
\begin{equation}
\textrm{logit}(\pi(z, \vX_i^g))= \gamma_0+ \gamma_z z+\boldsymbol{\gamma_x^T}\vX_i^g
\end{equation}
Parameters $\boldsymbol{\gamma}=[\gamma_0, \gamma_z,\boldsymbol{\gamma_x^T}]$ are estimated using a binomial logistic regression. 
In scenarios 3, where $G_i$ is defined as  the number of treated neighbors, we simply have:
\begin{equation}
G_i \sim Bin(N_i, \pi(z, \vX_i^g)) 
\end{equation}
\item Given estimated parameters $\widehat{\boldsymbol{\gamma}}$, for each unit we compute $\widehat{\Lambda}_i=\lambda(G_i; Z_i; \vX^g_i)$
%For example, if $G_i$ is defined as the number of treated units in the neighborhood $\mathcal{N}_i$, then we can assume a binomial distribution for $G_i$ and model the probability of being treated using a logistic regression weighted by the unit's degree $N_i$;
\item We use the observed data ($Y_i, Z_i, G_i, \vX^g_i$) and $\widehat{\Lambda}_i=\lambda(G_i; Z_i; \vX^g_i)$ to estimate the parameters of a model $Y_i(z,g)\mid \lambda(z; g; \vX^g_i) \sim f^Y(z, g, \lambda(g; z; \vX^g_i))$. In particular, in the simulation study we fit the following linear regression:
\begin{equation}
Y_i(z,g)= \beta_0 + \beta_z z + \beta_g g + \beta_{zg} zg +  \beta_{\lambda}\widehat{\lambda}_i(g; z; \vX^g_i) + \beta_{g\lambda}g\widehat{\lambda}_i(g; z; \vX^g_i)
\end{equation}
Alternative models could be used. In a more realistic scenario we recommend using the cubic polynomial function proposed by \citet{Hirano:Imbens:2004}.
\item  For a particular level of the joint treatment $(Z_i=z, G_i=g)$, for each unit $i \in B^g_j$ we predict the neighborhood propensity score evaluated at that level of the treatment, i.e., $\lambda(g; z; \vX^g_i)$, and use it to predict the potential outcome $Y_i(z,g)$.
\item To estimate the dose-response function $\mu_j(z,g)$ we average the potential outcomes over $\lambda(z; g; \vX^g_i)$
\[\widehat{\mu}_j(z,g)=\frac{\sum_{i\in B^g_j} \widehat{Y}_i(z,g)}{|B^g_j|}\]
\end{enumerate}
\item We derive the average dose-response function as follows:
\begin{equation*}
\widehat{\mu}(z,g)=\sum_{j=1}^J \widehat{\mu}_j(z,g) \pi^g_j
\end{equation*}
where $\pi^g_j=\frac{|B^g_j|}{v_g}$.

\end{enumerate}

%\cmntL{Should we say something on how we estimate overall effects?}

\subsection{{Statistical Inference and Interval Estimation}}
%\label{app:simplus}

%\subsection{Propensity Score-Based Estimator: Inference Results}

%\cmntL{I reported the bootstrap section here. We should perhaps add an introduction and fix some sentences.}

In this section, we focus on the derivation of standard errors and confidence intervals for our proposed propensity score estimator.
We propose the use of a bootstrap method with an independent resampling scheme with
replacement. After computing the neighborhood treatment $G_i$ and the neighborhood covariates for each unit, these are considered as node attributes.  
%Conditional on individual and neighborhood treatments and covariates units are considered independent and 
the bootstrap procedure resamples units independently. 
%according an hypothetical egocentric sampling.

%In this section, we focus on our proposed propensity score estimator.
In Table \ref{tab:est} we present the results of using this estimator to perform inference about main and spillover effects. For every scenario,  we simulated one dataset and applying the procedure described in Section \ref{sec:estimator} to derive point estimates, standard errors, and confidence intervals. Standard errors are the result of 1000 bootstrap replications. Resampling is performed at unit-level.
%\sout{in accordance with the absence of correlation between neighbors' outcomes}.
Also shown are bootstrapped $95\%$ confidence intervals based on normal approximation.

It is worth noting that standard errors are consistent with root mean square errors obtained through Monte Carlo simulations, reported in Tables \ref{tab:main}, \ref{tab:delta0} and  \ref{tab:delta1}.

\begin{table}
\caption{Estimated Effects, Standard Errors and Confidence Intervals}
%\centering
\hspace{-2.25cm}
{\small
\begin{tabular}{@{\!\!\!\!\!\!\!\!\!}c@{\!\!\!\!\!\!\!\!\!\!\!}c@{\,\,}|@{\,}c@{\!\,\,\,\,}c@{\!\,\,\,}c|@{\!\,\,\,}c@{\!\,\,\,\,}c@{\!\,\,\,}c@{\!\,\,\,}|@{\!\,\,\,}c@{\!\,\,\,}c@{\!\,\,\,}c}
\multicolumn{2}{c@{\!\!\!\!\!}}{Scenario}& &&&&&&&\\
& interference & $\widehat{\tau}$&SE&CI&$\widehat{\Delta(0)}$&SE&CI&$\widehat{\Delta(1)}$&SE&CI\\
\hline
\multirow{3}{4cm}{\centering 1\\($Z_i\ind G_i|\vX_i^{ind}$)}&low&-13.174 &0.095&[-13.356,-12.985]&-2.072 &0.075&[-2.219, -1.923]&-0.576&0.045&[-0.663,-0.487]\\
&medium&-13.133 &0.127&[-13.379,-12.878]&-3.572&0.078&[-3.729,-3.423]&-2.111&0.043&[-2.197,-2.026]\\
&high&-13.077 &0.171&[-13.405,-12.731]&-4.519&0.074&[-4.664,-4.375]&-3.093&0.044&[-3.179,-3.005]\\[0.3cm]
\multirow{3}{4cm}{\centering 2\\($Z_i\ind G_i|\vX_i^{ind},\vX_i^{neigh} $)}&low&-13.083 &0.056&[-13.192,-12.973]&-2.413& 0.071& [-2.553, -2.273]&-0.815& 0.046&[-0.904,-0.725]\\
&medium& -13.111&0.073&[-13.254,-12.968]&-4.097&0.072&[-4.239,-3.957]&-2.447&0.044&[-2.535,-2.360]\\
&high&-13.156 &0.081&[-13.314,-12.997]& -5.138&0.082&[-5.303,-4.982]&-3.506&0.046&[-3.598,-3.412] \\[0.3cm]
\multirow{3}{5cm}{\centering 3\\($Z_i\ind G_i|\vX_i^{ind},\mathtt{degree}_i$)}&low& -13.234&0.178&[-13.582,-12.885]&-1.140&0.100&[-1.333,-0.942]& 0.581&0.033&[0.518,0.646]\\
&medium&-13.121 &0.254&[-13.619,-12.623]& -1.979&0.097&[-2.180,-1.800]&-0.452&0.031&[-0.514,-0.392]\\
&high& -13.163&0.228&[-13.609,-12.716]&-3.661&0.111&[-3.885,-3.451]&-2.090&0.031&[-2.152.-2.029]\\[0.3cm]
 \multirow{3}{5cm}{\centering 4\\($Z_i\nind G_i|\vX_i^{z},\vX_i^{neigh}$)}&low& -13.085&0.073&[-13.229,-12.943]&-1.695&0.109&[-1.922,-1.493]&-1.289&0.089&[-1.462,-1.114]\\
&medium&-13.061&0.071&[-13.201,-12.924]&  -3.913 &0.117&[-4.151,-3.694]&-3.273 &0.092&[-3.456,-3.095]\\
&high& -13.119&0.079&[-13.271,-12.959]& -5.231&0.120&[-5.479,-5.008]&-4.761&0.088&[-4.934,-4.588]\\
\end{tabular}}
\label{tab:est}
\end{table}%

%\cmntL{I added the discussion here}
The variance of our estimator is assessed with respect to the distribution induced by random sampling using bootstrap methods. The properties of this variance estimator do depend on the sampling mechanism. In particular, the bootstrap resampling mechanism should match the one that was actually used to derive the sample. We have proposed a resampling scheme that matches an egocentric sampling mechanism \citep{Kolaczyk:2009, Perri:2018}, in that we take independent samples with replacement where sampled units carry with them both the neighborhood treatment $G_i$ and neighborhood covariates, regardless of whether their neighbors have been resampled. 
This approach is valid when the observed sample was actually obtained by an egocentric sampling mechanism.
%it is also when units' outcomes are independent given the observed covariates. 
In such a case, the properties of our variance estimator should be similar to the ones discussed in \citet{Imbens:Rubin:2015}. In the simulation study, the performance of our estimator has been assessed under outcomes independence. The results of our simulations are promising.  
However, the sampling mechanism is not independent 
%when either the sampling mechanism is not independent or units' outcomes are not independent, 
our solution is not ideal and further investigation is needed. 
In the case of clustered sampling, which sample separated
clusters of data where there are no links between clusters, a clustered bootstrap could be used. 
However, when individuals within clusters are organized in networks,
smaller clusters with higher correlation can also be defined using a community detection algorithm.
We will further investigate these issues in future work.

%In fact, since we treat the neighborhood treatment $G_i$ as if it were a second treatment assigned to unit $i$, we are allowed to resample individuals, regardless of their neighbors. However, standard errors derived with such a bootstrapping procedure are valid only if there is no correlation between neighbors' outcomes. If this is not the case, we would have to use a block bootstrap method  that attempts to replicate the correlation by resampling  blocks of data.
%
%Given the complex structure of social networks, where neighborhoods are not well separated and different units may share few neighbors, we could use graph partitioning methods as sampling procedures. Investigating these alternative sampling procedures, however, goes beyond the scope of this paper.

%%% %%% %%%
%%% %%% %%%
%%% %%% %%%

\section{Proofs}
\label{app:proofs}

\subsection{Proof of Theorem \ref{theo: identification}}

We want to prove that the adjusted observed mean $\overline{Y}^{obs}_{z, g}$ is equal to the quantity $\mu(z,g)$.
\begin{equation*}
\begin{aligned}
\overline{Y}^{obs}_{z, g}&=E[E\bigm[Y_i| Z_i=z, G_i=g, \vX_i=\vx, i \in V_g] | \, i \in V_g]\\
&= E[E\bigm[Y_i(z,g)| Z_i=z, G_i=g, \vX_i=\vx, i \in V_g] | \, i \in V_g]\\
&= E[E\bigm[Y_i(z,g)| \vX_i=\vx, i \in V_g] | \, i \in V_g]=E\big[Y_i(z,g)| \, i \in V_g\big]= \mu(z,g)
\end{aligned}
\end{equation*}
where the fist equality is the definition of the adjusted observed mean, expressed in terms of conditional mean, the second equality holds by both Assumption \ref{ass: consistency} (No Multiple Versions of Treatment) and Assumption \ref{ass:SUTNVA} (Neighborhood Interference), and the third equality holds by the Assumption \ref{ass: Totunconf} (Unconfoundedness).

\subsection{Proof of Theorem \ref{theo: biasA}}

\begin{equation*}
\begin{aligned}
\tau_{X^{\star}}^{obs}&=
\sum_{\vx \in \mathcal{X }^{\star}}\bigg(E[Y_i| Z_i=1,\vX^{\star}_i=\vx \, ]- E[Y_i| Z_i=0,\vX^{\star}_i=\vx \, ]\bigg) P(\vX^{\star}_i=\vx)\\
&=\sum_{\vx \in \mathcal{X }^{\star}}\bigg(\sum_{g\in \mathcal{G}}E[Y_i| Z_i=1, G_i=g,\vX^{\star}_i=\vx , i\in V_g]P(G_i=g|Z_i=1,\vX^{\star}_i=\vx)\\
&\quad- E[Y_i| Z_i=0, G_i=g,\vX^{\star}_i=\vx , i\in V_g ]P(G_i=g|Z_i=0,\vX^{\star}_i=\vx)\bigg) P(\vX^{\star}_i=\vx)
\end{aligned}
\end{equation*}
by iterated equations and by definition of $V_g$ leading to $P(G_i=g|Z_i=z,\vX^{\star}_i=\vx, i \notin V_g)=0$
\begin{equation*}
\begin{aligned}
&=\sum_{\vx \in \mathcal{X }^{\star}}\bigg(\sum_{g\in \mathcal{G}}E[Y_i(1,g)| Z_i=1, G_i=g,\vX^{\star}_i=\vx , i\in V_g  ]P(G_i=g|Z_i=1,\vX^{\star}_i=\vx)\\
&\quad- E[Y_i(0,g)| Z_i=0, G_i=g,\vX^{\star}_i=\vx  , i\in V_g]P(G_i=g|Z_i=0,\vX^{\star}_i=\vx)\bigg) P(\vX^{\star}_i=\vx)
\end{aligned}
\end{equation*}
by consistency
\begin{equation*}
\begin{aligned}
&=\sum_{\vx \in \mathcal{X^{\star} }}\bigg(\sum_{g\in \mathcal{G}}E[Y_i(1,g)| \vX^{\star}_i=\vx  , i\in V_g ]P(G_i=g|Z_i=1,\vX^{\star}_i=\vx)\\
&\quad- E[Y_i(0,g)| \vX^{\star}_i=\vx  , i\in V_g ]P(G_i=g|Z_i=0,\vX^{\star}_i=\vx)\bigg) P(\vX^{\star}_i=\vx)\\
 \end{aligned}
\end{equation*}
by the unconfoundedness assumption.

\subsection{Proof of Corollary \ref{cor: bias1}}

\begin{equation*}
\begin{aligned}
\tau_{X^{\star}}^{obs}-\tau&=\bigg\{\sum_{\vx \in \mathcal{X }^{\star}}\bigg(E[Y_i| Z_i=1,\vX^{\star}_i=\vx  ]- E[Y_i| Z_i=0,\vX^{\star}_i=\vx  ]\bigg) P(\vX^{\star}_i=\vx)\bigg\}\\
&\quad- \bigg\{\sum_{g\in \mathcal{G}}\bigg(E[Y_i(1, g)-Y_i(0, g)|\, i\in V_g]\bigg)P(G_i=g)\bigg\} \\
&=\bigg\{\sum_{\vx \in \mathcal{X }^{\star}}\bigg(\sum_{g\in \mathcal{G}}E[Y_i| Z_i=1, G_i=g,\vX^{\star}_i=\vx  , i\in V_g ]P(G_i=g|Z_i=1,\vX^{\star}_i=\vx)\\
&\quad- E[Y_i| Z_i=0, G_i=g,\vX^{\star}_i=\vx  , i\in V_g ]P(G_i=g|Z_i=0,\vX^{\star}_i=\vx)\bigg) P(\vX^{\star}_i=\vx)\bigg\}\\
&\quad- \bigg\{\sum_{\vx \in \mathcal{X }^{\star}}\sum_{g\in \mathcal{G}}\bigg(E[Y_i(1, g)-Y_i(0, g)|\vX^{\star}_i=\vx  , i\in V_g ]\bigg)P(G_i=g|\vX^{\star}_i=\vx)P(\vX^{\star}_i=\vx)\bigg\} \end{aligned}
\end{equation*}
by iterated equations
\begin{equation*}
\begin{aligned}
&=\bigg\{\sum_{\vx \in \mathcal{X }^{\star}}\bigg(\sum_{g\in \mathcal{G}}E[Y_i| Z_i=1, G_i=g,\vX^{\star}_i=\vx  , i\in V_g ]P(G_i=g|Z_i=1,\vX^{\star}_i=\vx)\\
&\quad- E[Y_i| Z_i=0, G_i=g,\vX^{\star}_i=\vx  , i\in V_g ]P(G_i=g|Z_i=0,\vX^{\star}_i=\vx)\bigg) P(\vX^{\star}_i=\vx)\bigg\}\\
&\quad- \bigg\{\sum_{\vx \in \mathcal{X }^{\star}}\sum_{g\in \mathcal{G}}\bigg(E[Y_i(1, g)|Z_i=1, G_i=g,\vX^{\star}_i=\vx  , i\in V_g ]-E[Y_i(0, g)Z_i=0, G_i=g,\vX^{\star}_i=\vx  , i\in V_g ]\bigg)\\&\qquad \qquad P(G_i=g|\vX^{\star}_i=\vx)P(\vX^{\star}_i=\vx)\bigg\} \end{aligned}
\end{equation*}
by the unconfoundedness assumption
\begin{equation*}
\begin{aligned}
&=\bigg\{\sum_{\vx \in \mathcal{X }^{\star}}\bigg(\sum_{g\in \mathcal{G}}E[Y_i| Z_i=1, G_i=g,\vX^{\star}_i=\vx  , i\in V_g ]P(G_i=g|Z_i=1,\vX^{\star}_i=\vx)\\
&\quad- E[Y_i| Z_i=0, G_i=g,\vX^{\star}_i=\vx  , i\in V_g ]P(G_i=g|Z_i=0,\vX^{\star}_i=\vx)\bigg) P(\vX^{\star}_i=\vx)\bigg\}\\
&\quad- \bigg\{\sum_{\vx \in \mathcal{X }^{\star}}\sum_{g\in \mathcal{G}}\bigg(E[Y_i|Z_i=1, G_i=g,\vX^{\star}_i=\vx  , i\in V_g ]-E[Y_i|Z_i=0, G_i=g,\vX^{\star}_i=\vx  , i\in V_g]\bigg)\\&\qquad\qquad \qquad P(G_i=g|\vX^{\star}_i=\vx)P(\vX^{\star}_i=\vx)\bigg\}=0
\end{aligned}
\end{equation*}
by consistency and by the assumption of independence between $Z_i$ and $G_i$.

\subsection{Proof of Corollary \ref{cor: bias2}}

We continue the previous proof without the assumption of independence between $Z_i$ and $G_i$ conditional on $\vX^{\star}$. Then
\begin{equation*}
\begin{aligned}
\tau_{X^{\star}}^{obs}-\tau&=\sum_{\vx \in \mathcal{X }^{\star}}\bigg(\sum_{g\in \mathcal{G}}E[Y_i|Z_i=1, G_i=g, \vX^{\star}_i=\vx  , i\in V_g ]-E[Y_i|Z_i=1, G_i=g', \vX^{\star}_i=\vx  , i\in V_g ]\bigg)\\
& \qquad \qquad\qquad \bigg( P(G_i=g|Z_i=1,\vX^{\star}_i=\vx)-P(G_i=g|\vX^{\star}_i=\vx)\bigg)P(\vX^{\star}_i=\vx)\\
&\quad -\sum_{\vx \in \mathcal{X }^{\star}}\bigg(\sum_{g\in \mathcal{G}}E[Y_i|Z_i=0, G_i=g, \vX^{\star}_i=\vx  , i\in V_g ]-E[Y_i|Z_i=0, G_i=g', \vX^{\star}_i=\vx  , i\in V_g ]\bigg)\\
& \qquad \qquad\qquad\bigg( P(G_i=g|Z_i=0,\vX^{\star}_i=\vx)-P(G_i=g|\vX^{\star}_i=\vx)\bigg)P(\vX^{\star}_i=\vx)
\end{aligned}
\end{equation*}
since $E[Y_i| Z_i=1, G_i=g', \vX^{\star}_i=\vx  , i\in V_g]$ and $E[Y_i| Z_i=0, G_i=g', \vX^{\star}_i=\vx  , i\in V_g ]$ do not depend on $g$. If there is no interaction between the individual treatment $Z_i$ and the neighborhood treatment $G_i$, the simplified expression of the bias formula is straightforward.

\subsection{Proof of Theorem \ref{theo: biasB}}

\begin{equation*}
\begin{aligned}
\tau_{X^{\star}}^{obs}-\tau&=\bigg\{\sum_{\vx \in \mathcal{X }^{\star}}\bigg(E[Y_i| Z_i=1,\vX^{\star}_i=\vx \, ]- E[Y_i| Z_i=0,\vX^{\star}_i=\vx \, ]\bigg) P(\vX^{\star}_i=\vx)\bigg\}\\
&\quad- \bigg\{\sum_{g\in \mathcal{G}}\bigg(E[Y_i(1, g)-Y_i(0, g)|i \in V_g]\bigg)P(G_i=g)\bigg\} \\
&=\bigg\{\sum_{\vx \in \mathcal{X }^{\star}}\bigg(\sum_{g\in \mathcal{G}}\sum_{\mathbf{u} \in \mathcal{U }}E[Y_i| Z_i=1, G_i=g,\mathbf{U}_i=\mathbf{u}, \vX^{\star}_i=\vx  , i\in V_g ]\\
&\qquad \qquad \qquad \qquad \qquad P(\mathbf{U}_i=\mathbf{u}|Z_i=1,G_i=g, \vX^{\star}_i=\vx)P(G_i=g|Z_i=1,\vX^{\star}_i=\vx)\\
&\qquad \qquad \qquad \qquad - E[Y_i| Z_i=0, G_i=g,\mathbf{U}_i=\mathbf{u}, \vX^{\star}_i=\vx  , i\in V_g ]\\
&\qquad \qquad \qquad \qquad \qquad P(\mathbf{U}_i=\mathbf{u}|Z_i=0,G_i=g, \vX^{\star}_i=\vx)P(G_i=g|Z_i=0,\vX^{\star}_i=\vx)\bigg) P(\vX^{\star}_i=\vx)\bigg\}\\
&\quad- \bigg\{\sum_{\vx \in \mathcal{X }^{\star}}\sum_{g\in \mathcal{G}}\sum_{\mathbf{u} \in \mathcal{U }}\bigg(E[Y_i(1, g)|\mathbf{U}_i=\mathbf{u}, \vX^{\star}_i=\vx  , i\in V_g ]-E[Y_i(0, g)|\mathbf{U}_i=\mathbf{u}, \vX^{\star}_i=\vx  , i\in V_g ]\bigg)\\
&\qquad \qquad \qquad \qquad \qquad P(\mathbf{U}_i=\mathbf{u}|G_i=g, \vX^{\star}_i=\vx)P(G_i=g|\vX^{\star}_i=\vx)P(\vX^{\star}_i=\vx)\bigg\}
\end{aligned}
\end{equation*}
by iterated equations and by definition of $V_g$
\begin{equation*}
\begin{aligned}
&=\bigg\{\sum_{\vx \in \mathcal{X }^{\star}}\bigg(\sum_{g\in \mathcal{G}}\sum_{\mathbf{u} \in \mathcal{U }}E[Y_i| Z_i=1, G_i=g,\mathbf{U}_i=\mathbf{u}, \vX^{\star}_i=\vx  , i\in V_g ]P(\mathbf{U}_i=\mathbf{u}|Z_i=1,G_i=g, \vX^{\star}_i=\vx)\\
&\qquad \qquad \qquad \qquad \qquad P(G_i=g|Z_i=1,\vX^{\star}_i=\vx)\\
&\qquad \qquad \qquad \qquad - E[Y_i| Z_i=0, G_i=g,\mathbf{U}_i=\mathbf{u}, \vX^{\star}_i=\vx  , i\in V_g ]P(\mathbf{U}_i=\mathbf{u}|Z_i=0,G_i=g, \vX^{\star}_i=\vx)\\
&\qquad \qquad \qquad \qquad \qquad P(G_i=g|Z_i=0,\vX^{\star}_i=\vx)\bigg) P(\vX^{\star}_i=\vx)\bigg\}\\
&\quad- \bigg\{\sum_{\vx \in \mathcal{X }^{\star}}\sum_{g\in \mathcal{G}}\sum_{\mathbf{u} \in \mathcal{U }}\bigg(E[Y_i|Z_i=1, G_i=g,\mathbf{U}_i=\mathbf{u}, \vX^{\star}_i=\vx  , i\in V_g]\\
&\hspace{4cm}-E[Y_i|Z_i=0, G_i=g,\mathbf{U}_i=\mathbf{u}, \vX^{\star}_i=\vx  , i\in V_g ]\bigg)\\
&\qquad \qquad \qquad \qquad \qquad P(\mathbf{U}_i=\mathbf{u}|G_i=g, \vX^{\star}_i=\vx)P(G_i=g|\vX^{\star}_i=\vx)P(\vX^{\star}_i=\vx)\bigg\}
\end{aligned}
\end{equation*}
by the unconfoundedness assumption conditional on $\vX^{\star}_i$ and $\mathbf{U}_i$ and by consistency
\begin{equation*}
\begin{aligned}
&=\sum_{\vx \in \mathcal{X^{\star} }}\sum_{g\in \mathcal{G}}\sum_{u\in \mathcal{U}} \bigg(E[Y_i|Z_i=1, G_i=g, \mathbf{U}_i=\mathbf{u},\vX^{\star}_i=\vx  , i\in V_g ]\\&\hspace{4cm}-E[Y_i|Z_i=1, G_i=g' ,\mathbf{U}_i=\mathbf{u}',\vX^{\star}_i=\vx  , i\in V_g ]\bigg)\\
& \qquad \qquad\qquad \qquad\qquad \bigg( P(\mathbf{U}_i=\mathbf{u}|Z_i=1,G_i=g, \vX^{\star}_i=\vx)P(G_i=g|Z_i=1,\vX^{\star}_i=\vx)\\
&\qquad \qquad\qquad\qquad\qquad -P(\mathbf{U}_i=\mathbf{u}|G_i=g, \vX^{\star}_i=\vx)P(G_i=g|\vX^{\star}_i=\vx)\bigg)P(\vX^{\star}_i=\vx)\\
&\quad -\sum_{\vx \in \mathcal{X^{\star} }}\sum_{g\in \mathcal{G}}\sum_{u\in \mathcal{U}} \bigg(E[Y_i|Z_i=0, G_i=g, \mathbf{U}_i=\mathbf{u},\vX^{\star}_i=\vx  , i\in V_g]\\&\hspace{4cm}-E[Y_i|Z_i=0, G_i=g', \mathbf{U}_i=\mathbf{u}',\vX^{\star}_i=\vx  , i\in V_g ]\bigg)\\
& \qquad \qquad\qquad \qquad\qquad\bigg( P(\mathbf{U}_i=\mathbf{u}|Z_i=0,G_i=g, \vX^{\star}_i=\vx)P(G_i=g|Z_i=0,\vX^{\star}_i=\vx)\\
&\qquad \qquad\qquad\qquad\qquad -P(\mathbf{U}_i=\mathbf{u}|G_i=g, \vX^{\star}_i=\vx)P(G_i=g|\vX^{\star}_i=\vx)\bigg)P(\vX^{\star}_i=\vx)
\end{aligned}
\end{equation*}
 by adding and subtracting two quantities $E[Y_i| Z_i=1, G_i=g', \mathbf{U}_i=\mathbf{u}',\vX^{\star}_i=\vx  , i\in V_g ]$ and $E[Y_i| Z_i=0, G_i=g', \mathbf{U}_i=\mathbf{u}',\vX^{\star}_i=\vx  , i\in V_g ]$ that do not depend on $g$ and $\mathbf{u}$. Now if there is no interaction between the individual treatment $Z_i$ and the neighborhood treatment $G_i$, the bias formula formula reduces to
\begin{equation*}
\begin{aligned}
&=\sum_{\vx \in \mathcal{X^{\star} }}\sum_{g\in \mathcal{G}}\sum_{u\in \mathcal{U}} \bigg(E[Y_i|Z_i=z, G_i=g, \mathbf{U}_i=\mathbf{u},\vX^{\star}_i=\vx  , i\in V_g ]\\&\hspace{4cm}-E[Y_i|Z_i=z, G_i=g', \mathbf{U}_i=\mathbf{u}',\vX^{\star}_i=\vx  , i\in V_g ]\bigg)\\
& \qquad \qquad\qquad\qquad \bigg( P(\mathbf{U}_i=\mathbf{u}|Z_i=1,G_i=g, \vX^{\star}_i=\vx)P(G_i=g|Z_i=1,\vX^{\star}_i=\vx) \\& \qquad \qquad\qquad\qquad\quad -P(\mathbf{U}_i=\mathbf{u}|Z_i=0,G_i=g, \vX^{\star}_i=\vx)P(G_i=g|Z_i=0,\vX^{\star}_i=\vx)\bigg)P(\vX_i=\vx)
\end{aligned}
\end{equation*}

\subsection{Proof of Corollary \ref{cor: bias3}}

Following the expression of the bias in the proof Theorem \ref{theo: biasB},
\begin{equation*}
\begin{aligned}
\tau_{X^{\star}}^{obs}-\tau&=\bigg\{\sum_{\vx \in \mathcal{X }^{\star}}\bigg(\sum_{g\in \mathcal{G}}\sum_{\mathbf{u} \in \mathcal{U }}E[Y_i| Z_i=1, G_i=g,\mathbf{U}_i=\mathbf{u}, \vX^{\star}_i=\vx  , i\in V_g ]\\
&\qquad \qquad \qquad \qquad\quad  P(\mathbf{U}_i=\mathbf{u}|Z_i=1,G_i=g, \vX^{\star}_i=\vx)P(G_i=g|Z_i=1,\vX^{\star}_i=\vx)\\
&\qquad \qquad \qquad \qquad \quad- E[Y_i| Z_i=0, G_i=g,\mathbf{U}_i=\mathbf{u}, \vX^{\star}_i=\vx  , i\in V_g ]\\
&\qquad \qquad \qquad \qquad  P(\mathbf{U}_i=\mathbf{u}|Z_i=0,G_i=g, \vX^{\star}_i=\vx)P(G_i=g|Z_i=0,\vX^{\star}_i=\vx)\bigg) P(\vX^{\star}_i=\vx)\bigg\}\\
&\!\!\!\!\!- \bigg\{\sum_{\vx \in \mathcal{X }^{\star}}\sum_{g\in \mathcal{G}}\sum_{\mathbf{u} \in \mathcal{U }}\bigg(E[Y_i|Z_i=1, G_i=g,\mathbf{U}_i=\mathbf{u}, \vX^{\star}_i=\vx  , i\in V_g ]\\&\hspace{4cm}-E[Y_i|Z_i=0, G_i=g,\mathbf{U}_i=\mathbf{u}, \vX^{\star}_i=\vx  , i\in V_g ]\bigg)\\
&\qquad \qquad \qquad \qquad \qquad P(\mathbf{U}_i=\mathbf{u}|G_i=g, \vX^{\star}_i=\vx)P(G_i=g|\vX^{\star}_i=\vx)P(\vX^{\star}_i=\vx)\bigg\}
\end{aligned}
\end{equation*}
which results from iterated equations, the unconfoundedness assumption conditional on $\vX^{\star}_i$ and $\mathbf{U}_i$ and consistency, the independence between $Z_i$ and $G_i$ given $\vX^{\star}_i$ leads to
\begin{align*}
&=\bigg\{\sum_{\vx \in \mathcal{X }^{\star}}\bigg(\sum_{g\in \mathcal{G}}\sum_{\mathbf{u} \in \mathcal{U }}E[Y_i| Z_i=1, G_i=g,\mathbf{U}_i=\mathbf{u}, \vX^{\star}_i=\vx , i\in V_g ]\\
&\qquad \qquad \qquad \qquad \qquad P(\mathbf{U}_i=\mathbf{u}|Z_i=1,G_i=g, \vX^{\star}_i=\vx)P(G_i=g|\vX^{\star}_i=\vx)\\
&\qquad \qquad \qquad \qquad - E[Y_i| Z_i=0, G_i=g,\mathbf{U}_i=\mathbf{u}, \vX^{\star}_i=\vx , i\in V_g ]\\
&\qquad \qquad \qquad \qquad P(\mathbf{U}_i=\mathbf{u}|Z_i=0,G_i=g, \vX^{\star}_i=\vx)P(G_i=g|\vX^{\star}_i=\vx)\bigg) P(\vX^{\star}_i=\vx)\bigg\}\\
&\quad- \bigg\{\sum_{\vx \in \mathcal{X }^{\star}}\sum_{g\in \mathcal{G}}\sum_{\mathbf{u} \in \mathcal{U }}\bigg(E[Y_i|Z_i=1, G_i=g,\mathbf{U}_i=\mathbf{u}, \vX^{\star}_i=\vx , i\in V_g ]\\&\hspace{4cm}-E[Y_i|Z_i=0, G_i=g,\mathbf{U}_i=\mathbf{u}, \vX^{\star}_i=\vx , i\in V_g ]\bigg)\\
&\qquad \qquad \qquad \qquad \qquad P(\mathbf{U}_i=\mathbf{u}|G_i=g, \vX^{\star}_i=\vx)P(G_i=g|\vX^{\star}_i=\vx)P(\vX^{\star}_i=\vx)\bigg\}
\end{align*}
where we can just marginalize over the distribution of $G_i$, yielding
\begin{align*}
&=\bigg\{\sum_{\vx \in \mathcal{X }^{\star}}\bigg(\sum_{\mathbf{u} \in \mathcal{U }}E[Y_i| Z_i=1, \mathbf{U}_i=\mathbf{u}, \vX^{\star}_i=\vx \, ]P(\mathbf{U}_i=\mathbf{u}|Z_i=1,\vX^{\star}_i=\vx)\\
&\qquad \qquad \qquad \qquad - E[Y_i| Z_i=0, \mathbf{U}_i=\mathbf{u}, \vX^{\star}_i=\vx \, ]P(\mathbf{U}_i=\mathbf{u}|Z_i=0, \vX^{\star}_i=\vx) \bigg) P(\vX^{\star}_i=\vx)\bigg\}\\
&\quad- \bigg\{\sum_{\vx \in \mathcal{X }^{\star}}\sum_{\mathbf{u} \in \mathcal{U }}\bigg(E[Y_i|Z_i=1,\mathbf{U}_i=\mathbf{u}, \vX^{\star}_i=\vx \, ]-E[Y_i|Z_i=0,\mathbf{U}_i=\mathbf{u}, \vX^{\star}_i=\vx \, ]\bigg)\\
&\qquad \qquad \qquad \qquad \qquad P(\mathbf{U}_i=\mathbf{u}| \vX^{\star}_i=\vx)P(\vX^{\star}_i=\vx)\bigg\}
\end{align*}
Now we can just proceed as we would do to derive the bias given by unmeasured confounders. By adding and subtracting two quantities $E[Y_i| Z_i=1, \mathbf{U}_i=\mathbf{u}',\vX^{\star}_i=\vx \, ]$ and $E[Y_i| Z_i=0, \mathbf{U}_i=\mathbf{u}',\vX^{\star}_i=\vx \, ]$ we get
\begin{equation*}
\begin{aligned}
&=\sum_{\vx \in \mathcal{X }^{\star}}\bigg(\sum_{\mathbf{u}\in \mathcal{U}}E[Y_i|Z_i=1, \mathbf{U}_i=\mathbf{u}, \vX^{\star}_i=\vx \, ]-E[Y_i|Z_i=1, \mathbf{U}_i=\mathbf{u}', \vX^{\star}_i=\vx \, ]\bigg)\\
& \qquad \qquad\qquad \bigg( P(\mathbf{U}_i=\mathbf{u}|Z_i=1,\vX^{\star}_i=\vx)-P(\mathbf{U}_i=\mathbf{u}|\vX^{\star}_i=\vx)\bigg)P(\vX^{\star}_i=\vx)\\
&\quad -\sum_{\vx \in \mathcal{X }^{\star}}\bigg(\sum_{\mathbf{u}\in \mathcal{U}}E[Y_i|Z_i=0, \mathbf{U}_i=\mathbf{u}, \vX^{\star}_i=\vx \, ]-E[Y_i|Z_i=0, \mathbf{U}_i=\mathbf{u}', \vX^{\star}_i=\vx \, ]\bigg)\\
& \qquad \qquad\qquad\bigg( P(\mathbf{U}_i=\mathbf{u}|Z_i=0,\vX^{\star}_i=\vx)-P(\mathbf{U}_i=\mathbf{u}|\vX^{\star}_i=\vx)\bigg)P(\vX^{\star}_i=\vx)
\end{aligned}
\end{equation*}
 If there is no interaction between the individual treatment $Z_i$ and $\mathbf{U}_i$, we can have a simplified expression of the bias formula:
\begin{equation*}
\begin{aligned}
&=\sum_{\vx \in \mathcal{X }^{\star}}\bigg(\sum_{\mathbf{u}\in \mathcal{U}}E[Y_i|Z_i=z, \mathbf{U}_i=\mathbf{u}, \vX^{\star}_i=\vx \, ]-E[Y_i|Z_i=z, \mathbf{U}_i=\mathbf{u}', \vX^{\star}_i=\vx \, ]\bigg)\\
& \qquad \qquad\qquad \bigg( P(\mathbf{U}_i=\mathbf{u}|Z_i=1,\vX^{\star}_i=\vx)-P(\mathbf{U}_i=\mathbf{u}|Z_i=0,\vX^{\star}_i=\vx)\bigg)P(\vX^{\star}_i=\vx)\\
\end{aligned}
\end{equation*}

\subsection{Proof of Corollary \ref{cor: bias4}}

\begin{equation*}
\begin{aligned}
\tau_{X^{\star}}^{obs}-\tau&=\bigg\{\sum_{\vx \in \mathcal{X }^{\star}}\bigg(E[Y_i| Z_i=1,\vX^{\star}_i=\vx \, ]- E[Y_i| Z_i=0,\vX^{\star}_i=\vx \, ]\bigg) P(\vX^{\star}_i=\vx)\bigg\}\\
&\quad- \bigg\{\sum_{g\in \mathcal{G}}\bigg(E[Y_i(1, g)Y_i(0, g)| i\in V_g]\bigg)P(G_i=g)\bigg\} \\
&=\bigg\{\sum_{\vx \in \mathcal{X }^{\star}}\bigg(E[Y_i| Z_i=1,\vX^{\star}_i=\vx \, ]- E[Y_i| Z_i=0,\vX^{\star}_i=\vx \, ]\bigg) P(\vX^{\star}_i=\vx)\bigg\}\\
&\quad-\big(E[Y_i(1)]-E[Y_i(0)]\big)\\
&=\bigg\{\sum_{\vx \in \mathcal{X }^{\star}}\bigg(E[Y_i| Z_i=1,\vX^{\star}_i=\vx \, ]- E[Y_i| Z_i=0,\vX^{\star}_i=\vx \, ]\bigg) P(\vX^{\star}_i=\vx)\bigg\}\\
&\quad-\big(E[Y_i(1)|Z_i=1]-E[Y_i(0)Z_i=0]\big)
\end{aligned}
\end{equation*}
by SUTVA and by definition of $V_g$, which results in $P(G_i=g)=P(G_i=g|\in V_g)P(i\in V_g)$. Now we can just proceed as we would do to derive the bias given by unmeasured confounders.
By iterated equations and applying the unconfoudedness assumption we get
\begin{equation*}
\begin{aligned}
&=\bigg\{\sum_{\vx \in \mathcal{X }^{\star}}\bigg(E[Y_i| Z_i=1,\vX^{\star}_i=\vx \, ]- E[Y_i| Z_i=0,\vX^{\star}_i=\vx \, ]\bigg) P(\vX^{\star}_i=\vx)\bigg\}\\
&\quad- \bigg\{\sum_{\vx \in \mathcal{X }^{\star}}\bigg(E[Y_i(1)Z_i=1]-E[Y_i(0)|Z_i=0]\bigg)P(\vX^{\star}_i=\vx)\bigg\} \\
&=\bigg\{\sum_{\vx \in \mathcal{X }^{\star}}\bigg(E[Y_i| Z_i=1,\vX^{\star}_i=\vx \, ]- E[Y_i| Z_i=0,\vX^{\star}_i=\vx \, ]\bigg) P(\vX^{\star}_i=\vx)\bigg\}\\
&\quad- \bigg\{\sum_{\vx \in \mathcal{X }^{\star}}\bigg(E[Y_i|Z_i=1]-E[Y_i|Z_i=0]\bigg)P(\vX^{\star}_i=\vx)\bigg\}
\end{aligned}
\end{equation*}
By adding and subtracting two quantities $E[Y_i| Z_i=1, \mathbf{U}_i=\mathbf{u}',\vX^{\star}_i=\vx \, ]$ and $E[Y_i| Z_i=0, \mathbf{U}_i=\mathbf{u}',\vX^{\star}_i=\vx \, ]$ we get
\begin{equation*}
\begin{aligned}
&=\sum_{\vx \in \mathcal{X }^{\star}}\bigg(\sum_{\mathbf{u}\in \mathcal{U}}E[Y_i|Z_i=1, \mathbf{U}_i=\mathbf{u}, \vX^{\star}_i=\vx \, ]-E[Y_i|Z_i=1, \mathbf{U}_i=\mathbf{u}', \vX^{\star}_i=\vx \, ]\bigg)\\
& \qquad \qquad\qquad \bigg( P(\mathbf{U}_i=\mathbf{u}|Z_i=1,\vX^{\star}_i=\vx)-P(\mathbf{U}_i=\mathbf{u}|\vX^{\star}_i=\vx)\bigg)P(\vX^{\star}_i=\vx)\\
&\quad -\sum_{\vx \in \mathcal{X }^{\star}}\bigg(\sum_{\mathbf{u}\in \mathcal{U}}E[Y_i|Z_i=0, \mathbf{U}_i=\mathbf{u}, \vX^{\star}_i=\vx \, ]-E[Y_i|Z_i=0, \mathbf{U}_i=\mathbf{u}', \vX^{\star}_i=\vx \, ]\bigg)\\
& \qquad \qquad\qquad\bigg( P(\mathbf{U}_i=\mathbf{u}|Z_i=0,\vX^{\star}_i=\vx)-P(\mathbf{U}_i=\mathbf{u}|\vX^{\star}_i=\vx)\bigg)P(\vX^{\star}_i=\vx)
\end{aligned}
\end{equation*}
 If there is no interaction between the individual treatment $Z_i$ and $\mathbf{U}_i$, we can have a simplified expression of the bias formula:
\begin{equation*}
\begin{aligned}
&=\sum_{\vx \in \mathcal{X }^{\star}}\bigg(\sum_{\mathbf{u}\in \mathcal{U}}E[Y_i|Z_i=z, \mathbf{U}_i=\mathbf{u}, \vX^{\star}_i=\vx \, ]-E[Y_i|Z_i=z, \mathbf{U}_i=\mathbf{u}', \vX^{\star}_i=\vx \, ]\bigg)\\
& \qquad \qquad\qquad \bigg( P(\mathbf{U}_i=\mathbf{u}|Z_i=1,\vX^{\star}_i=\vx)-P(\mathbf{U}_i=\mathbf{u}|Z_i=0,\vX^{\star}_i=\vx)\bigg)P(\vX^{\star}_i=\vx)\\
\end{aligned}
\end{equation*}

\subsection{Proof of Proposition \ref{prop:balance}}

We have to show that $P(Z_i=z, G_i=g| \vX_i, \psi(z; g; \vX_i))=P(Z_i=z, G_i=g|\psi(z; g; \vX_i))$. First consider the left hand side:
\[P(Z_i=z, G_i=g| \vX_i,  \psi(z; g; \vX_i))=P(Z_i=z, G_i=g| \vX_i)=\psi(z; g; \vX_i)\]
where the first equality follows because the joint propensity score is a function of $\vX_i$ and the second is by the definition of the joint propensity score. Second, consider the right hand side. By iterated equations we have that
\begin{equation*}
\begin{aligned}
P(Z_i=z, G_i=g| \psi(z; g; \vX_i))&=E_{\vX}[P(Z_i=z, G_i=g| \vX_i, \psi(z; g; \vX_i))|\psi(z; g; \vX_i)]\\
&=E_{\vX}[P(Z_i=z, G_i=g| \vX_i)|\psi(z; g; \vX_i)]\\&=E_{\vX}[\psi(z; g; \vX_i)|\psi(z; g; \vX_i)]=\psi(z; g; \vX_i)
\end{aligned}
\end{equation*}
where the second equality is also because the joint propensity score is a function of $\vX_i$, the third equality is by the definition of the joint propensity score and the fourth is equality is just trivial.

\subsection{Proof of Proposition \ref{prop:PSZuncon}}

We have to show that if Assumption \ref{ass: Totunconf} holds then $P(Z_i=z, G_i=g| Y_i(z,g),   \psi(z; g; \vX_i))=P(Z_i=z,G_i=g|  \psi(z; g; \vX_i))$.The proof proceeds by showing that both the left and the right hand sides of this equation are equal to the joint propensity score itself and, hence, they are also equal to each other. In doing so we will make use of the assumption of unconfoundedness \ref{ass: Totunconf}.
Notice that in the proof of the Balancing Property we have already shown
that the right hand side of the equation is equal to the propensity score, i.e., $P(Z_i=z, G_i=g|  \psi(z; g; \vX_i))= \psi(z; g; \vX_i)$. Now to prove that $P(Z_i=z, G_i=g| Y_i(z,g),  \psi(z; g; \vX_i))= \psi(z; g; \vX_i)$ we use the law of iterated equations:
\begin{equation*}
\begin{aligned}
P(Z_i=z,G_i=g|Y_i(z,g),   \psi(z; g; \vX_i))&=E_{\vX}[P(Z_i=z,G_i=g| \vX_i, Y_i(z,g),  \psi(z; g; \vX_i))|Y_i(z,g),  \psi(z; g; \vX_i)]\\
&=E_{\vX}[P(Z_i=z,G_i=g| Y_i(z,g),\vX_i)|Y_i(z,g),  \psi(z; g; \vX_i)]\\&=E_{\vX}[P(Z_i=z,G_i=g| \vX_i)|Y_i(z,g),  \psi(z; g; \vX_i)]\\&=E_{\vX}[ \psi(z; g; \vX_i)| Y_i(z,g), \psi(z; g; \vX_i)]= \psi(z; g; \vX_i)
\end{aligned}
\end{equation*}
where the second equality holds because $ \psi(z; g; \vX_i)$ is a function of $\vX_i$ and the third follows from assumption \ref{ass: Totunconf}.

\subsection{Proof of Proposition \ref{prop:2PSuncon}}

We have to show that if Assumption \ref{ass: Totunconf} holds given $\vX_i$ then
$P(Z_i=z, G_i=g| Y_i(z,g),   \lambda(z; g; \vX^g_i), \phi(1; \vX^z_i))=P(Z_i=z, G_i=g| \lambda(z; g; \vX^g_i), \phi(1; \vX^z_i))$.
First, by definition of $\vX^z_i$ and $\vX^g_i$ we can write $ \phi(1; \vX^z_i)=P(Z_i=1| \vX_i)=P(Z_i=1| \vX^z_i)$ and $\lambda(z; g; \vX^g_i)=P(G_i=g| Z_i=z, \vX_i)=P(G_i=g| Z_i=z, \vX^g_i)$. Now, the proof proceeds by showing that both the left and the right hand sides of this equation are equal to the joint propensity score itself and, hence, they are also equal to each other. Let us consider the right hand side of the equation. By iterated equations we have that
\begin{equation*}
\begin{aligned}
P(Z_i=z, G_i=g| &\lambda(z; g; \vX^g_i), \phi(1; \vX^z_i))\\
&=E_{\vX}[P(Z_i=z, G_i=g| \vX_i, \lambda(z; g; \vX^g_i), \phi(1; \vX^z_i))|\lambda(z; g; \vX^g_i), \phi(1; \vX^z_i)]\\
&=E_{\vX}[P(Z_i=z, G_i=g| \vX_i)|\lambda(z; g; \vX^g_i), \phi(1; \vX^z_i)]\\&=E_{\vX}[\psi(z; g; \vX_i)|\lambda(z; g; \vX^g_i), \phi(1; \vX^z_i)]=\psi(z; g; \vX_i)
\end{aligned}
\end{equation*}
where we have used the result that both $\lambda(z; g; \vX^g_i)$ and $\phi(1; \vX^z_i)$ can be seen as functions of $\vX_i$ and the second equality is given Assumption \ref{ass: Totunconf}. The third equality is straightforward, given that by factorization $\psi(z; g; \vX_i)=\lambda(z; g; \vX^g_i)\phi(z; \vX^z_i)$ and that the value of the propensity score $\phi(z; \vX^z_i)$ for the binary treatment $Z_i$ is defined given $\phi(1; \vX^z_i)=1-\phi(0; \vX^z_i)$.
Finally, to prove that the left hand side $P(Z_i=z, G_i=g| Y_i(z,g),  \psi(z; g; \vX_i))= \psi(z; g; \vX_i)$ we use the law of iterated equations:
\begin{equation*}
\begin{aligned}
P(Z_i=z,&G_i=g|Y_i(z,g),  \lambda(z; g; \vX^g_i), \phi(1; \vX^z_i))\\
&=E_{\vX}[P(Z_i=z,G_i=g| \vX_i, Y_i(z,g),  \lambda(z; g; \vX^g_i), \phi(1; \vX^z_i))|Y_i(z,g),  \lambda(z; g; \vX^g_i), \phi(1; \vX^z_i)]\\
&=E_{\vX}[P(Z_i=z,G_i=g| Y_i(z,g),\vX_i)|Y_i(z,g),  \lambda(z; g; \vX^g_i), \phi(1; \vX^z_i)]\\&=E_{\vX}[P(Z_i=z,G_i=g| \vX_i)|Y_i(z,g), \lambda(z; g; \vX^g_i), \phi(1; \vX^z_i)]\\&=E_{\vX}[ \psi(z; g; \vX_i)| Y_i(z,g), \lambda(z; g; \vX^g_i), \phi(1; \vX^z_i)]= \psi(z; g; \vX_i)
\end{aligned}
\end{equation*}
where the second equality holds because $\lambda(z; g; \vX^g_i)$ and  $\phi(1; \vX^z_i)$ are functions of $\vX_i$, the third follows from assumption \ref{ass: Totunconf}, and the last equality is given by the same argument as above.

\end{document}